\documentclass[twoside,leqno,twocolumn]{article}
\usepackage[letterpaper]{geometry}
\usepackage[T1]{fontenc}
\usepackage{thm-restate}

\usepackage{ltexpprt}
\usepackage{hyperref}
\usepackage{xspace}

\usepackage[subrefformat=simple,labelformat=simple]{subcaption}

\usepackage[section]{algorithm}
\usepackage{algorithmicx}
\usepackage{algpseudocode}
\usepackage{graphicx}
\usepackage{amsmath}
\usepackage{amsfonts}
\usepackage{amssymb}
\usepackage{graphicx}
\usepackage{tikz}
\usepackage{stmaryrd}
\usepackage{float}
\usepackage{caption}
\usepackage{enumerate}
\usepackage{autobreak}
\usepackage{cases}
\usepackage{placeins}
\usepackage{algorithm}
\usepackage{algpseudocode}
\usepackage{amsmath}
\usepackage{layouts}
\usepackage{physics}
\usepackage{dsfont}
\usepackage{mathtools}
\usepackage{enumitem}
\usepackage{booktabs}
\usepackage{tabularx}
\usepackage[capitalise, noabbrev]{cleveref}
\usepackage[switch]{lineno}
\linenumbers
\crefname{algocf}{Algorithm}{Algorithms}
\Crefname{algocf}{Algorithm}{Algorithms}
\crefname{@theorem}{Theorem}{Theorems}
\Crefname{@theorem}{Theorem}{Theorems}

\usepackage{crossreftools}
\usepackage[% These options define the look of numbers and units in the text:
group-separator = {,},  % use a ',' instead of space as separator
group-minimum-digits=4  % already add a separation for 4-digit numbers
]{siunitx}
\algdef{SE}[DOWHILE]{Do}{doWhile}{\algorithmicdo}[1]{\algorithmicwhile\ #1}%
\crefname{claim}{Claim}{Claims}
\DeclarePairedDelimiter{\ceil}{\lceil}{\rceil}
\definecolor{falured}{rgb}{0.5, 0.09, 0.09}
\definecolor{pinegreen}{rgb}{0.0, 0.47, 0.44}
\definecolor{forestgreen}{rgb}{0.0, 0.27, 0.13}

\newcommand{\R}{\mathbb{R}}
\newcommand{\basis}{B}
\newcommand{\nonbasis}{N}
\newcommand{\AB}{A_{\basis}}
\newcommand{\AN}{A_{\nonbasis}}

\newcommand{\alg}{\ensuremath{\mathcal{A}}}
\newcommand{\cost}{\ensuremath{\mathcal{C}}}
\newcommand{\OPolyLog}{\Tilde{\mathcal{O}}}
\newcommand{\contr}[1]{\mathrm{ctrl}_{#1}\mathrm{-}}
\DeclareMathOperator{\diag}{diag}
\DeclareMathOperator{\sign}{sign}
\newcommand{\qae}{\mathrm{QAE}}
\newcommand{\qaa}{\mathrm{QAA}}
\newcommand{\oaa}{\mathrm{OAA}}
\newcommand{\qpe}{\mathrm{QPE}}
\newcommand{\qft}{\mathrm{QFT}}
\newcommand{\qls}{\mathrm{QLS}}
\newcommand{\qlsa}{\mathrm{QLS}}
\newcommand{\qlsaFourier}{\mathrm{QLSA - Fourier}}
\newcommand{\lcu}{\mathrm{LCU}}

\newcommand{\simplexIter}{\mathrm{SimplexIter}}
\newcommand{\qmin}{\mathrm{QMin}}
\newcommand{\qsearch}{\mathrm{QSearch}}
\newcommand{\isOptimal}{\mathrm{IsOptimal}}
\newcommand{\findColumn}{\mathrm{FindColumn - Random}}
\newcommand{\isUnbounded}{\mathrm{IsUnbounded}}
\newcommand{\findRow}{\mathrm{FindRow}}
\newcommand{\steepestEdgeRule}{\mathrm{FindColumn - QStER}}
\newcommand{\steepestEdgeCompare}{\mathrm{QStEC}}
\newcommand{\dantzig}{\mathrm{FindColumn - QDanR}}
\newcommand{\steepestEdgeRuleNoMath}{FindColumn - QStER}
\newcommand{\findColumnNoMath}{FindColumn - Random}
\newcommand{\dantzigNoMath}{FindColumn - QDanR}

\newcommand{\qsinf}{QSearch$_\infty$}
\newcommand{\domk}{\mathcal{D}_{\kappa}}

\newcommand{\bor}{\ensuremath{\mathcal{P}_b}}

\newcommand{\old}[1]{{}}

\newlength{\arrow}
\newcommand{\easy}{\textsc{easy}}
\newcommand{\hard}{\textsc{hard}}
\newcommand{\netlib}{\textsc{NETLIB}}
\newcommand{\stochlp}{\textsc{STOCHLP}}
\newcommand{\miplib}{\textsc{MIPLIB}}
\newcommand{\misc}{\textsc{MISC}}

\begin{document}

% Default Copyright Statement
\fancyfoot[R]{\scriptsize{Copyright \textcopyright\ 2023\\
Copyright for this paper is retained by authors.}}

\newlist{researchQuestion}{enumerate}{10}
\setlist[researchQuestion]{label*=RQ\arabic*}

\crefname{researchQuestioni}{}{}
\Crefname{researchQuestioni}{}{}

\clubpenalty 100000
\widowpenalty 100000

%\author{N. Onymous}
\title{Realistic Runtime Analysis for Quantum Simplex Computation\thanks{This work was supported by 
the German Federal Ministry of Education and Research (BMBF), project QuBRA, and the German Federal Ministry for Economic Affairs and Climate Action (BMWK), project ProvideQ.
DG and PG are also supported by Germany's Excellence Strategy -- Cluster of Excellence Matter and Light for Quantum Computing (ML4Q)
EXC 2004/1 (390534769).
}
} %TODO Please add

%\titlerunning{Dummy short title} %TODO optional, please use if title is longer than one line

\author{Sabrina Ammann\thanks{{Department of Mathematics, TU Braunschweig, Germany. }{\{s.ammann,sebastian.stiller,t.de-wolff\}@tu-bs.de}{}}
\and{Sándor P. Fekete}\thanks{{Department of Computer Science, TU Braunschweig, Germany. }{\{s.fekete,m.perk\}@tu-bs.de}{}}
\and{Paulina L. A. Goedicke}\thanks{{Institute for Theoretical Physics, Universität zu Köln, Germany. }{\{paulina.goedicke,david.gross\}@thp.uni-koeln.de}{}}
\and{David Gross}\footnotemark[3]%{{Institute for Theoretical Physics, Universität zu Köln, Germany. }{david.gross@thp.uni-koeln.de}{}}
\and{Maximilian Hess}\thanks{{Supply Chain Innovation, Infineon Technologies AG, Germany. }{Maximilian.Hess@infineon.com}{}}
\and{Andreea Lefterovici}\thanks{{Department of Physics, Leibniz Universität Hannover, Germany. }{\\ \hspace*{1cm}\{andreea.lefterovici,tobias.osborne,debora.ramacciotti,antonio.rotundo,shawn.skelton\}@itp.uni-hannover.de}{}}
\and{Tobias J.\ Osborne}\footnotemark[5]%\thanks{{Department of Physics, Leibniz Universität Hannover, Germany. }{tobias.osborne@itp.uni-hannover.de}{}}
\and{Michael Perk}\footnotemark[2]%{{Department of Computer Science, TU Braunschweig, Germany. }{m.perk@tu-bs.de}{}}
\and{Debora Ramacciotti}\footnotemark[5]%\thanks{{Department of Physics, Leibniz Universität Hannover, Germany. }{debora.ramacciotti@itp.uni-hannover.de}{}}
\and{Antonio Rotundo}\footnotemark[5]%\thanks{{Department of Physics, Leibniz Universität Hannover, Germany. }{antonio.rotundo@itp.uni-hannover.de}{}}
\and{S. E. Skelton}\footnotemark[5]%\thanks{{Leibniz Universität Hannover, Germany. }{shawn.skelton@itp.uni-hannover.de}{}}
\and{Sebastian Stiller}\footnotemark[1]%\thanks{{Department of Mathematics, TU Braunschweig, Germany. }{sebastian.stiller@tu-bs.de}{}}
\and{Timo de Wolff}\footnotemark[1]%\thanks{{Department of Mathematics, TU Braunschweig, Germany. }{t.de-wolff@tu-bs.de}{}}
}
\nolinenumbers %uncomment to disable line numbering

\date{}

\newcommand\relatedversion{}

\date{}

\onecolumn
\maketitle

% Copyright Statement
% When submitting your final paper to a SIAM proceedings, it is requested that you include
% the appropriate copyright in the footer of the paper.  The copyright added should be
% consistent with the copyright selected on the copyright form submitted with the paper.
% Please note that "20XX" should be changed to the year of the meeting.

% Default Copyright Statement
\fancyfoot[R]{\scriptsize{
    Copyright \textcopyright\ 2023 by the authors\\
Unauthorized reproduction of this article is prohibited}}

\begin{abstract}
In recent years, strong expectations have been raised for the possible power of
quantum computing for solving difficult optimization problems, based on
theoretical, asymptotic worst-case bounds. Can we expect this to have
consequences for Linear and Integer Programming when solving instances of
practically relevant size, a fundamental goal of Mathematical Programming,
Operations Research and Algorithm Engineering?
Answering this question faces a crucial impediment: The lack of sufficiently
large quantum platforms prevents performing real-world tests for comparison
with classical methods.  

In this paper, we present a quantum analog for
classical runtime analysis when solving real-world instances of important
optimization problems.  To this end, we measure the \emph{expected practical
performance} of quantum computers by analyzing the expected \emph{gate
complexity} of a quantum algorithm.  The lack of practical quantum platforms
for experimental comparison is addressed by \emph{hybrid} benchmarking, in
which the algorithm is performed on a classical system, logging the expected
cost of the various subroutines that are employed by the quantum versions.  In
particular, we provide an analysis of quantum methods for Linear Programming,
for which recent work has provided asymptotic speedup through quantum
subroutines for the Simplex method.  We show that a practical quantum advantage
for realistic problem sizes would require quantum gate operation times that are
considerably below current physical limitations.


\end{abstract}
\twocolumn

%\clearpage

\section{Introduction}
\label{sec:intro}

Measuring the performance of computational methods is at the heart of
of quantitative science. On the one hand, analyzing and improving the asymptotic worst-case complexity
has been leading the way for \emph{theoretical} research;
on the other, measuring the computational performance for a set
of well-chosen benchmark instances has been a driving force 
for progress on the \emph{practical} side. These two approaches
often complement each other, but do not necessarily lead 
to the same conclusions, as illustrated by the classical problem
of Linear Programming: While the ground-breaking ellipsoid method
by Khachiyan~\cite{ellipsoid} 
was the first algorithm with polynomial-time worst-case complexity
(proving that such algorithms do exist), it is of little practical 
use. On the other hand, the Simplex method by Dantzig~\cite{dantzig} 
may have an exponential worst-case runtime, but is the method of
choice for many practical optimization problems, including
as a subroutine for the solution of NP-hard combinatorial
optimization problems that can be expressed as 
linear programs with integer variables. 

Since the early days of Dantzig and Khachiyan,
the solution of ever larger instances of optimization problems
has witnessed tremendous development, not just for instances 
of Linear Programming (for which 15 years of progress 
in both hardware and algorithms were sufficient 
to achieve a speedup of six orders of magnitude~\cite{bixby2002solving}),
but also for solving large instances of 
NP-hard optimization problems. However, such gains
remain elusive for many important problems, reflecting
the asymptotic worst-case behavior even at moderate instance sizes.

In recent years, the prospect of real-world quantum computing 
(based on exploiting fundamentally different computing paradigms)
has raised the hope of applicable progress for such difficult
optimization problems, fueling an increasing amount of theoretical
work for the development of algorithmic methods.
A key aspect has been the analysis of asymptotic complexity,
often with improvements in theoretical worst-case runtime.
At this time, running benchmark experiments does not
yet provide a realistic alternative, as existing hardware platforms
are still in their infancy. Developing them to the point
of practical usefulness will involve tremendous effort and
expenses, so estimating their future performance ahead of
time is highly desirable.

How can one gauge the performance of a hardware platform that has not yet been
realized?  In this paper, we employ \emph{hybrid benchmarking}, a fundamental
alternative to asymptotic worst-case analysis, whereby the practical
performance of a quantum algorithm is estimated on the basis of 
generous analytic estimates on the gate complexity of standard quantum algorithms.
%\todo{I have concerns with our charicterization as 'practical' or 'realistic'. As is established in sections3, 4, our analysis is exceedingly generous to qunatum algorithms. Maybe 'generous analytic estimates' or 'optimistic analytic estimates'?}
%analytic instance-specific lower bounds for quantum gate counts. 
This is similar in spirit to recent work by Cade et al.~\cite{cade2023quantifying,cade2022practical}.
Here we use hybrid benchmarking to achieve an analysis of a quantum method for Linear Programming,
for which a recent paper by Nannicini~\cite{nannicini2022fast}
has provided an asymptotic speedup through quantum subroutines for the Simplex method.

\medskip
\paragraph*{Our Results}
%\todo[inline]{'expected practical performance' feels a bit strong...perhaps 'expected performance on a fault tolerant device'? Similarly can we shift 'analyzing quantum methods for linear programming' to 'analyzing a quantum method for linear programming'}
%Sandor: No, this is not strong; we are talking about LOWER bounds, and of course that is still valid if the quantum computer has to deal with additional aspects that make it worse.
\begin{itemize}
\item We describe an approach for gauging the practical performance
of a fully fault tolerant quantum device for solving real-world instances of optimization problems.
\item We provide a specific demonstration of this technique 
by analyzing a quantum method for linear programming.
\item We perform a concrete study by evaluating the performance of the quantum Simplex algorithm
for an established library of benchmark instances.
\item We show that a practical quantum advantage for realistic problem sizes would
require quantum gate operation times that are considerably below current
physical limitations.
%\item We show that practical application of such quantum methods would require switching times for quantum gates that are roughly 4 orders of magnitude smaller than those realizable in the foreseeable future. 
\end{itemize}

This main part of our paper focuses on the algorithmic analysis; our overall approach 
combines methods from classical mathematical programming and quantum
computing, so we provide involved scientific details in 
\cref{sec:details-simplex} (for algorithmic and mathematical subroutines).
Further details of the underlying quantum subroutines and the ensuing
mathematical analysis form a separate paper from the realm of 
theoretical physics, which is enclosed as 
\cref{sec:details-quantum-subroutines}.

\paragraph*{Related Work}
Cade~et~al.~\cite{cade2023quantifying,cade2022practical} presented an approach
%published a remarkable work introducing hybrid benchmarking, 
for computing bounds on the expected and worst-case number of oracle calls in quantum subroutines.
Their technique was applied to synthetic benchmarks of \textsc{MAX-$k$-SAT} to estimate the resources required for 
quantum unstructured search and maximum finding which replace classical subroutines. 
This should be contrasted with the techniques of,  e.g., Tacla~et~al.~\cite{tacla2023majorizationbased}, which consider the real-world performance of distinct quantum hardware against basic gate sets. 
%proposed a quantum benchmarking technique under assumptions 
%about the architecture of quantum devices. % and the 'least expensive' $1,2$-qubit gates to implement.
%Other metrics such as \emph{quantum volume} \cite{Cross_2019} have been proposed to measure and 
%compare system-wide gate error rates.
Unlike hardware-based benchmarking techniques, our hybrid benchmarking assumes access to fully fault-tolerant hardware and computes a lower bound on the expected gate complexity. Further details are elaborated in Section~\ref{sec:lb}.
%\todo{Marker for SF: maybe build a bridge to Section 3.}
%We extend this approach to benchmark instances, and count the gate, rather than oracle query complexity.
%Attempts were made to minimize the circuit depth \cite{dalzell2022endtoend}.

Many quantum algorithms rely on quantum subroutines like
$\qsearch$~\cite{boyer1998tight,cade2023quantifying}, 
quantum amplitude estimation $\qae$~\cite{boyer1998tight}, quantum phase estimation $\qpe$ \cite{cleve1998quantum}, quantum linear solvers QLS such as HHL \cite{harrow2009quantum} or QLSA \cite{childs2017quantum} and quantum minimum finding $\qmin$~\cite{durr1999quantum,cade2023quantifying}
to achieve better worst-case scalings than their classical counterparts. 
Quantum algorithms for semidefinite programming such as 
\cite{augustino2022quantum,brandao2017quantum,van2020quantum} can be applied to linear system problems.
Quantum interior point methods (qIPM) \cite{casares2020quantum,dalzell2022endtoend}
have also been proposed.
Casares~et~al.~\cite{casares2020quantum} embeds Grover and a QLS as subroutines of an otherwise classical algorithm to obtain an 
asymptotic speed-up,
%in the number of cost variables at the expense of poorer $\epsilon$ scaling, 
while Dalzell~et~al.~\cite{dalzell2022endtoend} 
analyze the qIPM of earlier work \cite{Kerenidis_2020,Kerenidis_2021} for small %($n=10^2$) 
randomly selected portfolio optimization problems,
%The authors find no quantum advantage and identified several bottleneck steps, 
%including the tomography needed to extract a solution from their chosen qls for matrix $G$, and the the relative size of the 
%Frobenius condition number of the matrix used in qls.
aiming at a thorough estimation of the leading order resources required to run the quantum routine on real-world instances. %(historic stock data).
For other recent related work, see~\cite{campbell2019applying,babbush2021focus}.
%Unlike our analysis which lower bounds the gate complexity of each subroutine, \cite{dalzell2022endtoend} simulates each 
%instance of the qIPM and numerically estimates each parameter scaling with respect certain quantum resources measures.
Quantum subroutines for the primal Simplex algorithm have been proposed by Nannicini~\cite{nannicini2022fast}, who
proves an asymptotic speedup over the classical primal Simplex method. 
Importantly, Nannicini's method works for classical and quantum inputs, %and does not use ``black box'' matrix access oracles, 
and thus is well suited for an examination from a practical point of view.
With current hardware designs, the proposed algorithm would be vulnerable to a latency problem 
%in the switching step, %expected to scale with the system size. 
as well as to other challenges, such as preparing oracles to encode (classical) data on the circuit, decoherence 
and gate error~\cite{jordan2021implementing}. 
Presuming that future technologies will overcome the many challenges to creating fully fault tolerant quantum devices, we apply the
hybrid benchmarking technique to Nannicini's quantum subroutine and compare it
to classical Simplex solvers, which have made tremendous progress in practical
performance over many years~\cite{bixby2002solving}.

%Chew et al.~\cite{chew2022ultrafast} achieved 
The fastest isolated quantum gate operation to date is
taking $6.5\cdot 10^{-9}$s for two physically realized qubits~\cite{chew2022ultrafast}.
Because any quantum computer requires 
classical control hardware by arbitrary waveform generators, there is a practical
limit supplied by the bandwidth of these devices, with a current best resolution in the $1-100$GHz regime.
Thus, it is reasonable to expect gate times limited for the foreseeable future to
above the $10^{-10}$s timescale.

%Because of the nature of quantum computing, 
%these gate operations (as the analog to elementary operations
%in classical computing) cannot be carried out in isolation
%(in fact, they require proximity of the involved particles),
%so focusing on gate operations is a benevolent measure
%for quantum computing runtime.
%, and provides generous lower bounds for the runtime. 

%classical improvements to the Simplex algorithm, for example the dual Simplex, can produce solutions well under 
%this asymptotic worst case \cite{bixby2002solving}.
%For the sake of a clear and relatively simple comparison, we restrict our analysis to the Simplex algorithm of \cite{nannicini2022fast}.
%However, most of the algorithms discussed above use QLSA and/or Grover subroutines, and so our 
%analysis can find applications outside of analyzing Simplex solvers.

\section{Preliminaries}
\label{sec:prelim}
% This will be written last.
% Add additional items here if necessary

\subsection{The Simplex Algorithm}
We briefly recall the simplex algorithm to fix notation and improve cross-community readability. The simplex algorithm solves a linear optimization problem
of the form $\min c^T x$ with constraints $Ax=b, x\geq 0$,
where~${A\in \R^{m\times n}}$, ${c\in \R^{n}}$, ${b\in \R^{m}}$. 
%\PG{c should be in $R^{n}$, right?}
A subset of $m$ linearly independent columns of $A$ is called a \emph{basis} $\basis\subset \{1,\dots,n\}$ 
%\PG{That should be an m?}.
%$\basis(j)$ refers to the $j$-th column of $\basis$.
The set ${\nonbasis = \{1,\dots,n\}\setminus\basis}$ is called the set of \emph{nonbasic} variables.
We denote by $\AB\in \R^{m\times m}$ an invertible submatrix of $A$ with columns $\basis$, and by
$\AN$ the remaining submatrix. The term \emph{basis} may refer to the set of column indices $\basis$ or to the submatrix $\AB$, 
depending on context. 
The maximum number of nonzero entries in any column or row of $A$ is given by $d_c$ and $d_r$, respectively. 
We also denote ${d=\max\{d_c, d_r\}}$.
Based on $\|\AB \|_p = \max_{\|x\|_p = 1} \|\AB x \|_p$ the condition number of $\AB$ is ${\kappa = \|A_B\|_2 \cdot \|A_B^{-1}\|_2}$. 
%\SA{I added the definitions. If I understood you right, physics notation is $\| A \|_2 = \sqrt{ \operatorname{tr}\left( A^\dag A \right) }$, which is the Frobeniusnorm? And I thought of the Spectral Norm?} 
We assume $\kappa$ to be a constant for one iteration, and a function ${\kappa: \R^{n\times n} \to \R_{\geq 1}}$ in \cref{sec:exp_design}. 
%In \cref{sec:exp_design} we use $\kappa$ as a function $\kappa: \R^{n\times n} \to \R_{\geq 1}$. 

We sketch the simplex algorithm below, introducing some important terms along the way.
\begin{enumerate}
\item Choose any basic feasible solution and compute the current basis $\basis$, the nonbasic variables $\nonbasis$ and the current solution $x = \AB^{-1} b$. 
\item Compute the \textit{reduced costs} $\bar{c}_\nonbasis^\top = c_\nonbasis^\top -c_B^\top A_B^{-1} A_\nonbasis$. If $\bar{c}_\nonbasis\geq 0$, stop the algorithm and return the optimal solution. Otherwise choose the pivot column $k$ with $\bar{c}_k<0$. 
\item Compute the \textit{basis direction} $u\coloneqq A_B^{-1} A_k$. If $u\leq 0$, stop the algorithm and return that the problem is unbounded. 
\item Perform the \textit{ratio test} by computing $\theta^\ast$ as
${\theta^\ast = \min\limits_{i:u_i>0}{\frac{x_{B_i}}{u_i}} \eqqcolon \frac{x_{B_\ell}}{u_\ell}}$ with the pivot row $\ell\in [m]$.
\item Replace $\basis(\ell)$ by $k$ in the basis. The new solution is 
$y_{B_i}\coloneqq x_{B_i} - \theta^\ast u_i$. Go to step 2.
\end{enumerate}
%\todo[inline]{SESS: can we make it explicitly clear what the ratio test does? what are the possible outcomes?}
%\AL{to be discusses: do we use or not Dantzig}
Several different pivoting rules such as \textit{Dantzig's} and \textit{steepest edge}
have been proposed. In Dantzig's rule one chooses the column $k$ with the most
negative $\bar{c}_k$~\cite{dantzig}. Goldfarb~et~al.~\cite{forrest1992steepest} 
later introduced the computationally more
expensive \textit{steepest edge} pivoting rule that computes the basis
direction $u$ for every nonbasic variable with negative reduced cost and
chooses $k$ with the most negative reduced cost per additional unit~(${\frac{\bar{c}_k}{\lVert u_k \rVert} = \max_{i} \frac{\bar{c}_i}{\lVert u_i
\rVert}}$).

% describe the method in detail.
\subsection{Quantum Computing and Gate Complexity}
%We briefly introduce the salient aspects of quantum computing and gate complexity required in the sequel; we 
In this paper, we only consider the \emph{quantum circuit model}
 of quantum computation. In this model, a quantum algorithm consists of three phases, which are briefly elaborated below: 
(i) state preparation, (ii) unitary evolution, and (iii) measurement;
refer to~\cite{nielsen2002quantum} for more details. 

A quantum computer operates on a $2^n$-dimensional complex Hilbert space, given by the tensor product of $n$ 2-dimensional 
complex Hilbert spaces, which we call \emph{qubits}.
We assume the ability to efficiently prepare qubits in two states, 
\begin{equation}
\label{eq:qubit}
\ket{0}=\begin{pmatrix}
1\\0
\end{pmatrix}\,,\quad \ket{1}=\begin{pmatrix}
0\\1
\end{pmatrix}\,,
\end{equation}
which form an orthonormal basis for the qubit Hilbert space. 
We call this the \emph{computational basis}.%
%\SA{Next sentence maybe as a footnote and without "Notice that"? For me the sentence seems to be misplaced.} 
\footnote{Notice that in \cref{eq:qubit} we have used Dirac notation, as it is standard in quantum mechanics: $\ket{\cdot}$ denotes a column vector, and $\bra{\cdot}$ its adjoint $\bra{\cdot}=(\ket{\cdot})^\dagger$.}
More generally, a qubit can be in a linear combination $\ket{\psi}=\psi_0\ket{0}+\psi_1\ket{1}$,
which we call a \emph{superposition}.
Physical states have unit norm, i.e.\ for qubits, we require $\abs{\psi_0}^2+\abs{\psi_1}^2=1$.
We call the components $\psi_0$, $\psi_1$ of the vector the \emph{amplitudes}.

The first phase in a quantum algorithm is the preparation of $n$ qubits in a desired computational basis state, 
$\ket{x_1}\otimes\ket{x_2}\otimes\dots \otimes \ket{x_n}$, with $x_i\in \{0,1\}$. 
This state is typically 
%with the initial state being 
$\ket{0}\otimes\ket{0}\otimes\dots \otimes \ket{0}$.

In the second phase, the quantum computer
manipulates the state through a series of elementary operations, called \emph{quantum gates};
these are unitary matrices acting non-trivially only on one or two qubits.
%\footnote{It is known that any $2^n$-dimensional unitary can be approximated to arbitrary precision using sequences of elementary gates \cite{nielsen2002quantum}.}
%corresponding to 1-qubit and 2-qubit gates.
%An example 1-qubit gate is the Pauli $X$ gate 
%\begin{equation}
% X=\begin{pmatrix}
%  0 & 1\\
%  1 & 0
% \end{pmatrix}\,,
%\end{equation}
%which flips $\ket{0}$ to $\ket{1}$ and vice versa. 
%Let say we have $n=3$ qubits and the $X$ gate is applied to the second one, then the quantum computer acts with the unitary $\mathds{1}\otimes X\otimes \mathds{1}$.
%Here, $\mathds{1}$ denotes the 2-dimensional identity matrix. \SA{Do we need examples for gates in the preliminiaries? Probably everybody, who reads this, knows, what an unitary matrix is.}
%In the course of quantum processing, 
The gates transform the state 
of the quantum computer from the initial computational basis element 
into a superposition of different computational basis states. 

For the third phase, we assume that the quantum computer is able to efficiently measure the state in the computational basis.
% which works as follows.
%To understand how measurement works in quantum mechanics, 
Consider an example with $n=2$ qubits, and let the final state of the quantum computation be $\ket{\psi}=\psi_{00}\ket{0}\otimes\ket{0}+\psi_{01}\ket{0}\otimes\ket{1}+\psi_{10}\ket{1}\otimes\ket{0}+\psi_{11}\ket{1}\otimes\ket{1}$. 
Then the measurement outcome is one of 00, 01, 10, or 11, with probability given by the corresponding component of $\ket{\psi}$ squared. 
For example, we measure 00 with probability $\abs{\psi_{00}}^2$. 
After the measurement, the system is left in the measured computational basis state; all information concerning the original superposition is lost.

%To compare the complexity of different algorithms one decomposes the unitaries into elementary gates%

%A set of elementary gates sufficient to perform this approximation is called a \emph{universal set of gates}. %\SA{Do you use this term in this paper? You once used universal family of gates, but I am not sure, whether these two terms mean the same. If they do, please use the same term.}
We quantify the cost of a quantum algorithm by counting the
number of quantum gates required to implement it, which we call \emph{gate complexity}. 
This number depends 
%on the universal set of quantum gates that can be accessed,
%which ultimately depends 
%on the capabilities of the quantum hardware used.
on a choice of an ``basic'' gate set that we decompose quantum gates into. In general, this can be the set of gates native to a particular quantum hardware design, or can be chosen to be a theoretically conveniant model.
%Description:
%\begin{itemize}
%    \item will be written in the end
%\item Linear Programming (Notations for basis, obj function etc.)
%\item Simplex method
%\item Quantum computing
%\item Gate complexity
%\item Here $d_c$ denotes the column sparsity and $d$ the sparsity of the basis matrix,
%$\kappa$ is the condition number of the basis matrix
%\item Complexity
%\item Throughout this paper, $\OPolyLog$ will be used to suppress polylogarithmic factors in the input
%parameters, i.e., $\OPolyLog(f(x)) = O(f(x)poly(\log n, \log m, \log\frac{1}{\varepsilon}, \log \kappa \log d \log L)$
%\end{itemize}

\section{Hybrid Benchmarking}
\label{sec:lb}
In analogy to classical runtime analysis, we are benchmarking the expected computational effort in the
standard model for asymptotic quantum algorithm analysis. Thus we are not benchmarking the performance
of some assumed quantum device.
%Unlinke others~\cite{dalzell2022endtoend}, our implementations of the quantum subroutines are not optimized towards a low curcuit depth.
Instead, we focus on quantifying or estimating (under a very generous
set of assumptions) the runtime of a quantum algorithm by counting the minimum expected number of elementary 
gates required to run it. This allows 
us to compare quantum performance and classical empirical runtime; the
outcomes are benevolent estimates for the required clock frequency of 
practical quantum hardware to allow competitive performance.

The goal is to derive statements on the possible performance of quantum computers, making no strong assumptions on the quantum hardware in use.
To this end, we use a relatively coarse (and benevolent) model of gate complexity: 
we count \emph{any} 1-qubit and 2-qubit gates and assume that no errors occur
in the circuit. This extends 
the techniques of Cade~et~al.~\cite{cade2023quantifying} beyond simple query counts
%\todo{SESS: I'm concerned about this because I don't think we actually go significantly beyond query calls for the major routines we use}
which we apply to real-world, non-synthetic benchmarks. 

Given a classical algorithm, $\alg_c$, composed of several subroutines 
$\Sigma^c_{j}$, for $j=0,1,\dots$, hybrid benchmarking can be applied to 
evaluate the required resources for a quantum algorithm built from  $\alg_c$ by replacing some $\Sigma^c_{j}$ with 
quantum subroutines, $\Sigma^c_j\rightarrow \Sigma^q_j$.
The structure of $\alg_c$ is otherwise untouched; in particular, we require 
the quantum subroutines to have classical inputs and outputs $(i_j, o_j)$.
Without this assumption, the exponential size of the Hilbert space needed for quantum 
simulations would prevent us from classically computing the inputs and outputs of the subroutines. 
In hybrid benchmarking, we find functions $b_j(i_j)$ that provide lower bounds 
on the gate complexity of each $\Sigma^q_j$ with input $i_j$. 
We then select a set of practical benchmarks of interest and solve them with {$\alg_c$},
logging the $i_j$'s and the classical runtime of each instance.
Finally, we use the logged inputs $i_j$ in $b_j(i_j)$ to compute the bounds for 
specific instances and compare them to each classical runtime,
%, see~\cref{fig:hybrid-benchmarking}. 
allowing us to set estimates for the required
%runtime of elementary quantum gates
clock frequency of quantum hardware required to provide a speedup.
%for providing any speedup.
%beyond which the quantum algorithm doesn't provide a speedup.

\old{
\begin{figure}[t]
    \includegraphics[trim={0 18.5cm 0 0},clip,width=\linewidth]{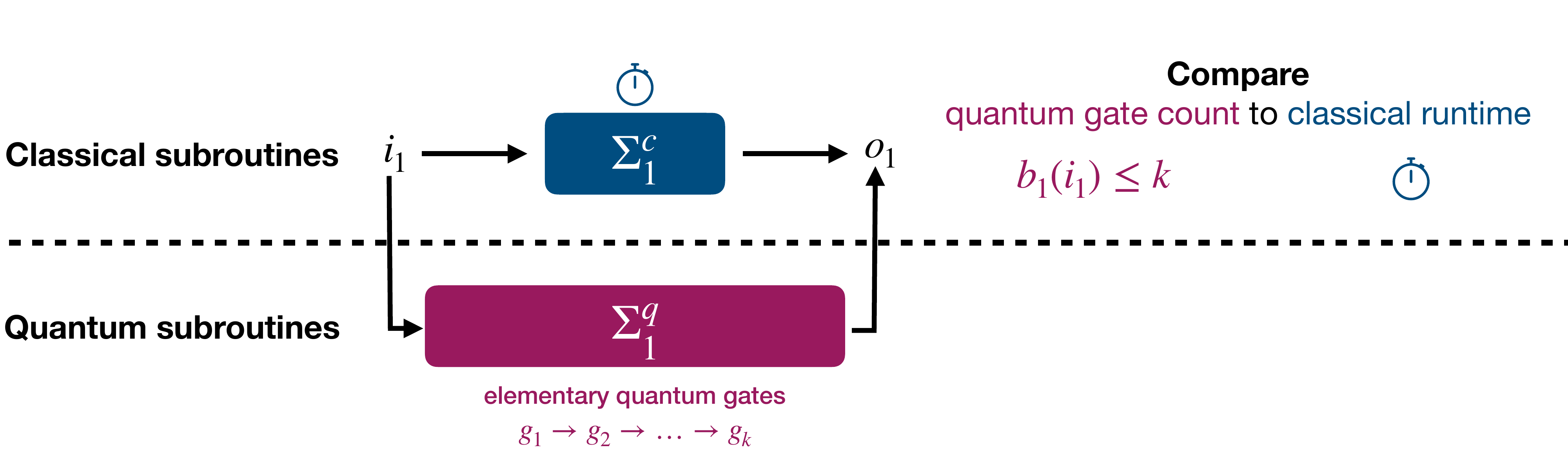}
    \caption{Schematic representation of hybrid benchmarking.}
    \label{fig:hybrid-benchmarking}
\end{figure}
}

%We presuppose that the inputs and outputs of the quantum subroutines are classical.

%For the same reason, we need to restrict to quantum algorithms with only polynomial speed-ups.
Moreover, the functions $b_j(i_j)$ provide lower bounds 
on the gate complexity of the specific quantum subroutines $\Sigma^q_j$ 
used to implement the algorithm. 
%Because ''lower bound'' has an established and conflicting usage in the literature, we sometimes refer to our bounds on the gate cost  as 'least bounds'.
We emphasize that our goal is not to estimate the best achievable performance \emph{in theory}, 
a notoriously hard problem, but possible performances \emph{in practice}, for quantum subroutines
currently standard to quantum computing literature.
Thus, our overall strategy is to identify how the quantum algorithm would perform 
under `the best of all possible worlds' assumptions, and identify ranges in which the 
quantum algorithm has a chance to be useful compared to its classical counterpart. 

The subroutines considered in this work have been chosen because they are accepted as the standard in the literature.
However, is entirely possible that other subroutines, possibly yet to be discovered, might be more efficient.
Because the bounds on gate count we have used are extremely generous and alternative subroutines have comperable scaling to our selections, it is in our opinion extremely unlikely that the qualitative results of this paper would change by simply swapping some quantum subroutine.
We also note that one can use our methods to investiagte the relative performance of competing quantum subroutines for real-world benchmarks without rerunning the classical experiments, by changing some of the functions $b_j(i_j)$.

The main advantage of the hybrid benchmarking technique is that it allows a large class of quantum 
algorithms to be benchmarked \emph{without} access to a fully fault-tolerant quantum computer.
%making our gate-counting methodology as device independent as possible. 
%Our restriction to quantum algorithms with proposed polynomial speed-ups and classical 
%inputs still encompasses many proposed applications, including within semi-definite 
This class includes algorithms for semi-definite
programming \cite{Gilyen2019SDP,brandao2017quantum,brandao2019quantum}, 
machine learning \cite{Li2019ML,wiebe2014quantum,Liu_2021}, 
the triangle problem \cite{magniez2005quantum}, and Monte Carlo methods \cite{Montanaro_2015}.

Furthermore, the quantum subroutines we use can have interesting applications in their own right, 
such as the quantum linear systems algorithm of \cite{childs2017quantum}.

%
%\old{
%We denote the complexity of an algorithm \alg\ as $\cost(\alg)$.
%We quantify the complexity by counting of how many one-qubit, two-qubit, and Toffoli 
%gates are required to run the algorithm.
%These are assigned respective costs $\cost_1$, $\cost_2$, and $\cost_T$. 
%We work at the level of logical qubits and gates, ignoring the quantum error correction overhead.
%}

\section{Fast Quantum Subroutines for the Simplex Method}
\label{sec:qsimplex}

To demonstrate and evaluate the usefulness of hybrid benchmarking, we 
consider the quantum Simplex iteration of Nannicini~\cite{nannicini2022fast},
which provides an asymptotic quantum speedup 
over the classical Simplex algorithm. 
This quantum algorithm replaces each Simplex 
iteration with a quantum subroutine that has the same 
inputs/outputs as the classical algorithm, allowing
a meaningful benchmarking with the proposed technique.

\subsection{Algorithmic Details}
In the classical Simplex method, an LU-factorization of the basis $A_B$ is
computed at every iteration.\footnote{For tuning practical performance, the factorization may
\emph{not} be
computed at every iteration.}  This accounts for a total
worst-case runtime of $O(d_{c}^{0.7} m^{1.9} + m^{2+\sigma (1)} + d_{c}
n)$~\cite{nannicini2022fast} when using the fast sparse matrix multiplication
algorithm of Yuster~and~Zwick~\cite{yuster2005fast}.
However, as long as one is able to identify the entering and leaving variables and 
if the current basis is optimal 
or unbounded, explicit knowledge about the solution vector 
or about any factorization of the basis matrix is not required.
Based on that, Nannicini proposes to replace the pivoting step at each Simplex iteration by a quantum algorithm that 
never computes any explicit representation of $\AB^{-1}$, $\AB^{-1}b$ or $\AB^{-1}\AN$, thus circumventing the issue of managing a factorization. The variable update is then done classically.
\begin{algorithm}
\caption{SimplexIter}\label{alg:simplex-iteration}
\begin{algorithmic}
\Function{SimplexIter}{Constraint Matrix $A$, basis $B$, cost $c$, precision $\varepsilon>0$, $\delta>0$}
\State Normalize $c$ and $A$, s. t.\ $\norm{c_{B}} =1$ resp. $\norm{A_{B}} \leq 1$
\State Apply IsOptimal($A$, $B$, $\varepsilon$). Output: 1, if current basis is optimal
\State Apply FindColumn($A$, $B$, $\varepsilon$). Output: index $k$ of a column with negative reduced cost
\State Apply IsUnbounded($A_{B}$, $A_{k}$, $\delta$). Output: 1, if the problem is unbounded
\State Apply FindRow($A_{B}$, $A_{k}$, $b$, $\delta$). Output: index $l$ of the row minimising the ratio test
\Return $(k,l)$
\EndFunction
\end{algorithmic}
\end{algorithm}

\Cref{alg:simplex-iteration} takes as input the matrix $A$, the basis variables $\basis$, the cost 
vector $c$, and precisions $\varepsilon$ and $\delta$. It outputs the indices of 
the pivot element.
%at the end of each iteration. 
The quantum subroutines $\isOptimal$ and $\isUnbounded$ check if the current solution is 
optimal or unbounded, see~\cref{sec:details-simplex} for details. 
%In the end of each simplex iteration, one has to do the variable update classically. 
The explicit computation of the solution 
vector $A_{\basis^*}^{-1}b$ for an optimal basis $\basis^*$ is only necessary in the
last iteration.
\begin{theorem}[\cite{nannicini2022fast}: Runtime of \cref{alg:simplex-iteration}]\label{theorem:nannicini-simplex-iteration}
There exist quantum subroutines%
\footnote{We give a different asymptotic runtime than that originally given in \cite{nannicini2022fast}. This is traceable to a difference in subroutines in the QLS.}
to identify if a basis is optimal, or determine
a column with negative reduced cost, with runtime\footnote{The notation $\OPolyLog$ is used to suppress polylogarithmic factors in the input
parameters, i.e., $$\OPolyLog(f(x)) = O(f(x)poly(\log n, \log m, \log\frac{1}{\varepsilon}, \log \kappa, \log d, \log L).$$}
\begin{multline*}
    \cost(\simplexIter) = \OPolyLog\left(\frac{\sqrt{n}}{\varepsilon} (d_{c}n + dm)\right) + \\
    \OPolyLog\left(\frac{\eta}{\varepsilon} d^{2} \kappa^{2} m^{1.5}\right) +
    \OPolyLog\left(\frac{\eta}{\delta} d^{2} \kappa^{2} m^{1.5}\right),
\end{multline*}where $\eta$ is the maximum norm of a column of $A$ or $b$.
\end{theorem}
The runtime of one quantum Simplex iteration scales better than the one of a classical Simplex iteration, with increasing system size. 
%The asymptotic scaling in $m$ and $n$ of a quantum simplex iteration is better than 
%the scaling of a classical iteration, 
This indicates a possible quantum advantage for large instances.
Classically, an efficient pivoting routine is crucial for the performance of the algorithm.
As the original Simplex $\mathrm{FindColumn}$-subroutine implements the \emph{random pivoting rule}, i.e.,
an update to a random basis that improves the objective function,
we replace it by $\steepestEdgeRule$, equivalent 
to a quantum version of the \emph{steepest edge rule}.
$\simplexIter$ is comprised of four subroutines, each of which in turn is comprised 
of different subroutines that can be broken down to basic quantum algorithms 
i.e., quantum search $\qsearch$~\cite{boyer1998tight,cade2023quantifying}, quantum amplitude estimation ($\qae$)~\cite{boyer1998tight}, quantum phase estimation ($\qpe$) \cite{cleve1998quantum},
the quantum linear solver algorithm, QLSA Fourier \cite{childs2017quantum}, and quantum minimum finding ($\qmin$)~\cite{durr1999quantum,cade2023quantifying}. 
A detailed review of algorithms and complexities 
can be found in~\cref{sec:details-quantum-subroutines}.

\subsection{Expected Gate Complexity}\label{sec:gate-counts}
The precise gate complexity of a quantum algorithm
depends on the selected hardware and on the way the problem instance is specified.
In the following, we provide lower bounds on the gate complexity (\cref{lem:qls,lem:isOptimal}) 
and the \emph{expected} gate complexity (\cref{lem:qsearch,lem:steepest,lem:cost_isUnbounded,lem:findRow})
for each subroutine 
of $\simplexIter$. To obtain a general non-asymptotic analysis, we 
make a number of very benevolent assumptions towards the quantum hardware 
and use generous lower bound estimates. As a consequence, the overall
estimates are already quite biased in favor of the quantum algorithms;
if anything, practical performance would be even worse than 
displayed in the following \cref{sec:experiments}.

We emphasize some of these simplifying assumptions. 
Firstly, it was beyond the scope of our analysis to compute the gate complexity
of preparing every unitary used, and so we have dropped (potentially practically relevant)
terms in order to bound the gate complexity.  Furthermore, while we can
straightforwardly lower bound the gate complexity of some quantum subroutines,
for others it is more meaningful to compute an expected gate complexity. 
The Simplex routines encorporate subroutines of each type, and so we combine
lower bound and expected gate complexities in a naive manner.  We interpret
this as a lower bound on the expected gate complexity of the routine.
In taking a measure of gate complexity that encorporates both expected gate
complexity and lower-bound gate complexity, we sacrifice adherence to a strict
definition of either.  In return, we obtain a measure that represents the lowest
expected gate complexity one would need to allocate to run a quantum algorithm.
Given the resource scarcity of current quantum hardware, this is a reasonable
heuristic for filtering out when a quantum algorithm has any potential to be
implemented.

Each subroutine of $\simplexIter$ is built on two basic quantum algorithms:
For  $\qsearch$ we can compute an expected gate complexity, and for
$\qlsaFourier$, we can then provide a bound on the fewest number of quantum gates
required.  Subsequently, we state the lower bounds on the expected number of
gates for Nannicini's subroutines $\isOptimal{}$,
$\steepestEdgeRule{}$, $\isUnbounded{}$ and $\findRow$.

The details of each subroutine and the proofs of the lemmas can be found in
\cref{sec:details-simplex} and \cref{sec:details-quantum-subroutines}.

%We also discuss lower bounds for $\steepestEdgeRule$.
\subsection{Quantum Subroutine Bounds}
$\qsearch$ is an algorithm that performs an unstructured search among a list of items; 
this is done by repeatedly acting with an operator $Q$. 
%$X$; 
The expected number of 
%iterations needed for $\qsearch$ to find one of the $t$ marked elements 
calls to $Q$ needed to find 
a marked element is given by \cref{lem:qsearch}.
A $\qls$ algorithm prepares a quantum state whose amplitude is equal to the solution of a linear system of equations.
Here, we consider the algorithm $\qlsaFourier$ from~\cite{childs2017quantum};
%takes as input a linear system of equations and returns an approximation of 
%the solution vector. 
its bounded expected gate complexity is given in \cref{lem:qls}.

\begin{restatable}
    [Iterations for \boldmath $\qsearch$]
    {lemma}{lemmaQSearch}
\label{lem:qsearch}
Let $X$ be a list of length $\abs{X}$, with $t$ marked items.
%, according to a function $\chi: X \longrightarrow \{ 0,1 \}$. 
The expected number $n_Q(\abs{X}, t)$ 
of iterations that $\qsearch$ needs to find a marked item is
\begin{align*}
n_Q(\abs{X}, t)&=\sum_{k=1}^{k_{max}}\frac{m_k}{2}\Bigl[\prod_{l=1}^{k-1}\frac{1}{2}+\frac{\sin(4(m_l+1)\theta)}{4(m_l+1)\sin(2\theta)}\Bigr], \\
k_{max}&=\Bigl\lceil \log_\lambda\frac{\abs{X}}{2\sqrt{\abs{X}-1}}\Bigr\rceil+4\,,
\end{align*}
with  $\sin^2\theta=t/\abs{X}$, $m_k = \lfloor \min(\lambda^k, \sqrt{\abs{X}})\rfloor$,  $\lambda=6/5$.
\end{restatable}

\begin{restatable}[\boldmath$\qlsaFourier$]{lemma}{lemmaQLS}
\label{lem:qls}
The gate complexity of $\qlsaFourier$ that solves the linear equation $Ax=b$ to
precision $\varepsilon$ has a lower bound of \begin{multline*}
    \cost[\qlsa\left(A, b, \varepsilon\right)]\ge 10 tw 
    \left(\frac{\pi}{2\arcsin(\alpha^{-1})}+1\right)\cdot\\
    (\norm{A}_1-d^2\gamma)
    \left(\left\lceil\log(\frac{\norm{A}_{1}}{\gamma}-d^2)\right\rceil -1\right),
\end{multline*}
with the smallest integer $w$ satisfying $\frac{e^w}{w^w} \leq \frac{\varepsilon_{seg}^2}{2}$ 
and $\alpha, \gamma, \Delta_z, K, t, \varepsilon_{seg}$ defined as

{\centering
\scalebox{.85}{
\begin{minipage}{\columnwidth}
\begin{alignat*}{4}
    \alpha &= 2\sqrt{\pi}\frac{\kappa}{\kappa+1}\sum_{k=-K}^{K}\abs{k}\Delta_z e^{-\frac{(k\Delta_z)^2}{2}}\,,  
    & t &= 2\sqrt{2}\kappa {\log(1+\frac{8\kappa}{\varepsilon})}\,,\\
    \Delta_z &=\frac{2\pi}{\kappa+1}\Bigl[\log\Bigl(1+\frac{8\kappa}{\varepsilon}\Bigr)\Bigr]^{-1/2}\,,
    &\varepsilon_{seg}&=\frac{\varepsilon}{90\gamma t d^2\left\lceil \frac{\norm{A}_{\max}}{\gamma}\right\rceil}\,,\\
    K &=\left\lfloor\frac{\kappa+1}{\pi}\log\Bigl(1+\frac{8\kappa}{\varepsilon}\Bigr)\right\rfloor\,,
    &\gamma &=\frac{\varepsilon}{ \sqrt{2}d^3t}\,.
\end{alignat*}
\end{minipage}
}
}

Here $\left|\left| A\right|\right|_{\max}\coloneqq \max_{i,j} \left|A_{ij}\right|$ is the largest element of $A$ in absolute value.
\end{restatable}
\subsection{Nannicinni Subroutines}
With these lemmas, we can now lower bound the expected gate complexity for each subroutine of $\simplexIter$.
The gate complexity of $\isOptimal$ can be bounded by~\cref{lem:isOptimal}.

\begin{restatable}[\boldmath$\isOptimal$]{lemma}{lemmaIsOptimal}\label{lem:isOptimal}
    The gate complexity of $\isOptimal$ has a lower bound of
    \begin{multline}\cost[\isOptimal(A,B,c,\varepsilon)] \geq \left(24\sqrt{n-m} -1 \right) \cdot\\
        \left(\frac{450\sqrt{6}\pi}{11\varepsilon} -1 \right) 
        \cost\left[\qls\left({A}_B,A_k,\frac{0.1 \varepsilon}{\sqrt{2}}\right)\right]\,.
        \label{eq:IsOptimal}
    \end{multline}
\end{restatable}

\begin{proof}
To determine the optimality of the current basis, $\isOptimal$ leverages quantum amplitude estimation. This procedure estimates the amplitude, denoted as $\tilde{\phi}$, corresponding to the scenario where the function CanEnterNFP yields an outcome of $1$.
Specifically, the function CanEnterNFP yields a value of $1$ when a designated column satisfies the criterion to potentially enter the basis.
%CanEnterNFP returns $1$, if a given column 
%(for $\isOptimal$ a superposition of all non-basic columns) 
%can enter the basis. 
Ultimately, the decision reached by $\isOptimal$ hinges upon whether the estimated amplitude $\tilde{\phi}$ is close to $1$ or $0$.
See \cref{sec:simplex_iter_common_subroutines,subsec:IsOptimal} for further implementation details.

The gate complexity of $\isOptimal$ involves applying quantum amplitude estimation (QAE) to CanEnterNFP and verifying ${\tilde{\phi}\in[0,\varepsilon_\qae)\cup(1-\varepsilon_\qae,1]}$ with precision parameters 
$\varepsilon_{\qae}=\frac{1}{4\sqrt{n-m}}$ and $\delta_{\qae} = \frac{1}{4}$.
We lower bound the gate complexity of $\isOptimal$ by only considering the gate complexity of QAE. 
    QAE and how to bound its gate complexity are explained in \cref{app:Quantum_phase_estimation}.
    The gate complexity of QAE can be lower bounded by
    \begin{align*}
        \cost[\qae(\alg, \chi_l,\varepsilon_\qae, \delta_\qae)] %&= \cost[\alg] + \cost[QPE(Q,\varepsilon_\qae, \delta_\qae)]\\
        %&\ge \cost[\alg] + (2^{n_{c}}-1)\cost[Q] \nonumber\\
        &\ge (2^{n_{c}+1}-1)\cost[\alg]\,,
    \end{align*}
    by combining 
    %both equations of 
    \cref{lem:cost_qae%
    %} and the result of \cref{
    ,lem:qpeCost} with  ${\alg = \mathrm{CanEnterNFP}}$,
    %$Q$ as the Grover operator 
    and
    \begin{equation*}
        n_{c}= \Bigl\lceil \log_2\frac{1}{\varepsilon_\qae}+\log_2\left(1+\frac{1}{2\delta_\qae}\right)\Bigr\rceil\,.
    \end{equation*}
    A lower bound for the gate complexity of CanEnterNFP is given by \cref{lem:cost_canenternfp}.
    Combining the bounds for CanEnterNFP and QAE concludes the proof.
\end{proof}

Bounds for choosing a pivot column based on the steepest edge pivoting rule can be found in~\cref{lem:steepest}. 
Results for other pivoting rules are given in~\cref{sec:proof_pivoting_rules}.
In~\cref{lem:cost_isUnbounded} we state a lower bound on the expected gate complexity for the routine that decides if the given linear program is unbounded.
A bound for choosing a pivot row is given in~\cref{lem:findRow}.
The proofs of \cref{lem:steepest,lem:cost_isUnbounded,lem:findRow} are omitted due to space constraints and can be found in~\cref{sec:details-simplex}.

\begin{restatable}[\boldmath$\steepestEdgeRule$]{lemma}{lemmaSteepestEdge}
    \label{lem:steepest}
     The expected gate complexity of $\steepestEdgeRule$ has a lower bound of
\begin{multline*}
    \cost[\mathrm{\steepestEdgeRule}(A, B, c, \varepsilon)] \geq 3 \left\lceil \log_3\frac{1}{\varepsilon} \right\rceil \,\cdot\\\left( \frac{40 \sqrt{3}\pi c_{\mathrm{max}}}{\varepsilon}-1 \right) \sum_{t=1}^{n-m-1} \frac{n_Q(n-m, t)}{t+1}\,\cdot\\
      \cost\left[ \qls\left(A_B, A_k, \frac{\varepsilon}{10 c_{\mathrm{max}} \sqrt{2}}\right)\right],
\end{multline*}
where $c_{\mathrm{max}}$ is the (absolute) maximum component in the cost vector $c$.
\end{restatable}

\begin{restatable}[\boldmath$\isUnbounded$]{lemma}{lemmaIsUnbounded}
    \label{lem:cost_isUnbounded}
    The expected gate complexity of $\isUnbounded$ has a lower bound of
    \begin{multline*}
        \cost[\text{IsUnbounded}(A_B, A_k, \delta)] \ge\;
        n_{Q}(m,t) \,\cdot\\\left(\frac{50 \sqrt{3}\pi}{18 \delta} - 1\right)
        \cost\left[\qls\left(A_B, A_k, \frac{\delta}{10}\right)\right],
    \end{multline*}
where $t$ is the number of positive components in $u = A_B^{-1}A_k$.
\end{restatable}

\begin{restatable}[\boldmath$\findRow$]{lemma}{lemmaFindRow}
    \label{lem:findRow}
    The expected gate complexity of $\findRow$ has a lower bound of
\begin{multline*}
    \cost[\findRow(A_B, A_k, b, \delta)] \ge\;n_Q(m, 0) \,\cdot\\ \left(\frac{\sqrt{3} \pi \|A_B^{-1} A_k\|}{2 \delta} - 1\right)
    \cost\left[\qls\left(A_B, b, \frac{\delta}{2}\right)\right].
\end{multline*}
\end{restatable}

\newcommand{\rqone}{Could \cref{alg:simplex-iteration} provide a speedup over the classical Simplex algorithm when quantum hardware is ready?}
\newcommand{\rqtwo}{What quantum gate operation times are required for \cref{alg:simplex-iteration} to provide a speedup over the classical primal Simplex algorithm?}
\newcommand{\rqthree}{Can we find LP instances that are easier to solve on quantum hardware?}

\section{Benchmarks and Evaluation}\label{sec:experiments}
In the previous section we derived lower bounds on the expected number of quantum gates
required for the subroutines of Nannicini's Simplex algorithm. In this section
we explicitly compute these bounds for various instances of linear programs using the 
steepest-edge pivoting rule on both the classical and quantum side.
For each of these instances, we track the classical runtime of a Simplex
iteration and compare these to the bounds for a quantum computer performing
Nannicini's algorithm, see~\cref{sec:lb,sec:gate-counts} for details.
From this comparison we answer the following three
questions. 
\begin{researchQuestion}[leftmargin=1cm]
    \item \rqone \label{rq:any-hope}
    \item \rqtwo \label{rq:how-fast}
    \item \rqthree \label{rq:easy-lp-instances}
\end{researchQuestion}

\subsection{Linear Programming Solvers} 
In order to apply the hybrid benchmarking technique,
the classical runtime of each Simplex iteration
and information regarding the iteration itself (e.g., the condition number)
need to be logged; see \cref{sec:lb}.
Choosing the proper LP solver is crucial for a fair comparison:
On one hand, the solver must be widely accepted; and, on the other, it must
implement a primal Simplex algorithm with the steepest edge
rule. To ensure transparent and straightforward evaluation,
we require the solver to be open source and easily extendable.
The survey by Gearhart~et~al.~\cite{gearhart2013comparison} compares the
performance of open-source LP solvers against the commercial solver
\emph{CPLEX}. The survey included \emph{COIN-OR Linear Programming} (CLP),
GNU Linear Programming Kit (GLPK), \emph{lp\_solve} and \emph{Modular In-core
Nonlinear Optimization System} (MINOS), with no open-source solver outperforming
CPLEX, meaning that classical performance may be even better with commercial
solvers. Among the open-source solutions, CLP and GLPK were found to be top
performing.  
More recently, \emph{Sequential object-oriented simPlex} (SoPlex) was made
 open source as part of the
\emph{SCIP} software package~\cite{gamrath2016scip} and can compete with 
commercial solvers like Gurobi or CPLEX. However, SoPlex does not
implement a primal Simplex but only a composite Simplex. We ultimately chose
GNU Linear Programming Kit (GLPK) for our experiments.

\begin{figure}[ht]
    \includegraphics[trim = 0mm 5mm 0mm 0mm, width=\linewidth]{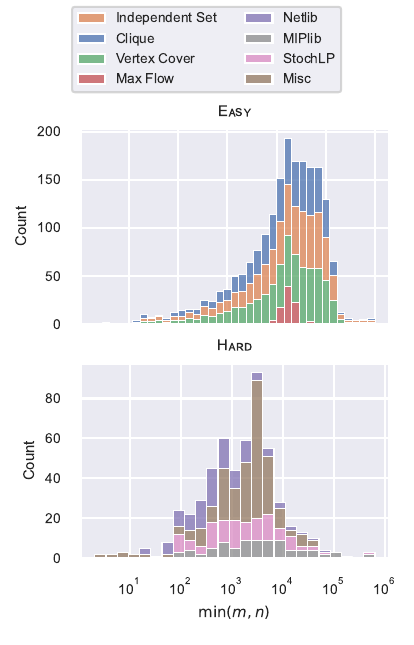}
    \caption{Size distribution of the used \easy{} (top) and \hard{} (bottom) instances.}
    \label{fig:instances}
\end{figure}

\subsection{Instances from Existing Benchmark Libraries}

We applied the hybrid benchmarking technique to a wide range of test instances.
The \hard{} instance set consists of \SI{532} instances from 
standard benchmark sets \netlib~\cite{Netlib}, \stochlp~\cite{Stochlp}, 
\misc~\cite{Misc} for LP solvers,
as well as LP-relaxations from the \miplib~\cite{Miplib} benchmark set.
The \easy{} instance set consists of self-generated LP formulations
that are well-suited for the quantum subroutine.
See \cref{tab:instance-details} and \cref{fig:instances} for more details.
The \easy{} instance set was generated in an attempt to answer \ref{rq:easy-lp-instances}.
Many instances in \hard{} are not sparse nor well-conditioned.
By \cref{theorem:nannicini-simplex-iteration}, \cref{alg:simplex-iteration} scales quadratically with
the number of nonzero elements in the matrix $A$ and the condition number $\kappa$.
Therefore, we constructed instances for sparse and well-conditioned problems,
giving the quantum algorithm an additional asymptotic benefit.
These are \emph{Maximum Flow},
\emph{Minimum Vertex Cover}, and \emph{Maximum Independent Set}.
The Maximum Flow formulations were generated from 
the Max-Flow-Min-Cut benchmark set~\cite{Jensen2022} and
the graphs for the other problems were either
(i) taken from the seconds DIMACS implementation challenge~\cite{dimacsCLQ} or
(ii) generated random Erd\H{o}s-R\`enyi graphs with up to \SI{1000} vertices; see~\cite{GenerateFastNetworks} for details.
%based on a work by Gilbert~\cite{Gilbert}. 
%We generate undirected graphs with up to \SI{1000} vertices and 
%an edge possibility between 0.1 and 0.9 using the largest (weakly) connected component.
%The first three problems are relaxations of Integer Programming formulations and the last problem is 
%the Linear Programming formulation of a M. 
%The first set of self-constructed instances are relaxations of Integer Programming 
%formulations for three well-known graph problems. Similar
%instances might be solved within branch-and-bound algorithms. Furthermore, we consider linear program formulations of the maximum flow problem. 
%Note that advanced solvers like Gurobi or CPLEX can solve these problems 
%%optimally in very short time.
%To ensure a fair comparison with other classical algorithm we also benchmark the algorithm on hard instances that were not designed to perform well on Nannicini's algorithm. 
%It is a collection of problems from various different sources and applications and the problem sizes range from ... to ... nodes and ... to ... edges. Additionally the capacity is ranged from ... to ... . For computational reasons we only considered problems with a maximum size of ... edges/ nodes, which results in (nodes + edges) constraints. 
%NETLIB
\addtocounter{footnote}{1}
\begin{table}
    \centering
    {\small
    \begin{tabularx}{\columnwidth}{cccr}
    \toprule
    Benchmark & group & \# Instances & Source\\
    \midrule
    Random Graphs\footnotemark[\value{footnote}] & \easy{} & \SI{1711}{} & self generated\\
    DIMACS Graphs\footnotemark[\value{footnote}] & \easy{} & \SI{101}{} & \cite{dimacsCLQ}\\
    Maximum Flow & \easy{} & \SI{88}{} & \cite{Verma_Batra} via \cite{Jensen2022}\\
    \netlib & \hard{} & \SI{109}{} & \cite{Netlib} \\
    \miplib & \hard{} & \SI{84}{} & \cite{Miplib}\\
    \stochlp & \hard{} & \SI{108}{} & \cite{Stochlp}\\
    \misc & \hard{} & \SI{231}{} & \cite{Misc}\\
    \bottomrule
    \end{tabularx}
    }
    \caption{Overview of the used benchmarking instances. Full overview can be found in~\cref{tab:instance-details-full}.\\
    {\footnotesize \protect\footnotemark[\value{footnote}]%
    Instances are LP relaxations of Min Vertex Cover, Max Independent Set and Max Clique.}}\label{tab:instance-details}
\end{table}
\subsection{Instance from Combinatorial Optimization}

Nannicini \cite{nannicini2022fast} notes that his approach is most likely to be
valuable for sparse instances with a low condition number. As this is rarely
the case for the instances from the benchmark libraries, we also use well-known LP relaxations
of graph problems, which ensure a low condition number. 
\begin{itemize}
    \item \textbf{Maximum Flow}:
	Given a directed weighted graph $(V,A)$ with a source vertex $s$, a
target vertex $t$ and a capacity function $c$, the Maximum Flow Problem deals
with finding the maximum amount of flow that can be sent through the edges
while not exceeding the maximum capacity of each edge. As relaxed linear
formulation we use the following.
\begin{equation*}
	\begin{array}{ll@{}llr}
\text{max}  & \displaystyle\sum\limits_{(u,t) \in A}  &x_{u,t} - \displaystyle\sum\limits_{(t,u) \in A}  x_{t,u} \\
\text{s.t.}& \displaystyle\sum\limits_{(u,v) \in A} &x_{u,v} -  \displaystyle\sum\limits_{(v,u) \in A}  x_{v,u}   = 0, \hfill\quad\forall~ u \in V\\
                 &&x_{u,v} \leq ~c(u,v), \hfill\forall~(u,v) \in A
    \end{array}
    \end{equation*}
 \item \textbf{Minimum Vertex Cover}:
    Given an undirected graph with a set of vertices $V$ and a set of edges $E$, a Minimum Vertex Cover is a smallest set of vertices that contains at least one vertex of every edge. As linear relaxation, we use the following.
    \begin{equation*}
	\begin{array}{ll@{}lr}
\text{minimize}  & \displaystyle\sum\limits_{v \in V}  &x_v &\\
\text{subject to}& \displaystyle x_u +  &x_v   \geq 1,  &\forall~\{u,v\}  \in E\\
                 &                     0\leq &x_{v} \leq 1, &\forall~~v \in V
\end{array}
\end{equation*}
\item \textbf{Maximum Independent Set}:
    A maximum independent set of an undirected graph $(V,E)$ is a largest set of
vertices that contains at most one vertex of every edge. This yields the following 
linear programming relaxation. 
\begin{equation*}
	\begin{array}{ll@{}lr}
\text{maximize}  & \displaystyle\sum\limits_{v \in V}  &x_v &\\
\text{subject to}& \displaystyle x_u +  &x_v   \leq 1,  &\forall~\{u,v\}  \in E\\
                 &              0\leq   &x_{v} \leq 1, &\forall~~v \in V
\end{array}
\end{equation*}
\item \textbf{Maximum Clique}:
    A maximum clique of an undirected graph $(V,E)$ is a largest set of
vertices that contains an edge between any two vertices. A linear
programming relaxation is the following.
\begin{equation*}
	\begin{array}{ll@{}lr}
\text{maximize}  & \displaystyle\sum\limits_{v \in V}  &x_v &\\
\text{subject to}& \displaystyle x_u +  &x_v   \leq 1,  &\forall~\{u,v\}  \notin E\\
                 &                0\leq &x_{v} \leq 1, &\forall~~v \in V
\end{array}
\end{equation*}
\end{itemize}
\begin{figure}[ht]
    \includegraphics[trim = 0mm 5mm 0mm 0mm, width=\linewidth]{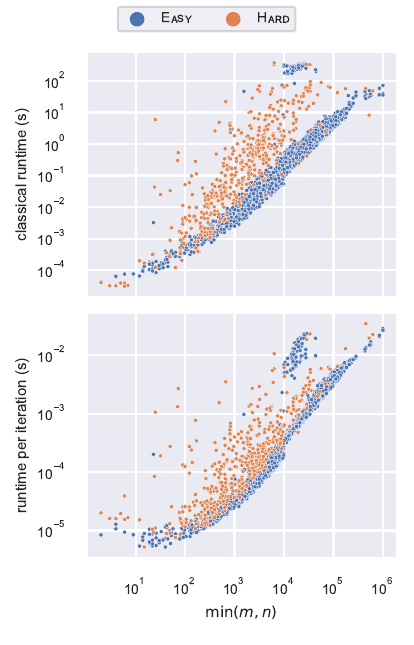}
    \caption{Pure runtime without logging, using GLPK. (Top) Total runtime. (Bottom) Average runtime per iteration.}
    \label{fig:runtime}
\end{figure}
\begin{figure*}
    \includegraphics[trim = 0mm 5mm 0mm 0mm, width=\linewidth]{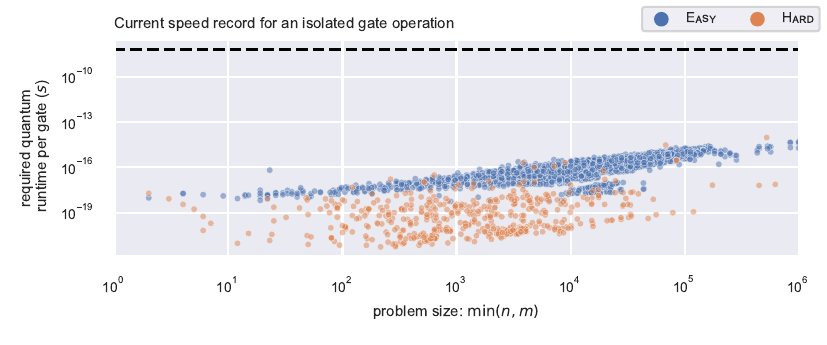}
    \caption{Minimum required gate evaluation time to outperform a classical iteration.
    Values are the average over all ratios between the (classical) 
    iteration runtime and the iteration's expected gate complexity bound, and are subject to a number of additional benevolent assumptions for the quantum algorithm. 
    The dashed line at $6.5 \times 10^{-9}$ is the best time achieved
for an isolated gate operation~\cite{chew2022ultrafast}.
%, with a realistic lower bound
%$10^{-10}$s in the foreseeable future.
%an estimate of the best performance a quantum computer is likely to achieve 
%providing a bound on the current status of classical control hardware
    %even under various benevolent assumptions for quantum computing.
	While the gap to the dashed line slowly narrows with increasing 
$min(m,n)$, the qualitative picture does not change for problem sizes 
that can realistically be treated.
    %For more details refer to~\cref{fig:quantum-gate-efficiency-all}.
}
    \label{fig:quantum-gate-efficiency}
\end{figure*}
\subsection{Experimental Design}
\label{sec:exp_design}
As the primal Simplex runs single-threaded, all
experiments were executed on a single CPU core of an 
Intel~Core~i7~6700K or Intel~Core~i7~4770K %($8\times \SIrange[]{3.8}{4.7}{\giga\hertz}$)
with \SI{16}{\giga\byte} DDR4 RAM and Ubuntu~22.04.2~LTS.
We adapted the standard GLPK implementation compiled by gcc~(v11.3.0) for our requirements and
introduced callbacks into the primal Simplex routine to log all important data.
The callbacks were written in Python~3.10 with native C elements using \mbox{ctypes}.
The source code is available on GitHub\footnote{URL not shown for double-blinding and available on authorized request.}. We distinguish between 
\emph{iteration-related} and \emph{instance-related} measures;
the latter include variable number $n$, constraint number $m$, 
the number of nonzero elements $d_n$, and the maximum 
weight $c_i$ of the objective function.

In each iteration, we save the total number of nonzeros in the basis $A_B$, the maximum 
number of nonzeros in its columns, its maximum absolute value, and
the norms $\|A_B\|_1, \|A_B\|_1^{-1}$, as well as the condition number $\kappa(A_B)$.
To compute the bound in \cref{lem:steepest}, we log the number of columns 
that would improve the objective function,
i.e., with reduced costs~$< 0$ for a minimization problem, 
as well as the absolute maximum reduced cost value.
Additionally, \cref{lem:cost_isUnbounded,lem:findRow} require the number of positive components in $u=A_B^{-1}A_k$
as well as~$\|u\|_2$.
Finally, we measure the time per iteration excluding the time for computing the above values.
Computing the exact value of the condition number $\kappa(A_B)$ in each iteration
comes with major drawbacks in efficiency.
Internally, GLPK computes the condition number $\kappa_1(A_B) = \|A_B\|_1 \cdot \|A_B^{-1}\|_1$
to print out a user warning when an ill-conditioned basis matrix is encountered.
Therefore, instead of calculating $\kappa(A_B)$ explicitly, we only save the value of $\kappa_1(A_B)$.
Given $\kappa(A_B) \geq 1$ and $\|A_B\|_2 \geq \frac{\|A_B\|_1}{\sqrt{m}}$, we can establish
$\kappa(A_B) \geq \max{\left(1, \frac{\kappa_1(A_B)}{m}\right)}$
as a lower bound for $\kappa(A_B)$.
This data is used to compute the lower bound for the expected number of gates per iteration. 
Therefore, we sum the individual costs of each subroutine of $\simplexIter$
(stated in
\cref{lem:isOptimal,lem:steepest,lem:cost_isUnbounded,lem:findRow}). All plots %in this section 
were generated
with the classical steepest edge pivoting rule as well as the quantum steepest edge bound, see~\cref{lem:steepest}. 
As \cref{alg:simplex-iteration} operates on a normalized matrix~$\hat{A}=\frac{A}{\|A\|_2}$, we lower bound the norms from \cref{lem:qls} with 
$\|\hat{A}_B\|_1\geq \frac{\|A_B\|_1}{d\|A_B\|_{max}}$ and
$\|\hat{A}_B\|_{max} \geq \frac{1}{d}$. 
% This might be commented out of we can compute the bounds in time.
%To allow for easier computation only the last term from the sum $\sum_{t=1}^{q-1} \frac{n_Q(q, t)}{t+1}$ is evaluated.
%
For all calculations we used precision parameters $\varepsilon=\delta=10^{-3}$;
as the performance of the quantum solver is linearly dependent on $\varepsilon$ and $\delta$,
a choice of $\varepsilon=\delta=10^{-6}$ (a default value for many classical solvers like CPLEX or Gurobi)
would further diminish the performance of the quantum solver by a factor of $10^{-3}$.
\subsection{Evaluation}
We ran the experiments on all instances from \cref{tab:instance-details} with a maximum time limit of \SI{1800}{\second}. 
Evaluations for other pivoting rules that Nannicini~\cite{nannicini2022fast} mentioned 
can be found in \cref{sec:additional-figures}.
\cref{fig:runtime} shows the runtime distribution
of the two classes.
It is clear that most of the instances from \hard{} needed more time, both in total and within
each iteration.
On the empirical basis of this study, we can answer the proposed research questions.
%In
%this framework the answers can only be given empirically.
\begin{figure*}
    \includegraphics[trim = 0mm 5mm 0mm 0mm, width=\linewidth]{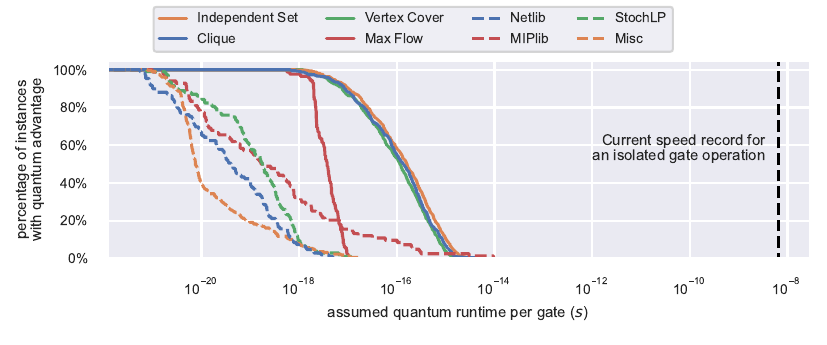}
    \caption{Percentage of instances (y-axis) that allow for a quantum advantage over the classical Simplex algorithm
     with the assumed given quantum gate evaluation time (x-axis). Despite the benevolent assumptions, none are even close to the realistic bound of $6.5 \times 10^{-9}$ seconds.}
    \label{fig:quantum-advantage-counts}
\end{figure*}

\vspace{.25cm}
\noindent\textbf{\ref{rq:any-hope}: \rqone}

Each dot in \cref{fig:quantum-gate-efficiency} shows the average ratio between the classical runtime and the lower bound of the expected gate complexity in that iteration. Thus, the y-values
give the time in which a quantum algorithm has to evaluate a single gate to
get an overall improvement over the classical Simplex iteration.
The black dotted line represents the fastest quantum gate that has currently been 
realized; the timescale are nanoseconds, based on 
the $6.5\cdot 10^{-9}$s from~\cite{chew2022ultrafast}. 
As described in \cref{sec:intro}, 
%Because any quantum computer requires the use of 
%classical control hardware in the form of arbitrary waveform generators there is a practical 
%limit supplied by the bandwidth of these devices. The current best resolution
%here is in the $1-100$GHz regime. 
it is reasonable to expect that gate times will be limited for the
foreseeable future to above the $10^{-10}$s timescale.
This suggests that even with a very optimistic view of future hardware improvements,
quantum computers are several orders of magnitude away
from being useful in practice for this specific algorithm. This holds for both
the \easy{} and \hard{} set of instances.\\

\noindent\textbf{\ref{rq:how-fast}: \rqtwo}
Our framework allows an empirical answer.
%In this framework the answer can only be given empirically. 
\cref{fig:quantum-advantage-counts} shows how many instances require a specific 
runtime per gate from a quantum computer to perform one Simplex iteration in the 
same runtime as a classical computer. To beat a classical computer on the \easy{} 
benchmark set, a quantum computer needs to evaluate one gate in about $10^{-18}$ seconds.
Further, a time of at most $10^{-21}$ seconds per gate is needed to outperform the classical computer on every instance. 
In addition, recall that we estimated the number of gates by a lower bound. 
Thus, the true number of gates could be even larger, and the time left for one gate even lower.\\

\noindent\textbf{\ref{rq:easy-lp-instances}: \rqthree}
Our \easy{} benchmark set aimed for instances that scale well with
the asymptotic runtime from \cref{theorem:nannicini-simplex-iteration}. Our instance
set is well-conditioned (with a condition number of $1$ in almost all iterations) and very
sparse, see~\cref{fig:sparsity-condition}.
\cref{fig:quantum-gate-efficiency} shows that the computed lower bounds, while being 
several orders of magnitude higher than for most of the \hard{} benchmark set, 
will still not allow for a quantum advantage over the classical methods.

\section{Conclusions}
\label{sec:conclusion}

We have presented an analysis for gauging the practical performance of 
quantum computers for solving large-scale, real-world instances of 
important optimization problems, resulting in estimates for
the physical gate-time requirements for future
hardware to deliver real-world quantum speed up.
We have evaluated the resulting hybrid benchmarking for
solving linear optimization problems for Nannicini's %cutting-edge
quantum version of the simplex method.

There are a number of important conclusions. Specifically, the asymptotic
advantage of a quantum method like the one of Nannicini appears to be unlikely
to play out in the practical dimensions that are relevant for realistic
applications (reflected by the considered large-scale benchmark instances): 
Even under very benevolent assumptions (e.g., ignoring 
error correction and other physical aspects of quantum computers)
and for purposefully constructed sparse and well-conditioned linear programs
that are better suited for such a quantum algorithm, the required gate efficiency
seems beyond what is physically possible. This will not fundamentally change for
even larger instances of practical dimensions.

More generally, our methods offer a way to provide meaningful estimates for the
practical performance of algorithms on future quantum devices. 
Many quantum
subroutines are built from a fixed pool of algorithms like unstructured quantum search or
QLS.
%\todo[inline]{SESS: changed to the most generic description of these quantum routines}
Deriving such runtime estimates %for these core building blocks 
can be used to analyze a variety of quantum algorithms beyond asymptotic complexity.
This offers a pathway for realistic perspectives of %possible and necessary
future progress, instead of just asymptotic worst-case runtime,
which may lead to unrealistic expectations on possible benefits.

%Further research can also be conducted to improve our bounds, e.g., by closely
%investigating how the oracles access the input matrices.

\clearpage
\bibliography{bibliography}
\clearpage
\appendix
\onecolumn
\section{Additional Figures and Tables}
\label{sec:additional-figures}

\addtocounter{footnote}{1}
\begin{table*}[ht]
    \centering
    \begin{tabularx}{\columnwidth}{cccccr}
    \toprule
    Benchmark & group & \# Instances & \# Variables & \# Constraints & Source\\
    \midrule
    Random Graphs\footnotemark[\value{footnote}] & \easy{} & \SI{1711}{} & \SIrange{6}{174345}{} & \SIrange{2}{173755}{} & self generated\\
    DIMACS Graphs\footnotemark[\value{footnote}] & \easy{} & \SI{101}{} & \SIrange{912}{1001836}{}  & \SIrange{702}{999836}{} & \cite{dimacsCLQ}\\
    Maximum Flow & \easy{} & \SI{88}{} & \SIrange{13792}{1112772}{} & \SIrange{890}{44032}{} & \cite{Verma_Batra} via \cite{Jensen2022}\\
    \netlib & \hard{} & \SI{109}{} & \SIrange{49}{243209}{} & \SIrange{24}{78862}{} & \cite{Netlib} \\
    \miplib & \hard{} & \SI{84}{} & \SIrange{263}{641857}{} & \SIrange{32}{624166}{} & \cite{Miplib}\\
    \stochlp & \hard{} & \SI{108}{} & \SIrange{178}{1298168}{} & \SIrange{80}{450047}{}  & \cite{Stochlp}\\
    \misc & \hard{} & \SI{231}{} & \SIrange{5}{1124162}{} & \SIrange{2}{34078}{} & \cite{Misc}\\
    \bottomrule
    \end{tabularx}
    
    \caption{Full overview of the used benchmarking instances.\\
    {\footnotesize\protect\footnotemark[\value{footnote}]%
    Instances are LP relaxations of Min Vertex Cover, Max Independent Set and Max Clique.}}\label{tab:instance-details-full}
\end{table*}

\begin{figure*}[ht]
    \centering
    \includegraphics[width=\linewidth]{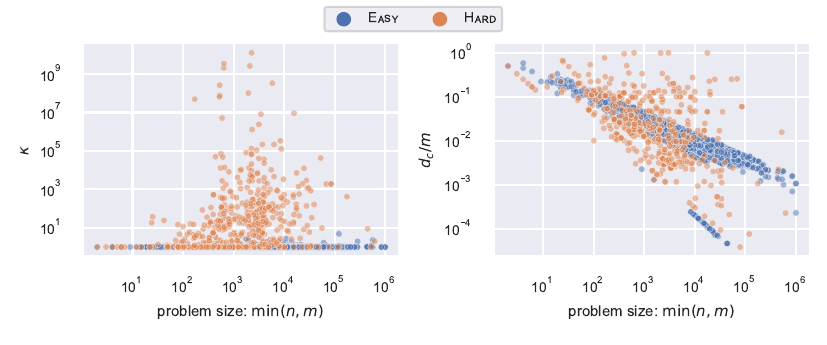}
    \caption{(Left) Average values for the condition number $\kappa(A_B)$. (Right) 
    the average fraction of nonzeros $d_c/m$ in the least sparse column of $A_B$.}
    \label{fig:sparsity-condition}
\end{figure*}

\begin{figure*}[ht]
    \includegraphics[width=\linewidth]{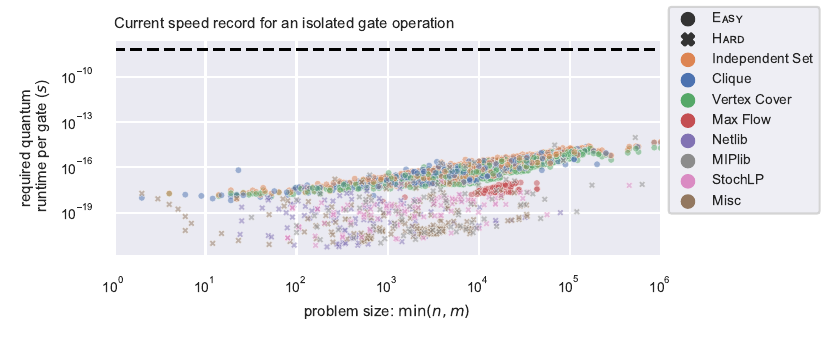}
    \caption{Extension of \cref{fig:quantum-gate-efficiency}. Minimum required gate evaluation time to outperform a classical iteration.
    Values are the average over all ratios between the (classical) 
    iteration runtime and the iteration's expected gate complexity bound, and are subject to a number of additional benevolent assumptions for the quantum algorithm.
    The black dotted line represents the fastest quantum gate that has currently been 
    realized.}
    \label{fig:quantum-gate-efficiency-all}
\end{figure*}

\begin{figure*}[ht]
    \includegraphics[width=\linewidth]{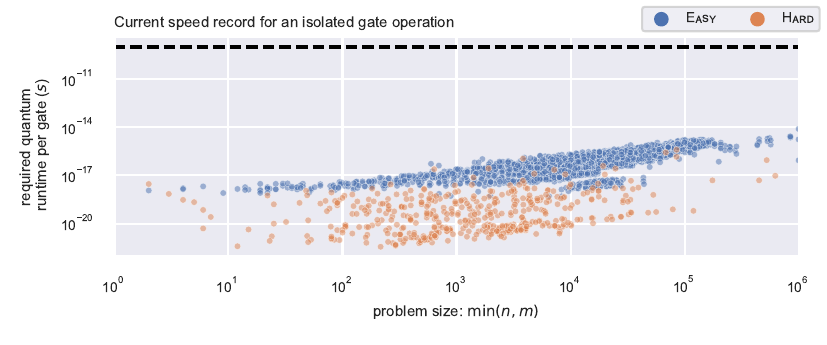}
    \caption{
        Analog to \cref{fig:quantum-gate-efficiency} for the classical Dantzig 
        pivoting rule and corresponding $\dantzig$ lower bounds from \cref{lem:cost_Dantzig}. 
    Minimum required gate evaluation time to outperform a classical iteration.
    Values are the average over all ratios between the (classical) 
    iteration runtime and the iteration's expected gate complexity bound, and are subject to a number of additional benevolent assumptions for the quantum algorithm.
    The black dotted line represents the fastest quantum gate that has currently been 
    realized.}
\end{figure*}

\begin{figure*}[ht]
    \includegraphics[width=\linewidth]{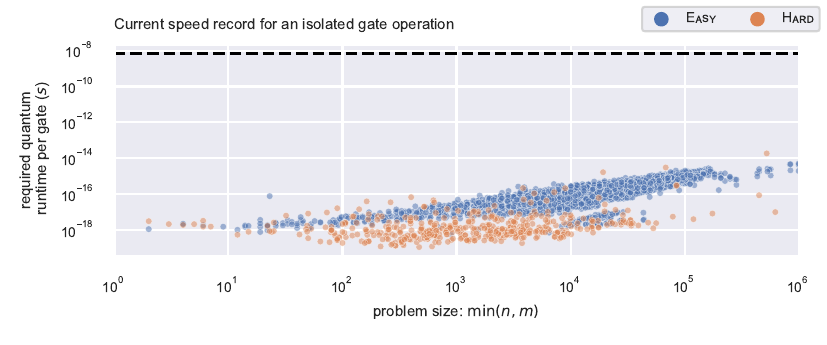}
    \caption{
        Analog to \cref{fig:quantum-gate-efficiency} for the classical steepest edge 
        pivoting rule and $\findColumn$ lower bounds from \cref{lemma:cost_FindColumn}.
        We compared against steepest edge, because a random pivoting rule is not available in GLPK.
        Minimum required gate evaluation time to outperform a classical iteration.
        Values are the average over all ratios between the (classical) 
        iteration runtime and the iteration's expected gate complexity bound, and are subject to a number of additional benevolent assumptions for the quantum algorithm. 
        The black dotted line represents the fastest quantum gate that has currently been 
        realized.}
\end{figure*}
\twocolumn

\clearpage
\section{Fast Subroutines for the Simplex Method}\label{sec:details-simplex}
\label{app:B}
In \cref{sec:simplex_iter_subroutines} we provide more details about the subroutines used in
\cref{alg:simplex-iteration}, in \cref{sec:simplex_iter_proofs}
we give proofs for the lemmas presented in \cref{sec:qsimplex}.

\subsection{Notation} 
In this section, we use the following conventions:
\begin{itemize}
\item Often we suppress Kronecker products when writing states, so that $\otimes_{i=1}^n\ket{x_i}$ is denoted $\ket{x_1x_2...x_n}$.
\item When denoting a quantum state whose amplitudes are given by a vector $v$, we write the quantum state as $\ket{v}$. For example, $\ket{\psi}=\ket{(0,1)}$ is a quantum state with $\psi_0=0$ and $\psi_1=1$. 
\item To implement several subroutines, we need to prepare superpositions of states $\ket{A_k}$ where $A_k$ are the columns of matrix $A$. We assume access to a unitary $U_{rhs}$ which can do this, namely $U_{rhs}\ket{k}\ket{0}=\ket{k}\ket{A_k}.$ 
\item Some subroutines need to apply a QLS to a superposition of linear systems $A_Bx=A_k$, for different values of $k$. We can do this by using $U_{rhs}$ as oracle $\mathcal{P}_b$ in QLS as described in \cref{sec:qlsa_fourier}.
We write this as $A_k=U_{rhs}\left(\cdot\right)$.
\item We always assume that the input matrix $A$ and cost vector $c$ are normalized, such that $\left|\left|A_B\right|\right|\leq 1, \left|\left|c_B\right|\right|=1$.
\item We denote by $\contr{n}{U}$ a controlled version of unitary $U$, controlled on $n$ qubits, see \cref{sec:controlled_unitaries} for more details.
\end{itemize}

\subsection{Subroutines of SimplexIter}\label{sec:simplex_iter_subroutines}
In the following, we explain in detail how the quantum subroutines used in \cref{alg:simplex-iteration} are implemented.
These are IsOptimal, FindColumn, IsUnbounded, and FindRow; they are built using
a number of core subroutines, which we summarize in
\cref{sec:simplex_iter_common_subroutines}, and several common quantum
algorithms, which are explained in \cref{sec:details-quantum-subroutines}.

\subsubsection{Core Subroutines}\label{sec:simplex_iter_common_subroutines}
IsOptimal, FindColumn, IsUnbounded, and FindRow share some core subroutines, which we explain in further detail.

\paragraph{RedCost}
The first core subroutine is RedCost, which is used by CanEnterNFN and CanEnterNFP, QStER, QDanR.
Given a basis $A_B$, a nonbasic column $A_k$, and a cost $c$ as inputs, RedCost prepares a state $\ket{\psi}$ with $\psi_0=\bar{c}_k$.
Here, $\bar{c}_k$ is the reduced cost of the column $k$. 
We first encode the solution of
\begin{equation*}
\begin{pmatrix}A_B & 0 \\ 
0 & 1\end{pmatrix}
\begin{pmatrix}x \\ 
y
\end{pmatrix} =
\begin{pmatrix}
A_k \\ 
c_k
\end{pmatrix}
\end{equation*}
in a quantum state, using a quantum linear solver (QLS). For an explanation of QLS see \cref{sec:qlsa_fourier}.
At this point, the state is $\ket{\hat{x}}=\ket{(A_B^{-1} A_k, c_k)}$, tensored with the success/failure flag of the QLS. 
We then consider a unitary that acts on $\ket{0}$ as $U_c\ket{0}=\ket{(-c_B, 1)}$. 
Acting with $U_c^\dagger$ on $\ket{\hat{x}}$, we produce a state whose zeroth amplitude is given by $c_k$, as wanted. 
More precisely, we have
\begin{equation*}
    \ket{\psi}\equiv U_c^\dagger \ket{\hat{x}}=\frac{1}{\sqrt{2}}\begin{pmatrix}
        -c_B & 1\\
        \dots & \dots
    \end{pmatrix}\cdot \frac{1}{\norm{\hat{x}}}\begin{pmatrix}
        A_B^{-1}A_k\\
        c_k
    \end{pmatrix}\,,
\end{equation*}
hence the zeroth amplitude of $\ket{\psi}$ is
\begin{equation}\label{eq:redcost_psi0}
    \psi_0=\frac{1}{\sqrt{2}\norm{\hat{x}}}\bar{c}_k\,.
\end{equation}
The prefactors $\sqrt{2}$ and $\norm{\hat{x}}$ are needed to mentain a proper normalization of the states. 
Moreover, RedCost also returns the success/failure flag of the QLS. This is used to ensure that the algorithm continues only if the former has been successful. 
\begin{algorithm}
\caption{RedCost}\label{alg:red_cost}
\begin{algorithmic}
\Function{RedCost}{basis $A_B$, nonbasic column $A_k$, cost $c$, tolerance $\varepsilon>0$}
\State $\hat{A}_B\gets \diag(A_B, 1)$, $\hat{b}\gets(A_k, c_k)^{T}$ 
\State Apply QLS($\hat{A}_B, \hat{b}, \varepsilon)$, let $\ket{\hat{x}}=\ket{(A_B^{-1} A_k, c_k)}$ be the solution
\State Compute $\ket{\psi}=U_c^{\dagger}\ket{\hat{x}}$
\Comment $U_c\ket{0}=\ket{(-c_B, 1)}$, result: $\psi_0=\bar{c}_k$
\State \Return $\ket{\psi}$ and the QLS flag
\EndFunction
\end{algorithmic}
\end{algorithm}

\paragraph{Interfere}
The second core subroutine is Interfere. This is used by SignEstNFN, SignEstNFP, QStER, and QDanR. 
In order to understand how Interfere works, let $U$ and $V$ be two unitaries that act on $\ket{0}$ as $U\ket{0}=\sum \psi_j\ket{j}$ and $V\ket{0}=\sum \beta_j\ket{j}$. 
With the ancilla initialized to the state $\ket{0}$, Interfere subsequently prepares the state
%Given such $U$ and $V$, and ancilla prepared in state $\ket{0}$, Interfere prepares the state
\begin{equation*}
\ket{\phi}=\frac{1}{2}\ket{0}_a \otimes \sum_j\left(\psi_j+\beta_j\right)\ket{j}-\frac{1}{2}\ket{1}_a \otimes \sum_j\left(\psi_j-\beta_j\right)\ket{j}
\end{equation*}
This is can be done following the steps in \cref{alg:interfere}.
\begin{algorithm}
\caption{Interfere}\label{alg:interfere}
\begin{algorithmic}
\Function{Interfere}{unitaries $U$, unitary $V$}
\State Apply $\mathrm{H}$ on an auxiliary qubit prepared in state $\ket{0}_a$
\State Act with $(\ketbra{0}\otimes \mathds{1}+\ketbra{1} \otimes U) \cdot(\ketbra{0}\otimes V+\ketbra{1} \otimes \mathds{1})$ \State Apply $\mathrm{H}$ to the ancilla
\EndFunction
\end{algorithmic}   
\end{algorithm}

\paragraph{SignEstNFN and SignEstNFP}
%SgnEstNFN and SgnEstNFP are two closely related subroutines, so we consider them together. 
Given the similar structure of SignEstNFN and SignEstNFP, we address both subroutines simultaneously.
SignEstNFN is used in CanEnterNFN, QStER, QDanR, IsUnbounded, and FindRow. SignEstNFP is used in CanEnterNFP. 
Both subroutines take as input a unitary $U$, an index $l$, and precision $\varepsilon$. 
Let $\ket{\psi}$ be the state generated by $U$ acting on $\ket{0}$, the goal of SignEstNFN and SignEstNFP is to decide whether the $l$-th amplitude of $\ket{\psi}$ is positive. 
More precisely, they should return 0 if $\psi_{l}<-\varepsilon$, and return 1 if $\psi_{l}\ge-\varepsilon$.
Both subroutines have a bounded probability of failure. 
The difference between the two is that SignEstNFN returns false negatives with small probability, while SignEstNFP returns false positives with small probability.
This affects only the last step of the algorithm, as we explain below.
%More precisely, if SgnEstNFN returns 0, we can conclude that $\psi_{l}<-\varepsilon$. 
%SgnEstNFN returns 0 if the $l$-th amplitude of $\ket{\psi}$ is negative. 
%More precisely, if SgnEstNFN returns 0, we can conclude that $\psi_{l}<-\varepsilon$. 

The first steps and the overall strategy are the same.
To implement SignEstNFN and SignEstNFP, we first add an ancilla and use Interfere, see \cref{alg:interfere}. 
%Interfere is another core subroutine explained in \cref{sec:simplex_iter_common_subroutines}.
The inputs of Interfere are the unitary $U$, same as the input of SignEst, and another unitary $V$, that acts on $\ket{0}$ as $V\ket{0}=\ket{l}$. 
Interfere generates a state with $(0,l)$-th amplitude equal to $(\psi_l+1)/2$. 
We then use quantum amplitude estimation (QAE) to estimate this amplitude. 
If the amplitude is smaller than 1/2, then we can conclude that $\psi_l$ is negative.
The way quantum amplitude estimation works is explained in \cref{app:Quantum_ampl_estimation}.
% The two subroutines differ in which precision parameters we use for QAE, and in exactly how we compare the amplitude in the last step.
The two subroutines differ in the precision parameters used for QAE, and in manner in which amplitude comparision is performed in the last step.

For SignEstNFN, we use QAE with precision parameters 
%In this case, we use it to estimate the $(0,l)$-th amplitude of the state prepared by Interfere, with precision parameters 
\begin{equation*}
    \varepsilon_{\qae}=\frac{\varepsilon}{\sqrt{3}\pi}\,,\quad \delta_{\qae} = \frac{1}{4}\,.
\end{equation*}
The output of QAE is a number $\tilde{\phi}\in[0,1]$, with probability greater than 3/4, $\varepsilon_{\qae}$-close to
\begin{equation*}
    \phi =\frac{1}{\pi}\arcsin(\frac{1+\psi_l}{2})\,.
\end{equation*}
From this expression, we notice that $\psi_l<0$ if $\phi\in[0,1/6]\cup[5/6,1]$. 
Since QAE works with finite precision, using this interval we would run the risk of having false positives and negatives. 
%QAE and all the other routines involved in SgnESTNFN work with finite precision. 
%Since in the simplex algorithm it is important to avoid picking a column with positive reduced cost by mistake, 
Therefore, SignEstNFN returns 1 if $\phi\in[1/6-2\varepsilon_{\qae},5/6+2\varepsilon_\qae]$.
As shown in \cite{nannicini2022fast} (see App.\ A.3), this ensures that SignEstNFN returns 1 if $\alpha\ge -\varepsilon$ with probability at least 3/4.
In other words, SignEstNFN returns "no false negative" with bounded probability.
%if SgnEstNFN returns 0, then $\psi_l<-\varepsilon$. 
%Notice that since we want to apply Grover to this oracle, it is important that also the comparison step is performed coherently.
\begin{algorithm}[ht]
\caption{SignEstNFN}\label{alg:SgnEstNFN}
\begin{algorithmic}
\Function{SignEstNFN}{unitary $U$, binary string $l$, tolerance $\varepsilon>0$}
\State $\alg \gets \text{Interfere}(U,V)$, $\chi_l(x) \gets \begin{cases}
1 \quad x=(0,l)\\
0 \quad \text{otherwise}
\end{cases}$
\Comment $V\ket{0}=\ket{l}$
\State $\varepsilon_{\qae}\gets \varepsilon/\sqrt{3\pi^2}$, $\delta_\qae\gets 1/4$
\State Run QAE$(\alg, \chi_l, \delta_\qae, \varepsilon_\qae)$, let $\tilde{\phi}$ be the estimate obtained
\Comment $\sin\left(\pi\tilde{\phi}\right)\approx (1+\psi_{l})/2$
\If{$\phi\in[1/6-2\varepsilon_{\qae},5/6+2\varepsilon_\qae]$}
\Return 1
\Else\,
\Return 0
\EndIf
\EndFunction
\end{algorithmic}
\end{algorithm}

For SignEstNFP, we use QAE with precision parameters 
%In this case, we use it to estimate the $(0,l)$-th amplitude of the state prepared by Interfere, with precision parameters 
\begin{equation*}
    \varepsilon_{\qae}=\frac{\varepsilon}{9\sqrt{3}\pi}\,,\quad \delta_{\qae} = \frac{1}{4}\,.
\end{equation*}
SignEstNFP returns 1 if $\phi\in(1/6-2\varepsilon_{\qae},5/6+2\varepsilon_\qae)$.
This makes sure that SignEstNFP returns 0 if $\alpha\le -\varepsilon$ with probability at least 3/4, see App.\ A.3 in \cite{nannicini2022fast}.
In other words, SignEstNFP returns "no false positive" with bounded probability.
\begin{algorithm}[h!]
\caption{{SignEstNFP}}
\label{alg:signestnfp}
\begin{algorithmic}
\Function{SignEstNFP}{unitary $U$, binary string $l$, tolerance $\varepsilon>0$}
\State $\alg \gets \text{Interfere}(U,V)$, $\chi_l(x) \gets \begin{cases}
1 \quad x=(0,l)\\
0 \quad \text{otherwise}
\end{cases}$
\Comment $V\ket{0}=\ket{l}$
\State $\varepsilon_{\qae}\gets \varepsilon/9\sqrt{3\pi^2}$, $\delta_\qae\gets 1/4$
\State Run QAE$(\alg, \chi_l, \delta_\qae, \varepsilon_\qae)$, let $\tilde{\phi}$ be the estimate obtained
\Comment $\sin\left(\pi\tilde{\phi}\right)\approx (1+\psi_{l})/2$
\If{$\phi\in(1/6-2\varepsilon_{\qae},5/6+2\varepsilon_\qae)$}
\Return 1
\Else\,
\Return 0
\EndIf
\EndFunction
\end{algorithmic} 
\end{algorithm}

\paragraph{CanEnterNFN and CanEnterNFP}
Given their similar nature, we give an explanation of both CanEnterNFN and CanEnterNFP subroutines together. 
The sole difference between these two is that CanEnterNFN uses SignEstNFN, while CanEnterNFP uses SignEstNFP.
%These are two similar core subroutines, so we explain them together. 
%The only difference between the two is that CanEnterNFN uses SgnEstNFN, while CanEnterNFP uses SgnEstNFP.

Given a basis $A_B$, nonbasic column $A_k$, cost $c$, and tolerance $\varepsilon$,  CanEnterNFN and CanEnterNFP return 1 if $\bar{c}_k<-\varepsilon\lVert\hat{x}\rVert$, and 0 otherwise.
Here, $\hat{x}=(A_B^{-1}A_k, c_k)$, see RedCost (\cref{alg:red_cost}) for details. 
Notice that this quantity changes in unpredictable ways during runtime. 
In particular, one might run into problems if $\hat{x}$ becomes large sometime during the simplex run.
Therefore, fixing the "tolerance" in the quantum algorithm is not as simple as in the classical algorithm. 
% For the moment, we set aside this sublety, and use the same value for $\varepsilon$ as we do in the classical algorithm.
For the moment, we set aside this sublety, and use the same value for $\varepsilon$ as that used in the classical algorithm.  
%In both the classical and quantum version of the simplex algorithm, it can happen that valid candidate columns (i.e.\ columns with negative reduced cost) are incorrectly discarded, when the reduced cost is negative but closer to zero than the tolerance of the algorithm. 
%The problem in the quantum algorithm is that the tolerance fluctuates depending on the value of $\lVert\hat{x}\rVert$, which is unknown. 
%In particular, if $\lVert\hat{x}\rVert$ becomes large during the simplex run, the quantum version of the algorithm might discard more valid non-basic columns than the classical algorithm and behave rather differently. 
%In practice, this means that in the quantum algorithm, we might need to consider smaller values of $\varepsilon$ to make up for the fluctuations of $\lVert\hat{x}\rVert$. 
%However, notice that the quantity entering in the gate count for the quantum algorithm is only $\varepsilon$, without $\lVert\hat{x}\rVert$. 
%Since so far we haven't found instances of LP on which the quantum algorithm is competitive with classical solvers, we have decided not to consider this possible overhead: we use the same tolerance in the quantum algorithm as we do in the classical one.

The oracle CanEnterNFN is implemented in two steps.
First, we use RedCost, see \cref{alg:red_cost}, to calculate the reduced cost of a given column and store it into the zeroth amplitude of a quantum state.
Second, we use SignEstNFN or SignEstNFP, see \cref{alg:SgnEstNFN} and \cref{alg:signestnfp}, to estimate the sign of this amplitude.
%Then we implement an algorithm that, given a unitary $U$ and an index $j$, computes the sign of the $j$-th amplitude of the state generated by $U$ (acting on $\ket{0}$): this is . 
We can then implement CanEnterNFN by applying SignEstNFN with inputs $U=\text{RedCost}$ and $j=0$. 
\begin{algorithm}[H]
\begin{algorithmic}
\caption{CanEnterNFN}\label{alg:canenternfn}
\Function{CanEnterNFN}{basis $A_B$, nonbasic column $A_k$, cost $c$, tolerance $\varepsilon>0$}
\State $\varepsilon_{\text{SE}}\gets \varepsilon \cdot 1.1/\sqrt{2}$, $\varepsilon_{\text{RC}}\gets \varepsilon \cdot 0.1/\sqrt{2}$
\State $U\gets \text{RedCost}(A_B, A_k, c, \varepsilon_{\text{RC}})$, $l \gets 0$
\State Apply SignEstNFN$(U, l, \varepsilon_{\text{SE}})$
\If{SignEstNFN returned 0 and flag `success'}
    	\Return 1
\Else\,
	\Return 0
\EndIf
\EndFunction
\end{algorithmic}
\end{algorithm}
Similarly, CanEnterNFP is implemented by applying SignEstNFP with inputs $U=\text{RedCost}$ and $j=0$.
\begin{algorithm}[H]
\caption{{CanEnterNFP}}
\label{alg:canenternfp}
\begin{algorithmic}
\Function{CanEnterNFP}{basis $A_B$, nonbasic column $A_k$, cost $c$, tolerance $\varepsilon>0$}
\State $\varepsilon_{\text{SE}}\gets \varepsilon \cdot 1.1/\sqrt{2}$, $\varepsilon_{\text{RC}}\gets \varepsilon \cdot 0.1/\sqrt{2}$
\State $U\gets \text{RedCost}(A_B, A_k, c, \varepsilon_{\text{RC}})$, $l \gets 0$
\State Apply SignEstNFP$(U, l, \varepsilon_{\text{SE}})$
\If{SignEstNFP returned 0 and flag `success'}
     \Return 1
\Else\,
 \Return 0
\EndIf
\EndFunction
\end{algorithmic}
\end{algorithm}
%Notice that this is the same as Algorithm \ref{alg:canenternfn} apart from the use of SignEstNFP instead of SignEstNFN. 
%Note that IsOptimal is the same algorithm as FindColumn (explained in Section \ref{alg:find_column}) apart from two differences:
%it uses SignEstNFP instead of SignEstNFN, and quantum amplitude estimation instead of QSearch.
%Since IsOptimal and FindColumn only differ by SignEst, refer to Section \ref{subsec:FindColumn} for a detailed explanation of each subroutine.

\subsubsection{IsOptimal}\label{subsec:IsOptimal}
%We explain how IsOptimal is implemented. 
Given the matrix $A$, the basis $B$, the cost $c$ and some precision $\varepsilon$ as inputs, IsOptimal determines if the current basis is optimal. 
%If it returns false, FindColumn is called, in order to determine which column should enter the basis. 
To implement IsOptimal, we use an oracle CanEnterNFP, see \cref{alg:canenternfp}, which determines if an input column $k$ has negative reduced cost. 
Here NFP stands for "no false positive", because the oracle is built such that with bounded probability it returns no false positive. 
%can enter the basis while avoiding false positives. 
We act with CanEnterNFP on a superposition of all nonbasic columns, which we prepare by acting with $U_{rhs}$ on $\ket{\psi_N}\equiv1/\sqrt{\abs{N}}\sum_{k\in N}\ket{k}$.
Here $N$ is the set of nonbasic columns and $\abs{N}=n-m$.
We then use quantum amplitude estimation (QAE) to estimate the amplitude with which CanEnterNFP returns 1.
If this amplitude is zero, we can conclude that no nonbasic column has negative reduced cost and the solution is optimal.
QAE is explained in \cref{app:Quantum_ampl_estimation}.
Let $\varepsilon_\qae$ and $\delta_\qae$ be the precision parameters of QAE, and $\psi_1$ be the amplitude we want to estimate.
QAE outputs an estimate $\tilde{\phi}$ that with probability at least $1-\delta_\qae$ is $\varepsilon_\qae$-close to $\phi\equiv 1/\pi\arcsin\psi_1$, i.e.\ $\abs{\tilde{\phi}-\phi}\le \varepsilon_\qae$.
In the case of IsOptimal, we need to choose $\varepsilon_\qae$ in a such a way that we are able to decide if no column has negative reduced cost.
When only one column has negative reduced cost, $\psi_1=1/\sqrt{\abs{N}}$; therefore, we need to choose $\varepsilon_\qae\sim 1/\sqrt{\abs{N}}$. 
We choose precision parameters $\varepsilon_\qae=1/(4\sqrt{\abs{N}})$ and $\delta=1/4$. 
Our choice of $\varepsilon_\qae$ guarantees that if $\tilde{\phi}\in[0,\varepsilon_\qae)\cup(1-\varepsilon_\qae,1]$, then $\psi_1< 1/\sqrt{\abs{N}}$, and we can conclude that the solution is optimal.
%Once we have this oracle, we can use quantum amplitude estimation to find out if the set of possible columns is empty or not. It returns 1 when this set is empty, that is, when the basis is optimal.
\begin{algorithm}[H]
\caption{IsOptimal}\label{alg:is_optimal}
\begin{algorithmic}
\Function{Isoptimal}{matrix $A$, basis $B$, cost $c$, tolerance $\varepsilon>0$}
\State $\ket{\psi_N}\gets\frac{1}{\sqrt{\abs{N}}}\sum_{k\in N}\ket{k}$
\Comment Prepare superposition nonbasic columns
\State $\alg \gets \text{CanEnterNFP}(A_B, A_k=U_{r h s}(\psi_N), c, \varepsilon)$, $\chi\gets \mathds{1}_2$
\Comment $U_{rhs}\ket{k} \ket{0}=\ket{k}\ket{A_k}$
\State $\delta_\qae \gets 1/4$, $\varepsilon_\qae\gets 1/(4\sqrt{\abs{N}})$
\State Run QAE$(\alg, \chi, \varepsilon_\qae, \delta_\qae)$, let $\tilde{\phi}$ be the result
\If{$\tilde{\phi}\in[0,\varepsilon_\qae)\cup(1-\varepsilon_\qae,1]}$
    \Return 1
\Else \;\Return 0
\EndIf
%\Comment Find probability of CanEnterNFP = 1
\EndFunction
\end{algorithmic}
\end{algorithm}

\subsubsection{FindColumn: Random, QStER and QDanR}\label{subsec:FindColumn}
In the following we list three algorithms that can be used for determining the pivot column:
Random -- corresponding to the random pivot rule, QStER -- corresponding to a quantum steepest edge rule  and QDanR -- corresponding to the quantum version of Dantzig`s rule.
We first explain how $\findColumn$ is implemented. 

\paragraph{\findColumnNoMath}

Given a basis $B$, a cost $c$, and a tolerance $\varepsilon>0$ as inputs, $\findColumn$ outputs the index $k$ of a column with negative reduced cost.
This is done in two steps.
First, we implement an oracle, CanEnterNFN, that takes as input the index of a column $k$, and returns 1 if the reduced cost of the column is negative and 0 otherwise. 
Here NFN stands for "no false negative", because the oracle is built in such a way that it returns no false negative with bounded probability. 
Once we obtain this oracle, we can run QSearch (see \cref{app:QSearch}) on it to find a column with negative reduced cost.
\begin{algorithm}[ht]
\caption{$\findColumn$}\label{alg:find_column}
\begin{algorithmic}
\Function{$\findColumn$}{matrix $A$, basis $B$, cost $c$, tolerance $\varepsilon>0$}
\State $\chi \gets \text{CanEnter}(A_B, A_k=U_{r h s}(\cdot), c, \varepsilon)$
\Comment $U_{rhs}\ket{k} \ket{0}=\ket{k}\ket{A_k}$
\State Run QSearch$(\chi, \varepsilon)$
\EndFunction
\end{algorithmic}
\end{algorithm}

%In particular, notice that for RedCost, we have  $l=0$, and $\psi_0$ is given by \eqref{eq:redcost_psi0}. 
%The resulting state is $\ket{\phi}=(\psi_0+1)/2\ket{0} \otimes\ket{0}+\dots$, where the ellipsis indicates terms orthogonal to $\ket{0}\otimes \ket{0}$. 
\paragraph{\steepestEdgeRuleNoMath}
%Similar to the classical case, one can implement more sophisticated pivoting rules than just randomly picking the first index which %has negative reduced cost. 
%These make the overall search for the optimal solution more efficient as fewer iterations are needed.
%Prominent examples include Dantzig's rule, i. e. finding the column with minimum reduced cost, and the \emph{steepest edge rule} %introduced by Goldfarb and Forrest. 
%Nannicini suggests quantum equivalents to 
We explain how to implement the quantum steepest edge rule and afterwards the quantum version of Dantzig's rule. 
Both routines apply the \emph{quantum minimum finding algorithm} to a subroutine that compares the reduced costs of two columns.

%We begin by explaining how to implement the quantum steepest edge rule. The subroutine applies the \emph{quantum minimum finding algorithm} to a subroutine that compares the reduced costs of two columns. 
The algorithm $\steepestEdgeRule$ takes as input the simplex tableau $A$, the cost vector $c$ and a precision parameter $\varepsilon$ and applies $\qmin$ to a subroutine $\steepestEdgeCompare$ that returns 1 if 
\begin{align*}
\frac{\bar{c}_k}{\norm{A_{B}^{-1} A_k}} \leq (1- \varepsilon) \frac{\bar{c}_j}{\norm{A_{B}^{-1} A_j}} - \varepsilon,
\end{align*}
where $j$ is the reference index as determined by $\qmin$, and 0 else. 
Here we interpret $\steepestEdgeCompare$ as an oracle function $\chi_j (i) : c \longrightarrow \{ 0,1 \}$ with
\[ 
 \chi_j(i) = \left\{
\begin{array}{ll}
  1, & \text{if} \ \  \steepestEdgeCompare(A_B, A_j, A_k, c, \varepsilon) = 1\\
   0,  & \text{else.}  
\end{array} 
\right.
\]

\begin{algorithm}[ht]
\caption{$\steepestEdgeRule$}
\begin{algorithmic}
\Function{$\steepestEdgeRule$}{$A$, $B$, $c$, $\varepsilon$} \Comment{$||A_B|| \leq 1$, $||c_B ||= 1$}
\State Apply $\qmin$($\chi$, $|c|= n-m$) to the cost vector $c$ interpreted as list and the oracle $\chi$:
\Function{$\steepestEdgeCompare$}{$A_B$, $A_j$, $A_k$, $c$, $\varepsilon$} 
\State Compute an estimate of the norms $\Tilde{d}_j =||(A_{B}^{-1} A_j , c_j )||$, $\Tilde{d}_k = ||(A_{B}^{-1} A_k, c_k)||$ with relative error $\varepsilon/4$.
\State Compute an estimate of the norms $\Tilde{e}_j =||A_{B}^{-1} A_j ||$, $\Tilde{e}_k = ||A_{B}^{-1} A_k||$ with relative error $\varepsilon/4$.
%\State $\varepsilon_{\text{RC}}\gets 0.1\varepsilon/ (\sqrt{2} \mathrm{max}_{l \in N} |c_l|)$ 
\State Let $U$ be the unitary that implements RedCost($A_B$, $A_j$, $c$, $\varepsilon / \max_{l \in [N]}|c_l|$) and then multiplies  $\ket{0^{\lceil \mathrm{log}m\rceil}}$ with $\frac{\Tilde{d}_j}{\Tilde{e}_j \mathrm{max}_{l \in N} |c_l| }$.
\State Let $V$ be the unitary that implements RedCost($A_B$, $A_k$, $c$, $\varepsilon / \max_{l \in [N]}|c_l|$) and then multiplies $\ket{0^{\lceil\mathrm{log}m\rceil}}$ with  $\frac{\Tilde{d}_k}{\Tilde{e}_k \mathrm{max}_{l \in N} |c_l| }$.
\State Let $W$ be the unitary that implements Interfere($U$, $V$). 
\State Apply  SignEstNFN($W$, $0^{\lceil \mathrm{log}m \rceil +1}$, $\frac{\varepsilon}{8 \mathrm{max}_{l \in N} |c_l|})$.
\If  {SignEstNFN($W$, $0^{\lceil \mathrm{log}m \rceil +1}$, $\frac{\varepsilon}{8 \mathrm{max}_{l \in N} |c_l|})$ returns 0}
\Return 1
\Else{}
\Return 0
\EndIf  
\EndFunction
\State Let $\chi_{k}(j)=0 $ for all $j \in [n-m]$
\Return $k$
\EndFunction
\end{algorithmic}
\end{algorithm}
\textbf{Remark:}
It should be noted that we have employed a different precision parameter than Nannicini in \cite{nannicini2022fast}. In order to argue our choice, we reconstruct the relevant part of the proof of Theorem 10 in \cite{nannicini2022fast} in the following. 
For simplicity, we abbreviate the tilde above the $\Tilde{e}_j$ and $\Tilde{d}_j$ symbols. 
We want to show that, if \cref{eq:SteepestEdgeCompare} holds, then SignEstNFN returns $0$.
In other words, we want to prove that if
\begin{equation}\label{eq:SteepestEdgeCompare}
\frac{\bar{c}_k}{\norm{A_{B}^{-1} A_k}} \leq (1- \varepsilon) \frac{\bar{c}_j}{\norm{A_{B}^{-1} A_j}} - \varepsilon,
\end{equation} holds, then the following has to be true
\begin{align*}
    \frac{1}{2}(\beta^{\prime}_0 - \alpha^{\prime}_0) \leq - \frac{\varepsilon}{8c_{max}} , 
\end{align*}
where $\alpha^{\prime}_{0} = \alpha_0 \cdot d_j/e_j c_{max}$ and $\beta^{\prime}_{0} = \beta_0 \cdot d_k/e_k c_{max}$ correspond to the amplitudes encoded by the unitaries $U$ and $V$. 
According to the RedCost subroutine $\alpha_0$ and $\beta_0$ have to satisfy
\begin{align}\label{eq:precisionalphaandbeta}
    \beta_0 \leq \frac{\bar{c}_k}{d_k} +   \frac{\varepsilon}{10c_{max}} \quad \text{and} \quad
    \alpha_0 \leq \frac{\bar{c}_j}{d_j} +  \frac{\varepsilon}{10c_{max}}.
\end{align}
Moreover, as the norms have relative error of $\epsilon /4 $, we know that
\begin{align*}
    \left(1 - \frac{\epsilon}{4}\right) \norm{A_{B}^{-1} A_j} &\leq  e_j \leq \left(1 + \frac{\epsilon}{4}\right)\norm{A_{B}^{-1} A_j}\\
    \left(1 - \frac{\epsilon}{4}\right) \norm{(A_{B}^{-1} A_j, c_j)} &\leq d_j \leq \left(1+ \frac{\epsilon}{4}\right) \norm{(A_{B}^{-1} A_j, c_j)}.
\end{align*}
Putting everything together, we estimate $\beta^{\prime}_{0}$ as follows
\begin{equation*}
    \beta^{\prime}_{0} \leq \left(\frac{\bar{c}_k}{d_k} +  \frac{\varepsilon}{10 c_{max}} \right) \frac{d_k}{c_{max} e_k} \left( \frac{1+ \frac{\varepsilon}{4}} {1- \frac{\varepsilon}{4}} \right)
\end{equation*}
which leads to
\begin{align*}
    \beta^{\prime}_{0} \leq \frac{\bar{c}_k}{e_k c_{max}} \left( \frac{1+ \frac{\varepsilon}{4}} {1-\frac{\varepsilon}{4}} \right) +  \frac{\varepsilon}{10 c_{max}} \frac{d_k}{c_{max} e_k} \left( \frac{1+ \frac{\varepsilon}{4}} {1- \frac{\varepsilon}{4}} \right) .
\end{align*}
Since
\begin{align*}
    \frac{1+ \frac{\varepsilon}{4}} {1- \frac{\varepsilon}{4}} \leq 1 + \frac{\epsilon}{2} + \frac{\epsilon ^{2}}{4}.
\end{align*}
We conclude that
\begin{align*}
    \beta^{\prime}_{0} \leq &\frac{\bar{c}_k}{e_k c_{max}} \left( 1 + \frac{\epsilon}{2} + \frac{\epsilon ^{2}}{4} \right) \nonumber\\
 &+  \frac{\varepsilon}{10 c_{max}} \frac{d_k}{c_{max} e_k} \left( 1 + \frac{\epsilon}{2} + \frac{\epsilon ^{2}}{4} \right) .
\end{align*}
Assuming that \cref{eq:SteepestEdgeCompare} holds true, we find
\begin{align*}
    \beta^{\prime}_{0} \leq &\frac{\bar{c}_j}{c_{max} e_j} \left( 1 + \frac{\epsilon}{2} + \frac{\epsilon ^{2}}{4} \right) \left( 1 - \varepsilon \right)   \varepsilon \frac{d_k}{10 (c_{max})^{2} e_k}  \nonumber\\
 &\cdot \left( 1 + \frac{\epsilon}{2} + \frac{\epsilon ^{2}}{4} \right)  - \left( 1 + \frac{\epsilon}{2} + \frac{\epsilon ^{2}}{4} \right)\frac{\varepsilon}{c_{max}}.
\end{align*}
Knowing that
\begin{align*}
   \frac{d_k }{e_k c_{max}} \leq 1 \quad \text{and} \quad
    \left( 1 + \frac{\epsilon}{2} + \frac{\epsilon ^{2}}{4} \right) \left( 1 - \epsilon \right) \leq 1 ,
\end{align*}
we obtain
\begin{align*}
    \beta^{\prime}_{0} \leq &\frac{\bar{c}_j}{c_{max} e_j}  +  \frac{\varepsilon}{10 c_{max}} \left( 1 + \frac{\epsilon}{2} + \frac{\epsilon ^{2}}{4} \right) \nonumber \\
 &- \frac{\varepsilon}{c_{max}} \left( 1 + \frac{\epsilon}{2} + \frac{\epsilon ^{2}}{4} \right).
\end{align*}
Now using
\begin{align*}
    1 + \frac{\epsilon}{2} + \frac{\epsilon ^{2}}{4} \leq 2 \quad \text{and} \quad
    -1 - \frac{\epsilon}{2} - \frac{\epsilon ^{2}}{4} \leq -1
\end{align*}
we obtain
\begin{align*}
     \beta^{\prime}_{0} \leq \frac{\bar{c}_j}{c_{max} e_j}   +  \frac{\varepsilon}{5c_{max}}  - \frac{\varepsilon}{c_{max}} \leq  \alpha^{\prime}_{0} -  \frac{\varepsilon}{4c_{max}} 
\end{align*}
which is equivalent to
\begin{align*}
    \frac{1}{2}\left(  \beta^{\prime}_{0}-  \alpha^{\prime}_{0} \leq \right) \leq - \frac{\varepsilon}{8c_{max}}.
\end{align*}
However, we obtained this result only by including an additional factor $1/c_{max}$ in the precision of the amplitudes $\alpha_0$ and $\beta_0$ (i.e. in \cref{eq:precisionalphaandbeta}).

%Neglecting the cost for norm estimation, the cost of QStER is given by: 
%\begin{align}
%\mathcal{C}[\mathrm{QStER}] = \langle n_{QMin}\rangle \times \mathcal{C}[\steepestEdgeCompare],
%\end{align}
%where $\langle n_{QMin}\rangle$ denotes the expected number of oracle calls of the quantum minimum finding algorithm (see \cref{lem:nqmin}) and $\mathcal{C}[\steepestEdgeCompare]$ denotes the cost of $\steepestEdgeCompare$, which is discussed in detail in the proof of \cref{lem:steepest}.

\paragraph{\dantzigNoMath}
In a similar fashion one can design a quantum subroutine that implements Dantzig's rule by applying $\qmin$ to a quantum oracle that compares two reduced cost vectors. The algorithm $\dantzig$ takes as input the simplex tableau $A$, the cost vector $c$ and a precision parameter $\varepsilon$ and applies $\qmin$ to a subroutine that returns 1 if 
\begin{align}\label{eq:qdan}
\bar{c}_k\leq (1- \varepsilon) \bar{c}_j - \varepsilon,
\end{align}
where $j$ is the reference index as determined by $\qmin$,
and 0 else.
\begin{algorithm}[ht]
\caption{$\dantzig$}
\begin{algorithmic}
\Function{$\dantzig$}{$A$, $B$, $c$, $\varepsilon$} \Comment{$||A_B|| \leq 1$, $||c_B ||= 1$}
\State Apply $\qmin$($\chi$, $|c|=n-m$) to the cost vector $c$ interpreted as list and the oracle $\chi$:
\Function{RedCostCompare}{$A_B$, $A_j$, $A_k$, $c$, $\varepsilon$} 
\State Compute an estimate of the norms $\Tilde{d}_j =||(A_{B}^{-1} A_j , c_j )||$, $\Tilde{d}_k = ||(A_{B}^{-1} A_k, c_k)||$ with relative precision $\varepsilon/4$.
\State Let $U$ be the unitary that implements RedCost($A_B$, $A_j$, $c$, $\varepsilon/\mathrm{max}_{l \in N} |c_l| \Tilde{e}_j$) and then multiplies  $\ket{0^{\lceil \mathrm{log}m\rceil}}$ with $\frac{\Tilde{d}_j}{\Tilde{e}_j \mathrm{max}_{l \in N} |c_l |}$.
\Comment{$\Tilde{e}_j =||A_{B}^{-1} A_j ||$} 
\State Let $V$ be the unitary that implements RedCost($A_B$, $A_k$, $c$, $\varepsilon/\mathrm{max}_{l \in N} |c_l| \Tilde{e}_k$) and then multiplies $\ket{0^{\lceil\mathrm{log}m\rceil}}$ with  $\frac{\Tilde{d}_k}{\Tilde{e}_k \mathrm{max}_{l \in N} |c_l| }$. 
\Comment{$\Tilde{e}_k = ||A_{B}^{-1} A_k||$}
\State Let $W$ be the unitary that implements Interfere($U$, $V$). 
\State Apply  SignEstNFN($W$, $0^{\lceil \mathrm{log}m \rceil +1}$, $\frac{\varepsilon}{8 \mathrm{max}_{l \in N} |c_l|})$.
\If  {SignEstNFN($W$, $0^{\lceil \mathrm{log}m \rceil +1}$, $\frac{\varepsilon}{8 \mathrm{max}_{l \in N} |c_l|})$ returns 0}
\Return 1
\Else{}
\Return 0
\EndIf  
\EndFunction
\State Let $\chi_{k}(j)=0 $ for all $j \in [n-m]$
\Return $k$
\EndFunction
\end{algorithmic}
\end{algorithm}
We interpret the oracle $\chi$ as a function $\chi_j (i) : c \longrightarrow \{ 0,1 \}$ with
\[ 
\chi_j(i) = \left\{
\begin{array}{ll}
  1, & \text{if} \ \  \text{RedCostCompare}(A_B, A_j, A_k, c, \varepsilon) = 1\\
  0,  & \text{else.}  
\end{array} 
\right.
\]
The complexity of this algorithm scales the same as $\steepestEdgeRule$ (neglecting the cost of norm estimation):
\begin{align*}
\mathcal{C}[\mathrm{QDanR}] = \langle n_{QMin}\rangle \times \mathcal{C}[\mathrm{RedCostCompare}],
\end{align*}
where $ \langle n_{\qmin}\rangle$ is the number of calls from the quantum minimum finding algorithm to the oracle $\chi$.

\subsubsection{IsUnbounded}

%\paragraph{Explanation of the algorithm in words.}

An instance is found to be unbounded if and only if 
$u \coloneqq A_B^{-1} A_k < 0$. 
Hence the quantum algorithm IsUnbounded is a Grover search 
for a positive amplitude of $u$. 
The Grover oracle $Q \coloneqq \text{SignEstNFN}^+(\text{QLS}(A_B, A_k, \delta/10))$ comprises the following two steps:
\begin{itemize}
    \item Solve the linear system $A_B x = A_k$ with precision $\frac{\delta}{10}$ by applying the QLS unitary, obtain the solution state $\ket{x^*}$.
    \item Apply SignEstNFN with precision $\frac{9 \delta}{10}$ to determine whether a specified amplitude of $\ket{x^*}$ is positive.
\end{itemize}
If no positive amplitude is found, the algorithm returns $1$, otherwise, it returns $0$.

\begin{algorithm}[H]
\caption{IsUnbounded}\label{alg:is_unbounded}
\begin{algorithmic}
\Function{IsUnbounded}{normalized basis $A_B$ ($||A_B||\leq 1$), 
non-basic column $A_k$ to enter the basis, 
precision $\delta > 0$}
\State $\chi \gets \text{SignEstNFN}^+(\qls(A_B, A_k, \frac{\delta}{10}))$
\State Run $\qsearch(\chi, \frac{9\delta }{10})$
\EndFunction
\end{algorithmic}
\end{algorithm}

\subsubsection{FindRow}

\begin{algorithm}
\caption{FindRow}\label{alg:find_row}
\begin{algorithmic}
\Function{FindRow}{normalized basis $A_B$ ($||A_B||\leq 1$), 
non-basic column $A_k$ to enter the basis, 
constraint vector $b$, precision $\delta > 0$}
\State $\varepsilon_Q \gets -2 \frac{\delta}{\|A_B^{-1}A_k\|} + 
\frac{\delta}{2}$
\State $\varepsilon_S \gets \frac{\delta}{\|A_B^{-1}A_k\|}$
\State $\chi_r \gets \text{SignEstNFN}
(\qls(A_B, b-rA_k, \varepsilon_Q), \varepsilon_S)$
\State $\alg_r \gets 1 \text{ if } 
\qsearch(\chi_r)$ finds a marked element, $0$ otherwise.
\State Binary search for $r^* > 0$ such that $\alg_{r^*}$ returns $1$ 
and $\alg_{r^*-\frac{\delta}{2\kappa ||A_k||}}$ returns $0$
\Return the index $j$ corresponding to the marked element
\EndFunction
\end{algorithmic}
\end{algorithm}

%\paragraph{Explanation of the algorithm in words.}
\paragraph{Description of the Algorithm.}

The heart of this algorithm is solving the linear system 
$A_Bx = b - r A_k$ for $r \geq 0$. 
The solution is 
$$x_r^* = A_B^{-1}b - r A_B^{-1}A_k = x - ru.$$ 
For each basis index $j$ which satisfies $u_j > 0$, 
we write $r_j := \frac{x_B(j)}{u_j} \geq 0$. 
Observe that $(x_{r_j}^*)_j = x_{B(j)} - \frac{x_B(j)}{u_j} u_j = 0$. 
The goal of the algorithm is to find the smallest $r_j$ and 
we write $r^* := \min_{j} r_j$. 
This value is characterized by the fact that $x^*_{r^* - \delta} > 0$ 
for all (especially small) $\delta > 0$ but 
$x^*_{r^*} \ngtr 0$.
The quantum version of the logic described above is a quantum algorithm $\mathcal{A}_r$, 
where $r > 0$, which 
\begin{itemize}
    \item uses a QLS algorithm to solve the linear system $A_B x = b - r A_k$, preparing a quantum state $\ket{x^*}$ proportional to the solution vector $x^* = A_B^{-1}(b - r A_k)$,
    \item then applies SignEstNFN (like in FindColumn) to find out if a specific amplitude of the QLS solution state $\ket{x^*}$ is negative
    \item Applies the Grover search QSearch to the two preceding steps to determine 
    if any amplitude of $\ket{x^*}$ is negative. 
    More concretely, the Grover oracle is given by 
    $Q_r \coloneqq \text{SignEstNFN}(\text{QLS}(A_B, b-r A_k, \delta'))$, 
    where $\delta'$ is the QLS precision which we expand on in the paragraph below.
\end{itemize}
FindRow then performs a binary search on $r$ to find $r^*$ 
such that $\mathcal{A}_{r^*}$ returns $1$ and $\mathcal{A}_{r^*-\delta}$ 
returns $0$ for a specific small choice of $\delta$, 
chosen to control the overall error.

\paragraph{Error Analysis}
We now briefly discuss how to choose the precision for the 
QLS and sign estimation subroutines. 
We denote these precision values by $\varepsilon_Q$ and $\varepsilon_S$, 
respectively.
The sign estimation subroutine is applied to the 
QLS unitary to determine if a specified component of the 
solution vector $x_r^* = A_B^{-1}b - r A_B^{-1}A_k$ is $<\delta/2$. 
The QLS precision $\varepsilon_Q$ determines that 
\begin{equation*}
    \abs{\alpha_k - x_{r,k}^*} < \varepsilon_Q,
\end{equation*}
where $x_{r,k}^*$ is the $k$-th component of the 
solution vector $x_r^*$ and $\alpha_k$ is the corresponding amplitude 
of the quantum state prepared by the QLS algorithm.
According to Prop 3 of \cite{nannicini2022fast} 
the sign estimation procedure applied to the 
$k$-th amplitude returns $0$ with high probability if 
$\alpha_k < -2\varepsilon_S$. 
For SignEstNFN to still correctly indicate whether this the case, 
we therefore require that
\begin{equation}\label{eq:error_relation_findrow}
-2 \varepsilon_S = -\frac{\delta}{2} + \varepsilon_Q.
\end{equation}
Nannicini suggests
$\varepsilon_S \coloneqq \frac{\delta}{\|A_B^{-1}A_k\|}$, 
which would in turn determine 
$\varepsilon_Q = -2 \frac{\delta}{\|A_B^{-1}A_k\|} + \frac{\delta}{2}$ 
according to \cref{eq:error_relation_findrow}.

\subsection{Proofs of Lemmas from \cref{sec:qsimplex}}\label{sec:simplex_iter_proofs}
In this section, we collect the proofs of \cref{lem:isOptimal,lem:cost_isUnbounded,lem:findRow,lem:steepest}, presented in \cref{sec:qsimplex} and additionally provide the results for $\findColumn$ and $\dantzig$.
Notice that the proofs of \cref{lem:qsearch} and \cref{lem:qls} are instead given in \cref{sec:details-quantum-subroutines}.
As explained in \cref{sec:simplex_iter_subroutines}, IsOptimal, FindColumn, IsUnbounded, and FindRow rely on a number of core subroutines, see \cref{sec:simplex_iter_common_subroutines}.
So, we first prove lower bounds for these core subroutines in \cref{sec:proof_core_subroutines}.

\subsubsection{Lower Bounds for Core Subroutines}\label{sec:proof_core_subroutines}
We prove lower bounds for the cost of the core subroutines presented in \cref{sec:simplex_iter_common_subroutines}.
These are used in the proofs of the other lemmas presented in this section.

\begin{lemma}[Cost of RedCost]
\label{lemma:cost_redcost}
The cost of RedCost is lower bounded by
\begin{equation*}
\cost[\text{RedCost}(A_B, A_k, c, \varepsilon)]\ge\cost[\qls({A}_B, A_k, \varepsilon)].
\end{equation*}
A lower bound for the cost of implementing a version of QLS is provided in \cref{lem:qls}.
\end{lemma}
\begin{proof}
We lower bound the cost of RedCost by only considering the second step in \cref{alg:red_cost}.
\begin{equation*}
\cost[\text{RedCost}(A_B, A_k, c, \varepsilon)]\ge\cost[\qls(\hat{A}_B, \hat{b}, \varepsilon)]\,,
\end{equation*}
where
\begin{equation*}
    \hat{A}_B=\begin{pmatrix}
    A_B & 0\\
    0 & 1
    \end{pmatrix}\,,\quad
    \hat{b} = \begin{pmatrix}
    A_k\\
    c_k
    \end{pmatrix}\,.
\end{equation*}
For simplicity, we lower bound the cost of solving this system of equations with a simpler system of equations, where we use $A_B$ instead of $\hat{A}_B$ and $A_k$ instead of $\hat{b}$. 
The difference between the two is a small overhead, which we neglect.
We conclude that
\begin{equation}\label{eq:cost_redcost}
\cost[\text{RedCost}(A_B, A_k, c, \varepsilon)]\ge\cost[\qls({A}_B, A_k, \varepsilon)].
\end{equation}
\end{proof}

\begin{lemma}[Cost of Interfere]
\label{lem:cost_interfere}
The cost of Interfere is lower bounded by
\begin{equation*}
    \cost[\text{Interfere}(U,V)]\ge \cost[U]+\cost[V]\,,
\end{equation*}
where $\cost[U]$ and $\cost[V]$ are the cost of implementing the input unitaries, $U$ and $V$.
\end{lemma}
\begin{proof}
The cost of Interfere is dominated by the controlled action of $U$ and $V$,
\begin{equation*}
    \cost[\text{Interfere}(U,V)]\ge \cost[\contr{1}U]+\cost[\contr{1}V]\,.
\end{equation*}
To deal with the control, we consider the simple lower bounds $\cost[\contr{1}U]\ge \cost[U]$ and $\cost[\contr{1}V]\ge \cost[V]$, leading to
\begin{equation}\label{eq:cost_interfere}
    \cost[\text{Interfere}(U,V)]\ge \cost[U]+\cost[V]\, .
\end{equation}
\end{proof}

\begin{lemma}[Cost of SignEstNFN]\label{lem:cost_signestnfn}
The cost of SignEstNFN is lower bounded by
\begin{equation*}
    \cost[\text{SignEstNFN}(U,l,\varepsilon)]\ge \left(\frac{5\sqrt{3}\pi}{\varepsilon}-1\right) \cost[U]\,.
\end{equation*}
\end{lemma}
\begin{proof}
The cost of SignEstNFN$(U, l, \varepsilon)$ is given by the cost of implementing QAE, and the cost of checking whether the result of QAE is in $[1/6-2\varepsilon_\qae, 5/6+2\varepsilon_\qae]$, see \cref{alg:SgnEstNFN}
We lower bound the cost of SignEstNFN by only considering the first, i.e.\ the cost of QAE.
Let $\varepsilon_{\qae}=\varepsilon/\sqrt{3\pi^2}$, $\delta_\qae=1/4$, $\alg=\mathrm{Interfere}(U, \mathds{1})$, and $\chi_l$ a function that selects the $(0,l)$-th component of the vector, i.e.\
\begin{equation*}
    \chi_l(x)= \begin{cases}
1 \quad x=(0,l)\\
0 \quad \mathrm{otherwise}
\end{cases}\,,
\end{equation*}
then
\begin{equation*}
    \cost[\text{SignEstNFN}(U,l,\varepsilon_{\text{SE}})]\ge\cost[\qae(\alg,\chi_l,\varepsilon_\qae, \delta_\qae)]\,.
\end{equation*}
The cost of quantum amplitude estimation can be lower bounded by
\begin{equation}\label{eq:qae_simple_low_bound}
    \cost[\qae(\alg, \chi_l,\varepsilon_\qae, \delta_\qae)] \ge (2^{n_{c}+1}-1)\cost[\alg]\,,
\end{equation}
where
\begin{equation*}
    n_{c}=\log_2\frac{1}{2\varepsilon_\qae}+\log_2\left(1+\frac{1}{\delta_\qae}\right)\,.
\end{equation*}
This can be shown combining \cref{lem:cost_qae} and \cref{lem:qpeCost}, and using $\cost[Q]\ge 2\cost[\alg]$ ($Q$ is the Grover operator).
A lower bound for the cost of $\alg=\text{Interfere}(U, \mathds{1})$ is given in \cref{lem:cost_interfere}.
In this case, $V=\mathds{1}$, so we have
\begin{equation*}
    \cost[\text{Interfere}(U,\mathds{1})]\ge \cost[U]\,.
\end{equation*}
Plugging this into \cref{eq:qae_simple_low_bound}, and using the definition of $\varepsilon_\qae$ and $\delta_\qae$, we get
\begin{equation}\label{eq:cost_signestnfn}
    \cost[\text{SignEstNFN}(U,l,\varepsilon)]\ge \left(\frac{5\sqrt{3}\pi}{\varepsilon}-1\right) \cost[U]\,.
\end{equation}
\end{proof}

\begin{lemma}[Cost of SignEstNFP]
\label{lem:cost_signestnfp}
The cost of SignEstNFP is lower bounded by
\begin{equation*}
    \cost[\text{SignEstNFP}(U,l,\varepsilon)]\ge \left(\frac{45\sqrt{3}\pi}{\varepsilon}-1\right) \cost[U]\,.
\end{equation*}
\end{lemma}
\begin{proof}
The proof is identical to the one of SignEstNFN, see \cref{lem:cost_signestnfn}.
The only difference is that now $\varepsilon_{\qae}=\varepsilon/9\sqrt{3\pi^2}$, which leads to a different prefactor.
\end{proof}

\begin{lemma}[Cost of CanEnterNFN]\label{lem:cost_canenternfn}
The cost of CanEnterNFN is lower bounded by
\begin{align*}
    \cost[\mathrm{CanEnterNFN}&(A_B, A_k,c,\varepsilon)]\ge \left(\frac{50\sqrt{6}\pi}{11\varepsilon}-1\right) \nonumber \\ \cdot &\cost\left[\qls\left({A}_B, A_k,\frac{0.1\varepsilon}{\sqrt{2}}\right)\right].
\end{align*}

A lower bound for the cost of implementing a version of QLS is provided in \cref{lem:qls}.
\end{lemma}
\begin{proof}
The cost of running CanEnterNFN is given by the cost of SignEstNFN and the if statement, see \cref{alg:canenternfn}.
To lower bound the cost of CanEnterNFN, we only consider the first, i.e.\ the cost of SignEstNFN.
The cost of SignEstNFN is given by \cref{lem:cost_signestnfn}.
Let $U= \text{RedCost}(A_B, A_k, c, \varepsilon_{\text{RC}})$, with $\varepsilon_{\text{RC}}= \varepsilon \cdot 0.1/\sqrt{2}$, $l=0$, and $\varepsilon_{\mathrm{SE}}=1.1\varepsilon/\sqrt{2}$, then
\begin{equation*}
\cost[\mathrm{CanEnter}(A_B, A_k,c,\varepsilon)\ge\cost[\mathrm{SignEstNFN}](U,l,\varepsilon_{\mathrm{SE}})\,.
\end{equation*}
Combining the results of \cref{lemma:cost_redcost} and \cref{lem:cost_signestnfn}, we obtain
\begin{align}\label{eq:cost_canenternfn}
    \cost[\mathrm{CanEnterNFN}&(A_B, A_k,c,\varepsilon)]\ge \left(\frac{50\sqrt{6}\pi}{11\varepsilon}-1\right) \nonumber \\ \cdot &\cost\left[\qls\left({A}_B, A_k,\frac{0.1\varepsilon}{\sqrt{2}}\right)\right].
\end{align}
\end{proof}

\begin{lemma}[Cost of CanEnterNFP]
\label{lem:cost_canenternfp}
The cost of CanEnterNFP is lower bounded by
\begin{align*}\label{eq:cost_canenternfp}
    \cost[\mathrm{CanEnterNFP}&(A_B, A_k,c,\varepsilon)]\ge \left(\frac{450\sqrt{6}\pi}{11\varepsilon}-1\right) \nonumber \\\ \cdot &\cost[\qls({A}_B, A_k,0.1\varepsilon/\sqrt{2}].
\end{align*}
A lower bound for the cost of implementing a version of QLS is provided in \cref{lem:qls}.
\end{lemma}
\begin{proof}
The proof is identical to the one of CanEnterNFN, see \cref{lem:cost_canenternfn}.
The only difference is that now we use SignEstNFP, which leads to a different prefactor.
\end{proof}

\subsubsection{Proofs of FindColumn with Different Pivoting Rules}\label{sec:proof_pivoting_rules}
\paragraph{Cost of Quantum Random Pivoting Rule}
\begin{restatable}[Cost of $\findColumn$]{lemma}{lemmaFindColumn}
\label{lemma:cost_FindColumn}
The cost of $~\findColumn$ is lower bounded by
\begin{align*}
\cost[&\mathrm{\findColumn}(A, B, c, \varepsilon)] \ge n_Q(n-m, t)   \\
\cdot &\left(\frac{50\sqrt{6}\pi}{11\varepsilon}-1\right) \cost\left[\qls\left({A}_B, A_k, \frac{0.1\varepsilon}{\sqrt{2}}\right)\right].
\end{align*}
\end{restatable}

\begin{proof}
Let $\chi=\mathrm{CanEnterNFN}(A_B, A_k=U_{rhs}(\cdot),c,\varepsilon)$, with $U_{rhs}\ket{k}\ket{0}=\ket{k}\ket{A_k}$, then the cost of $\findColumn$ is given by the cost of running QSearch on $\chi$:
\begin{equation*}
    \cost[\findColumn(A, B, c, \varepsilon)]=\cost[\qsearch(\chi, \varepsilon)]
\end{equation*}
A lower bound for the cost of running QSearch is given by \cref{lem:qsearch}.
In this case, $\abs{X}=n-m$, i.e.\ the number of nonbasic columns, and the number of marked items is the number of nonbasic columns with negative reduced price.
Up to a small overhead that we here neglect, the cost of one application of $Q$ is equal to one query to $\chi$
In this case, $\chi=\mathrm{CanEnterNFN}$.
A lower bound for the cost of CanEnterNFN is given by \cref{lem:cost_canenternfn}.
Combining the bounds for CanEnterNFN and QSearch, we arrive to the wanted result.
\end{proof}

\paragraph{Proof of \cref{lem:steepest} (QStER)}
\lemmaSteepestEdge*
%\begin{lemma}[Cost of Quantum Steepest Edge Rule] A lower bound for the gate count of $\steepestEdgeRule$ is given by
%\begin{align}
%    \mathcal{C}[\steepestEdgeRule(A, c, \varepsilon)] \geq 3\biggl( \frac{40 \sqrt{3}\pi c_{\mathrm{max}}}{\varepsilon}-1 \biggr)  \lceil s_{\mathrm{max}} \rceil \sum_{t=0}^{q-1} \frac{ n_Q(q,t)}{t+1} \mathcal{C}[ \mathrm{QLS}(\hat{A}_B, \hat{b}_k, \varepsilon / c_{\mathrm{max}}10 \sqrt{2})].
%\end{align}
%Here $q$ denotes the number of nonbasic columns, $ s_{\mathrm{max}} = \mathrm{log}_{3} 1 / \varepsilon$ and 
%Here $n_Q (q,t)$ denotes the number of times we need to apply the Grover operator in $\qsearch$:
%\begin{equation}\tag{\cref{eq:N_Q}}
%     n_Q(\abs{q},t)=\sum_{k=1}^{k_{max}}\frac{m_k}{2}\Bigl[\prod_{l=1}^{k-1}\frac{1}{2}+\frac{\sin(4(m_l+1)\theta)}{4(m_l+1)\sin(2\theta)}\Bigr]\,,
%\end{equation}
%with  $\sin^2\theta=t/\abs{q}$, $m_k = \lfloor \min(\lambda^k, \sqrt{\abs{q}})\rfloor$,  $\lambda=6/5$, and 
%\begin{equation}
%    k_{max}=\Bigl\lceil \log_\lambda\frac{\abs{q}}{2\sqrt{\abs{q}-1}}\Bigr\rceil+4\,.
%\end{equation} 
%Finally, 
%$ \mathcal{C}[ \qls(\hat{A}_B, \hat{b}_k, \varepsilon /c_{\mathrm{max}}10 \sqrt{2})]$ 
%is the cost of a quantum linear solver used with precision $\varepsilon/c_{\mathrm{max}}10\sqrt{2}$, and for
%\begin{equation}
%    \hat{A}_B=\begin{pmatrix}
%    A_B & 0\\
%    0 & 1
%   \end{pmatrix}\,,\quad
%   \hat{b}_k = \begin{pmatrix}
%   A_k\\
%    c_k
%    \end{pmatrix}\,.
%\end{equation}
%\end{lemma}
%%%%%%%%%%%%%%%%%%%%%%%%%%%%%%%%%%%%%%%%
\begin{proof}
The complexity of $\steepestEdgeRule$ is given by
\begin{align}\label{Eq. QStER Complexity}
\mathcal{C}[\steepestEdgeRule] = \langle n_{\qmin} \rangle \times \mathcal{C}[\steepestEdgeCompare],
\end{align}
where $\langle n_{\qmin} \rangle$ denotes the expected number of oracle calls of the quantum minimum finding algorithm and $\mathcal{C}[\steepestEdgeCompare]$ denotes the cost of the SteepestEdgeCompare subroutine. The latter can be lower bounded by the cost of $\mathrm{SignEstNFN}(W, 0, c, \frac{\varepsilon}{8 c_{max}})$ which itself can be lower bounded by the cost for quantum amplitude estimation
\begin{align*}
    \mathcal{C}[\steepestEdgeCompare] &\geq \mathcal{C}[\mathrm{SignEstNFN}(W, 0, c, \varepsilon/8 c_{max})] \nonumber \\ &\geq \mathcal{C}[\mathrm{QAE}(W, \chi_{0}, \varepsilon/8 c_{max}, \delta^{\prime})].
\end{align*}
The cost of quantum amplitude estimation can be divided into the cost of implementing the unitary $W$ and the cost of quantum phase estimation as
\begin{align*}
    \mathcal{C}[\qae(W, \chi_{0}, \varepsilon/8 c_{max}, \delta^{\prime})] = &\mathcal{C}[\mathrm{Interfere}(U, V)] \nonumber \\
&+ \mathcal{C}[\qpe(Q, \delta^{\prime}, \varepsilon/8 c_{max})].
\end{align*}
Here $Q$ is the Grover operator used by the $\qae$ subroutine. 
The parameter $\delta^{\prime}$ bounds the success probability $p$ via $p \geq 1 - \delta^{\prime}$. 
Analogous to the derivation of a lower bound for $\mathrm{QAE}$ in the proof of \cref{lem:cost_signestnfn}, we can now do the following estimation
\begin{align*}
    \mathcal{C}[\steepestEdgeCompare] \geq \biggl( 2^{n_{\varepsilon^{\prime}, \delta^{\prime}} +1} - 1 \biggr)  \mathcal{C}[\mathrm{Interfere}(U, V)],
\end{align*}
where 
\begin{align*}
n_{\varepsilon^{\prime}, \delta^{\prime}}= \mathrm{log}_{2} \frac{1}{2 \varepsilon^{\prime}} + \mathrm{log}_{2} \biggl( 1 + \frac{1}{\delta ^{\prime}} \biggr)
\end{align*}
and $\varepsilon^{\prime} = \varepsilon_{\qpe}/8 c_{max} = \varepsilon/8\sqrt{3}\pi c_{max} $.
The cost of $\mathcal{C}[\mathrm{Interfere}(U, V)]$ can be lower bounded by the cost of $V_r$, and the cost of $U_r$ (see discussion on \cref{alg:interfere}). Hence, we get the following estimation
\begin{align*}
    \mathcal{C}[\steepestEdgeCompare]  \geq \biggl( 2^{n_{\varepsilon^{\prime}, \delta^{\prime}} +1} - 1 \biggr) \cdot \biggl( \mathcal{C}[V_r] + \mathcal{C}[U_r] \biggr)
\end{align*}
where
\begin{align*}
     \mathcal{C}[V_r] &=  \mathcal{C}[\mathrm{RedCost} (A_B, A_k, \varepsilon /c_{\mathrm{max}}10 \sqrt{2})] \\ 
&\geq  \mathcal{C}[ \qls({A}_B, A_k, \varepsilon /c_{\mathrm{max}}10 \sqrt{2})],\\
     \mathcal{C}[U_r] &=  \mathcal{C}[\mathrm{RedCost}(A_B, A_j, \varepsilon /c_{\mathrm{max}}10 \sqrt{2})] \\ 
&\geq  \mathcal{C}[ \qls({A}_B, A_j, \varepsilon /c_{\mathrm{max}}10 \sqrt{2})].
\end{align*}
With that, we get the estimation
\begin{align*}
    \mathcal{C}[\steepestEdgeCompare] \geq& \biggl( 2^{n_{\varepsilon^{\prime}, \delta^{\prime}} +1} - 1 \biggr) \cdot \\
    &\mathcal{C}[ \qls(A_B, {A}_k, \varepsilon /c_{\mathrm{max}}10 \sqrt{2})]. 
\end{align*}
From \cref{lem:nqmin}, the expected number of calls from $\qmin$ with cutoff for a list of length $n-m$ is given by
\begin{align*}
    \langle n_{\qmin_{finite}}\rangle = 3 \lceil s_{\mathrm{max}} \rceil \sum_{s=0}^{n-m-1} \frac{ n_Q(n-m,s)}{s+1},
\end{align*}
where $s = \mathrm{sin}^{2} (\theta) |n-m|$, $n_Q(n-m,s)$ denotes the number of times we need to apply the Grover operator in QSearch and $ s_{\mathrm{max}} = \mathrm{log}_{3} 1 / \varepsilon$.
Putting everything into \cref{Eq. QStER Complexity}, we finally obtain that
\begin{align*}	
    \mathcal{C}[\mathrm{QStER}] \geq &\biggl( 2^{n_{\varepsilon^{\prime}, \delta^{\prime}} +1} - 1 \biggr) 3 \lceil s_{\mathrm{max}} \rceil \\
    &\sum_{s=0}^{n-m-1} \frac{ n_Q (n-m,s)}{s+1} \nonumber \\
 \cdot &\mathcal{C}[ \qls(A_B, A_k, \varepsilon /c_{\mathrm{max}}10 \sqrt{2})].
\end{align*}
Now setting $\delta^{\prime} = 1/4$ and $\varepsilon^{\prime} \rightarrow \varepsilon/ 8 \sqrt{3} \pi c_{\mathrm{max}}$ we obtain
\begin{align*}
    \mathcal{C}[\mathrm{QStER}] \geq &\biggl( \frac{40 \sqrt{3} \pi c_{\mathrm{max}}}{\varepsilon} - 1 \biggr) 3 \lceil  s_{\mathrm{max}} \rceil \cdot \\
    &\sum_{s=0}^{n-m-1} \frac{n_Q (n-m,s)}{s+1} \cdot \\
 &\mathcal{C}[ \qls({A}_B, A_k, \varepsilon /c_{\mathrm{max}}10 \sqrt{2})],
\end{align*}
completing the proof.
\end{proof}

\paragraph{Cost of Quantum Danztig's Pivoting Rule}
\begin{lemma} [Cost of $\dantzig$] A lower bound for the gate count of \linebreak $\dantzig$ is given by
\label{lem:cost_Dantzig}
\begin{multline}
    \mathcal{C}[\dantzig (A, c, \varepsilon)] \geq \\
    3\biggl(\frac{40 \sqrt{3}\pi c_{\mathrm{max}}}{\varepsilon}-1 \biggr)  \lceil s_{\mathrm{max}}\rceil
\cdot \sum_{t=0}^{n-m-1} \frac{n_Q (n-m,t)}{t+1} \\
\mathcal{C}\left[ \qls \left(A_B, A_k, \frac{\varepsilon}{\|A_{B}^{-1}A_k\| c_{\mathrm{max}} 10\sqrt{2}}\right)\right]
\end{multline}
Here $ s_{\mathrm{max}} = \mathrm{log}_{3} 1 / \varepsilon$ and $n_Q(\abs{n-m},t)$ denotes the number of times we need to apply the Grover operator in $\qmin$. 
%\begin{equation}\tag{\cref{eq:N_Q}}
%    n_Q(\abs{q},t)=\sum_{k=1}^{k_{max}}\frac{m_k}{2}\Bigl[\prod_{l=1}^{k-1}\frac{1}{2}+\frac{\sin(4(m_l+1)\theta)}{4(m_l+1)\sin(2\theta)}\Bigr]\,,
%\end{equation}
%with  $\sin^2\theta=t/\abs{q}$, $m_k = \lfloor \min(\lambda^k, \sqrt{\abs{q}})\rfloor$,  $\lambda=6/5$, and 
%\begin{equation}
%    k_{max}=\Bigl\lceil \log_\lambda\frac{\abs{q}}{2\sqrt{\abs{q}-1}}\Bigr\rceil+4\,.
%\end{equation} 
%Finally, $ \mathcal{C}[  \qls(\hat{A}_B, \hat{b}_k, \varepsilon /c_{\mathrm{max}}10 \sqrt{2})]$ is the cost of a quantum linear solver used with precision $\varepsilon/c_{\mathrm{max}}10\sqrt{2}$, and for \begin{equation}
%   \hat{A}_B=\begin{pmatrix}
%    A_B & 0\\
%   0 & 1
%    \end{pmatrix}\,,\quad
%    \hat{b}_k = \begin{pmatrix}
%   A_k\\
%   c_k
%   \end{pmatrix}\,.
%\end{equation}
\end{lemma}
\begin{proof}
The proof is analogous to the one for \cref{lem:steepest}. Only the precision parameter in the cost for QLS has to be adjusted to $\varepsilon_{QLS} \rightarrow \varepsilon/ 10 \sqrt{2} \norm{A_{B}^{-1}A_k} c_{\mathrm{max}}$.
\end{proof}

\subsubsection{Proof of \cref{lem:cost_isUnbounded} (IsUnbounded)}\label{sec:proof_is_unbounded}
\lemmaIsUnbounded*
\begin{proof}
Recall that the subroutine IsUnbounded is a Grover search
for a basic index $l$ that $(A_B^{-1}A_k)_l > 0$,
up to a precision $\delta$.
In order to bound the cost of the Grover oracle
\begin{equation*}
    Q = \text{SignEstNFN}^+(\qls(A_B, A_k, \frac{\delta}{10}), \frac{9 \delta}{10})
\end{equation*}
we follow the procedure starting at \cref{lem:cost_signestnfn}. The precision required
for the sign estimation routine is $\frac{9 \delta}{10}$, which we plug into
\cref{lem:cost_signestnfn} to receive the bound
\begin{equation*}
    \cost(Q) \geq (\frac{50 \sqrt{3}\pi}{18 \delta} - 1)
    \cost[\text{QLS}(A_B, A_k, \frac{\delta}{10})].
\end{equation*}
Putting this together with the $\qsearch$ cost bound from \cref{lem:qsearch},
we obtain the desired result.
\end{proof}

\subsubsection{Proof of \cref{lem:findRow} (FindRow)}\label{sec:proof_find_row}
\lemmaFindRow*
\begin{proof}

The sub-algorithm FindRow is a binary search on the quantum oracle $U_r$, 
looking for a value approximating $r^*$, satisfying the ratio test.
Hence, the cost of the algorithm splits up as 
$$\cost[\text{FindRow}(A_B, A_k, b, \delta)] = S \cdot \cost(U_r),$$
where $S$ is the expected number of steps in the binary search.
Each step of the binary search is an application of $\qsearch$, which is looking
for a negative component of $A_B^{-1}(b-r A_k)$. At least one step in the binary search
yields that the Grover search in $U_r$ did not find any marked elements 
(this is the case for small values of $r$, especially $0$).
We get the bound 
\begin{equation*}
    S \cdot \cost(U_r) \geq \cost(U_0).
\end{equation*}
The quantum oracle $U_r$ solves the linear system $A_B x = b - r A_k$ and 
proceeds to search for a negative amplitude using the sign estimation subroutine. 
Hence, its cost decomposes as 
\begin{equation*}
    \cost[U_r] = \cost[\text{QSearch}(Q_r)] \leq n_Q(m, t_r) \cost[Q_r]
\end{equation*}
where the Grover oracle is given by 
$Q_r = \text{SignEstNFN(QLS}(A_B, b-r A_k, \delta))$ and 
$t_r$ is the number of marked elements in the Grover search corresponding to $r$.
In the case of $r=0$, we have $t_r = 0$.
To bound the cost of the combination of sign estimation
and a QLS we follow the proof of \cref{lem:cost_signestnfn}.
The precision of the sign estimation required in $U_0$ is given by 
$\varepsilon_S = \frac{\delta}{||A_B^{-1}A_k||}$. 
Plugging this into \cref{lem:cost_signestnfn} yields
\begin{equation*}
    \cost(U_0) \geq (\frac{\sqrt{3} \pi ||A_B^{-1} A_k||}{2 \delta} - 1) 
    \cost[\qls(A_B, b, \frac{\delta}{2})].
\end{equation*}
Here $\frac{\delta}{2} \geq \frac{\delta}{2} - 2 \frac{\delta}{||A_B{-1} A_k||} = \varepsilon_Q$
is an upper bound to the QLS precision derived in \cref{eq:error_relation_findrow}.
\end{proof}

\clearpage
\section{Quantum Subroutines}\label{sec:details-quantum-subroutines}
%Herein we state common quantum routines with lower bounds for the gate cost. 
%We also provide brief explanations of each subroutines; see original papers for more detail.
We provide further physics-based foundations by
presenting common quantum subroutines with 
corresponding bounds on the lowest possible gate cost. 
Additionally, we provide concise explanations of each subroutine. 
For further technical details, we refer the reader to respective original papers.

%In some cases we use a quantum oracle, i.e.\ we can compute in superposition a function $\chi:X\rightarrow \{0,1\}$.
\subsection{Notation}
In this section, we use the following conventions:
\begin{itemize}
\item The cost of a $1$-qubit, $2$-qubit, Toffoli gate is respectively $\cost_1, \cost_2, \cost_T$, 
\item The cost of a unitary operation $U$ is denoted $\cost[U]$,
\item A unitary operation $U$ controlled on $n$ qubits is denoted  $\contr{n}U$,
\item A \textit{state preparation oracle} $\mathcal{P}_{\psi}$ is a unitary operation preparing some $n$-qubit state $\ket{\psi}$, i.e.\ $\mathcal{P}_{\psi}\ket{0^n}=\ket{\psi}.$ We do not take into account the costs of implementing state preparation oracles, and instead set $\cost\left[\mathcal{P}_{\psi}\right]=1$.
\end{itemize}

\subsection{Controlling Unitaries}\label{sec:controlled_unitaries}
We always assume that $U$ is applied when all the controlling qubits are in state $\ket{1}$.
In equation, 
\begin{equation*}
\contr{n}{U}=\ketbra{N-1}\otimes U+ \sum_{\tau=0}^{N-2}\ketbra{\tau}\otimes \mathds{1}\,,
\end{equation*}
where $N=2^n$ and $\ket{N-1}=\ket{11\dots1}$. 

As explained in chapter 4 from \cite{nielsen2002quantum}, we can implement $\contr{n}{U}$ by first calculating the AND of all the controlling qubits in an ancilla qubit, and then act with $\contr{1}{U}$. Computing the AND of the controlling qubits requires $n-1$ ancilla qubits and $2(n-1)$ Toffoli gates (the factor 2 is due to the uncomputation step). The cost of $\contr{1}{U}$ can be bounded by the cost of $U$. So finally we have
\begin{equation}\label{eq:cost_ctrtU}
\cost[\contr{n}{U}]\ge  2(n-1)\cost_T+\cost[U]\,.
\end{equation} 

\subsection{QSearch}\label{app:QSearch}
\cref{alg:qsearch} is a modified version of QSearch from \cite{boyer1998tight}, which is an extension of Grover algorithm. It can be used to identify a marked item within a  list. 
Assume a set $X$ and function $\chi$, where $\chi$ splits $X$ in a set of \emph{good} solutions, $G=\{x\in X\vert \chi(x)=1\}$, and \emph{bad} solutions, $B=\{x\in X\vert \chi(x)=0\}$. 
Often $X=\{0,1\}^n$; otherwise, we set $$n=\lceil \log_2 \abs{X} \rceil .$$  
QSearch outputs a marked (good) item, $x\in G$, or "No marked item" , with error probability upper bounded by $\varepsilon$.
\begin{algorithm}[h!]
\caption{QSearch}\label{alg:qsearch}
\begin{algorithmic}
\Function{QSEARCH}{oracle $\chi:X\rightarrow\{0,1\}$, $\varepsilon >0$}
\State $\lambda \gets 6/5$, $m \gets \lambda$
\State $s_{max}\gets\lceil\log_3{1/\varepsilon}\rceil$, $k_{max}\gets k_*+4$ 
\Comment{$
    k_*=\Bigl\lceil \log_\lambda\frac{\abs{X}}{2\sqrt{\abs{X}-1}}\Bigr\rceil
$}{}
\State $k\gets 1$, $s\gets 1$
\While{$s\le s_{max}$}
\While{$k\le k_{max}$}
\State Prepare $\ket{\psi_0}$
\Comment{$\ket{\psi_0}\propto\sum_{x=0}^{\abs{X}-1}\ket{x}$}{}
\State Choose $j$ uniformly at random in $[0,\lfloor m\rfloor]$
\State Apply $Q^j$
\Comment{$Q$ is the Grover operator}{}
\State Measure the register, let $x_*$ be the outcome
\If{$\chi(x_*)=1$}
    \Return $x_*$
\Else
\State $k\gets k+1$
\State $m \gets \min(\lambda m, \sqrt{\abs{X}})$
\EndIf
\EndWhile
\State $s\gets s+1$
\EndWhile\\
\Return "No marked item"
\EndFunction
\end{algorithmic}
\end{algorithm}
\lemmaQSearch*
%\begin{lemma}[Cost of QSearch]
%\label{lem:cost_QSearch}
%The expected cost of QSearch can be lower bounded by the expected cost of any QSearch iteration.
%%over $s_{max}$ iterations is
%%\begin{equation}\label{eq:cost_QSearch}
%%    \langle\cost[\text{QSearch}, s_{max}]\rangle=s_{max}\langle\cost[\text{QSearch}]\rangle.
%%\end{equation}
%This is
%\begin{equation}
%   \langle\cost[\text{QSearch}]\rangle= n_Q(\abs{X}, t) \cost[Q]+\langle n_{iter}\rangle\bigl( \cost[\psi_0]+\cost[meas]\bigr)\,,
%\end{equation}
%where the expected number of times we need to apply $Q$ in the inner while loop, i.e.\ for every value of $l$, is   
%\begin{equation}
%     n_Q(\abs{X}, t)=\sum_{k=1}^{k_{max}}\frac{m_k}{2}\Bigl[\prod_{l=1}^{k-1}\frac{1}{2}+\frac{\sin(4(m_l+1)\theta)}{4(m_l+1)\sin(2\theta)}\Bigr]\,,
%     \label{eq:N_Q}
%\end{equation}
%with
%\begin{align}
%m_k &= \lfloor \min(\lambda^k, \sqrt{\abs{X}})\rfloor \quad \text{and} \\ 
%k_{max}&=\le \Bigl\lceil \log_\lambda\frac{\abs{X}}{2\sqrt{\abs{X}-1}}\Bigr\rceil+4
%\end{align} 
%The cost of one action of $Q$ is 
%\begin{equation}
%\cost[Q]=\cost[\chi]+2(n+1)\cost_1+\cost_2+2(n-1)\cost_T\,,
%\end{equation} 
%the average number of iterations is (i.e.\ the times we increase $m$ by a factor $\lambda$)
%\begin{equation}
%    \langle n_{iter}\rangle = \sum_{k=1}^{k_{max}}\prod_{l=1}^{k-1}(1-\langle p_{j_l}\rangle)\,,
%\end{equation}
%the cost of preparing the initial state is
%$\cost[\psi_0]=n\cost_1$, and finally the cost of measuring is one further query to the oracle (either classical or quantum).
%\end{lemma}
\begin{proof}
We follow closely the notation and reasoning of \cite{cade2022grover}. 
Let $X$ be a list of size $\abs{X}$, with $t$ elements marked by function $\chi:X\rightarrow \{0,1\}$. 
Let $G\equiv\{x\in X\vert \chi(x)=1\}$ and $B\equiv\{x\in X\vert \chi(x)=0\}$ be the spaces of marked (good) and non-marked (bad) elements in $X$.
Our goal is to find a marked element. 
In Grover's algorithm, we first prepare a uniform superposition of all elements in the list, 
\begin{equation*}
\ket{\psi_0}=\frac{1}{\sqrt{\abs{X}}}\sum_{x=0}^{\abs{X}-1}\ket{x}\,.
\end{equation*}
We then rotate this initial guess to a state which is approximately a superposition only of marked items. 
This is done by repeatedly applying the operator $Q=R_0 R_B$, made out of two reflections: one around the state $\ket{\psi_0}$, $R_0=2\ketbra{\psi_0}-\mathds{1}$, and the other around the space of non-marked items, $R_B=2\Pi_B-\mathds{1}$. 
Here $\Pi_B$ is the projector on the space of non-marked items.
 %$B=\{x\in X\vert \chi(x)=0\}$ 

The optimal number of times we should apply $Q$ depends on the ratio between $t$ and $\abs{X}$, which is typically unknown. 
The algorithm of \cite{boyer1998tight} sidesteps this problem by randomly guessing a value for $t$ in a range which is gradually extended.  
Following \cite{cleve1998quantum}, we denote this algorithm as \qsinf. 
%\begin{algorithm}
%\label{QSearch_inner}
%\caption{\qsinf \SA{QSearch TWO}}
%\begin{algorithmic}
%\Function{QSEARCH$_{\infty}$}{List $X$, $\varepsilon >0$, oracle $\chi$}
%\State $\lambda \gets 6/5$, $m \gets \lambda$
%\While{true}
%\State Prepare $\ket{\psi_0}$
%\State Choose $j$ uniformly at random in $[0,\lfloor m\rfloor]$
%\State Apply $Q^j$
%\State Measure the register, let $x_*$ be the outcome
%\If{$\chi(x_*)=1$}
%    \Return $x_*$
%\Else
%\State $m \gets \min(\lambda m, \sqrt{\abs{X}})$
%\EndIf
%\EndWhile
%\EndFunction
%\end{algorithmic}
%\end{algorithm}
%\newline
This algorithm always outputs a solution if $t\ge 1$. 
We set a limit to the number of times we increase $m$ by a factor $\lambda$ to stop the algorithm if a solution is not found, $k_{max}$, after this limit is reached, the algorithm returns `no marked item'. 
This allows the algorithm to manage $t=0$ cases. 
There is a finite probability of  outputting `no marked item' even when $t\ge 1$, however this failure probability can be bounded. 

Let 
\begin{equation*}
    \ket{\psi_G}\equiv\frac{1}{\sqrt{t}}\sum_{x\in G}\ket{x}\,,\quad\ket{\psi_B}\equiv\frac{1}{\sqrt{\abs{X}-t}}\sum_{x\in B}\ket{x}\,,
\end{equation*}
be the superpositions of good and bad states. 
Our initial guess can be expressed as 
\begin{equation*}
    \ket{\psi_0}=\sin{\theta}\ket{\psi_G}+\cos{\theta}\ket{\psi_B}\,,
\end{equation*}
where $\sin^2{\theta}=t/\abs{X}$. 
The operator $Q$ generates translations in $\theta$. 
Let $\ket{\psi_j}=Q^j\ket{\psi_0}$, then one can show that
\begin{equation*}
    \ket{\psi_j}=\sin[(2j+1)\theta]\ket{\psi_G}+\cos[(2j+1)\theta]\ket{\psi_B}\,.
\end{equation*}
From this follows that, after $j$ applications of $Q$, the probability of finding a good state as the outcome of the measurement is
\begin{equation*}
    p_j=\sin^2[(2j+1)\theta]\,.
\end{equation*}
%This is the only thing we need to know to find the number of times we need to apply $Q$, on average, to find a marked element. 

The algorithm proceeds as follows. 
Let $m_k = \lfloor \min(\lambda^k, \sqrt{\abs{X}})\rfloor$.
We pick a uniformly random number $j_k\in[0,m_k]$, and apply $Q^{j_k}$. 
%With probability $p_{j_k}$, we return a marked item; otherwise, we set $k\rightarrow k+1$ and repeat; after $k_{max}$ iteration, if a marked item has not been yet found, we return "no marked item". Graphically, we have
%\begin{equation}\label{eq:qsinfGraph}
%\begin{split}
%    &j_1\fixedrightarrow{p_{j_1}} \text{return marked item} \\
%    &\phantom{j_1} \fixedrightarrow{1-p_{j_1}} j_2\fixedrightarrow{p_{j_2}} \text{return marked item}\\
%    &\phantom{j_1\fixedrightarrow{1-p_{j_1}} j_2} \fixedrightarrow{1-p_{j_2}} j_3\fixedrightarrow{p_{j_3}} \text{return marked item}\\
%    &\phantom{j_1\fixedrightarrow{1-p_{j_1}} j_2 \fixedrightarrow{1-p_{j_2}} j_3}\fixedrightarrow{1-p_{j_3}} \dots
%    \\
%    &\phantom{j_1\fixedrightarrow{1-p_{j_1}} j_2 \fixedrightarrow{1-p_{j_2}} j_3\fixedrightarrow{1-p_{j_3}}} \dots
%    \\
%    &\phantom{j_1\fixedrightarrow{1-p_{j_1}} j_2 \fixedrightarrow{1-p_{j_2}} j_3\fixedrightarrow{1-p_{j_3}}}j_{k_{max}}\fixedrightarrow{p_{j_{k_{max}}}}\text{return marked item}\\
%    &\phantom{j_1\fixedrightarrow{1-p_{j_1}} j_2 \fixedrightarrow{1-p_{j_2}} j_3\fixedrightarrow{1-p_{j_3}}j_{k_{max}}}\fixedrightarrow{1-p_{j_{k_{max}}}}\text{return "no marked item"}
%\end{split}
%\end{equation}
If $t=0$, all the $p_j=0$ and we correctly output "no marked item". 
If $t\ge 1$, we still have a finite probability of outputting "no marked item", which for $t\ge 1$ is incorrect. 
Averaging over the randomly chosen $j_k$, we find that the probability of outputting "no marked item"  is given by
\begin{equation}
\label{pfail}
    \langle p_{fail}\rangle = \prod_{k=1}^{k_{max}}(1- \langle p_{j_k}\rangle)\,,
\end{equation}
where 
\begin{align}\label{eq:avgSuccessProb}
    \langle p_{j_k}\rangle &= \frac{1}{m_k+1}\sum_{j_k=0}^{m_k}p_{j_k}\\&=\frac{1}{2}-\frac{\sin[4(m_k+1)\theta]}{4(m_k+1)\sin(2\theta)}\,.\nonumber
\end{align}
To upper bound $\langle p_{fail}\rangle$, we use lower bounds on $\langle p_{j_k}\rangle$ from \cite{boyer1998tight, cade2022grover}.
The bound provided by \cite{boyer1998tight} follows directly from \cref{eq:avgSuccessProb}: we have that $\langle p_{j_k}\rangle>1/4$ if $m_k+1\ge 1/\sin(2\theta)$. 
For $\theta\rightarrow \pi/2$, this quantity diverges, and the bound becomes useless; we have to consider the cases $\theta<\pi/6, \theta\geq \pi/6$ separately for technical reasons. 
For $\theta<\pi/6$, rewrite $1/\sin(2\theta)$ as
\begin{equation*}
    \frac{1}{\sin(2\theta)}=\frac12\sqrt{\frac{\abs{X}}{t}}\Bigl(1-\frac{t}{\abs{X}}\Bigr)^{-1/2} \le \sqrt{\frac{\abs{X}}{2}}\,.
\end{equation*}
The inequality is obtained by using $t\ge1$, and set $\abs{X}=2$ in the parentheses. 
From this follows that there exists a critical $k$ for which $m_k+1\ge1/\sin(2\theta)$ and the bound applies. 
A sufficient condition for this to happen is that $k\ge k_c$, where
\begin{equation}\label{eq:criticalk}
    k_c=\lceil-\log_\lambda\sin(2\theta)\rceil\,.
\end{equation}
%For $\theta\in[\pi/6,\pi/2]$, we can use an alternative bound, which is proved in \cite{cade2022grover} (see App.\ A1),
%\begin{equation}
%    p_j\ge \frac{1}{2}-\frac{1}{2\pi}\Bigl(1-\frac{\pi^2}{96}\Bigr)^{-1}>\frac{1}{4}\,,
%\end{equation}
%so, for $\theta\in[\pi/6,\pi/2]$, we have $\langle p_{j_k}\rangle>1/4$ for every $k$. 
%fir $\theta<\pi/6$ we have $\langle p_{j_k}\rangle\ge1/4$ if $k\ge k_c$, while if $k<k_c$, the bound doesn't apply and we can only say $\langle p_{j_k}\rangle\ge0$; 
From \cite{cade2022grover}, for $\theta\in[\pi/6,\pi/2]$, $\langle p_{j_k}\rangle>1/4$ for every $k$. 
Since our $\theta$ is not known, we must use the bounds resulting in the weakest upper bound to $\langle p_{fail}\rangle$, 
which turns out to be the $\theta<\pi/6$ case discussed above.
For the same reason, we use the worst-case value of $k_c$. This is found by setting $t=1$ in \cref{eq:criticalk},
\begin{equation*}
    k_c\le \Bigl\lceil \log_\lambda\frac{\abs{X}}{2\sqrt{\abs{X}-1}}\Bigr\rceil\equiv k_*\,.
\end{equation*}
Putting everything together, we find that for $t\ge1$ and for any $\theta$, 
\begin{equation*}
    \langle p_{fail}\rangle \le \Bigl(\frac34\Bigr)^{k_{max}-k_*}\,.
    \label{avg_p_fail_innerloop}
\end{equation*}
The value $k_{max}=k_*+4$ is obtained by requiring that $\langle p_{fail}\rangle\le 1/3$. We can boost this probability to $\varepsilon>0$ repeating the algorithm $s_{max}=\log_3{1/\varepsilon}$. The final result is algorithm \cref{alg:qsearch}. 

We use a slightly different time out from the one considered in \cite{cade2022grover}, which tracks the number of times $Q$ has been applied. 
We instead set a limit to the number of times we increase $m$ by a factor $\lambda$, i.e.\ stop the algorithm when $k$ reaches a certain value. 
As in  \cite{cade2022grover}, we set $\lambda=6/5$ to optimize the expected number of queries to $G$. 
In our version of QSearch, it is much easier to calculate the expectation value of $\langle n_Q\rangle$, which we need below to compute a bound on the expected gate cost of the algorithm. 
Ultimately we define this as a function of $\abs{X}, t$.

For the moment, we neglect the outer while loop required to boost the failure probability to $\varepsilon$. 
We then split the average cost of QSearch as 
\begin{equation*}
    \langle\cost[\text{QSearch}]\rangle=\langle n_Q\rangle \cost[Q]+\langle n_{iter}\rangle\bigl( \cost[\psi_0]+\cost[meas]\bigr)\,.
\end{equation*}
Above, $\langle n_Q\rangle$ is the average number of times we need to apply the operator $Q$, $\cost[Q]$ is the cost of one application of $Q$, 
$\langle n_{iter}\rangle$ is the average number of iterations (i.e.\ number of times we increase $m$ by a factor $\lambda$), 
$\cost[\psi_0]$ is the cost of preparing $\ket{\psi_0}$, and $\cost[meas]$ the cost of performing the measurement. 
Typically the first term is the dominant one.

To find an expression for the average number of times we need to apply $Q$, consider again the iteration over iterated sets of $j$'s.
Let $J_1=(j_1,j_2,j_3,\dots, j_{k_{max}})$ be a random draw for the $j$'s, and $J_2=(j_2, j_3, \dots, j_{k_{max}})$, the same draw with $j_1$ dropped. The average number of applications of $Q$ given a draw $J_1$, which we denote $\langle n_Q\rangle_{J_1}$ can be found through 
\begin{equation*}
    \langle n_Q\rangle_{J_1}=j_1+(1-p_{j_1})\langle n_Q\rangle_{J_2}\,.
\end{equation*}
We can define $J_3$ by dropping $j_2$ from $J_2$ and find a similar expression for $\langle n_Q\rangle_{J_2}$, ad so on. 
The resulting recursion relation can be expanded to 
\begin{equation*}
    \langle n_Q\rangle_{J_1}=\sum_{k=1}^{k_{max}}\Bigl[j_k\prod_{l=1}^{k-1}(1-p_{j_l})\Bigr]\,.
\end{equation*}
We are interested in averaging this expression over $J_1$. The result is
\begin{equation}\label{eq:qsinfExp}
    n_Q(\abs{X}, t)=\sum_{k=1}^{k_{max}}\Bigl[\langle j_k\rangle\prod_{l=1}^{k-1}(1-\langle p_{j_l}\rangle)\Bigr]\,,
\end{equation}
where
\begin{equation}\label{eq:avgj}
    \langle j_k\rangle = \frac{1}{m_k+1}\sum_{j_k=0}^{m_k}j_k=\frac{m_k}{2}\,,
\end{equation}
and $\langle p_{j_l}\rangle$ is given by \cref{eq:avgSuccessProb}. 
The expression for $n_Q(\abs{X}, t)$ is difficult to handle analytically, but can be easily estimated numerically for a given  $\theta$. 
To decompose the costs applying $Q$, consider $R_B$, $R_0$ separately. 
For $R_B$, we assume that the function $\chi$ can be implemented as a quantum oracle (also denoted by $\chi$) with action $\chi\cdot\ket{x}\ket{a}=\ket{x}\ket{a\oplus \chi(x)}$.
Here, $a=0,1$, and $\oplus$ denotes addition mod 2. 
Then we can implement $R_\chi$ with a phase query to $\chi$, 
$$\chi\cdot\ket{x}\ket{-}=(-1)^{\chi(x)}\ket{x}\ket{-},$$ 
so after tracing out the ancilla, we are left with the same action as $R_B$, with cost
\begin{equation}
\cost[R_B]=\cost[\chi]+\cost_1+\cost_2.
\label{eq:qsearch_Rbcost}
\end{equation}
The reflection around $\ket{\psi_0}$ is $R_{\psi_0}=\alg R_0\alg$, where $R_0=2\ketbra{0}-\mathds{1}$, and $\alg$ is a unitary that prepares $\ket{\psi_0}$, i.e.\ $\alg\ket{0}=\ket{\psi_0}$.
For QSearch, $\alg$ is implemented by a layer of Hadamard gates such that $\cost[\alg]=n\cost_1$. 
We state $\cost[R_{\psi_0}]$ in this more general form in anticipation of later Grover-oracle based routines. 

For $R_0$, we use the same strategy as before, now with an oracle mapping $x=0$ to $0$ and all other $x$'s to $1$.
This can be implemented as a Toffoli gate controlled on $x=0$ and targeting the $\ket{a}$, then finally apply $X$ on $\ket{a}$, at cost
\begin{equation}
\cost[R_{\psi_0}]=2\cost[\alg]+\cost_1+2(n-1)\cost_T.
\label{eq:qsearch_Rpsicost}
\end{equation}

Let $n=\lceil \log_2\abs{X} \rceil$, and we find that altogether
\begin{equation}
\cost[Q]=\cost[\chi]+2(n+1)\cost_1+\cost_2+2(n-1)\cost_T\,.
\end{equation} 
Here, $\cost[\chi]$ is the cost of one query to $\chi$.

Beside acting with $Q$, at every iteration, we need to prepare the state $\ket{\psi_0}$, and check that the outcome of the measurement is a good solution. 
The first requires $n$ Hadamard gates. 
For the second, we typically need to apply the classical oracle $\chi$ to the outcome of the measurement, or, in some other cases, we might need to use the quantum oracle instead.\footnote{For example, in Nannicini's algorithm, the quantum oracle queried by Grover has not a direct classical counter part, since it uses a quantum linear solver. 
%We have two options. We can use some classical linear solver to calculate the reduced cost for the column found by QSearch without fully inverting the matrix $A_B$, using, for example, Krylov subspace methods. 
%Otherwise, we can query again the quantum algorithm CanEnter. 
%In practice, the cost of measurement gives a negligible contribution to the cost of QSearch, so we can set it to zero.
} 
The average number of iterations is 
\begin{equation*}
    \langle n_{iter}\rangle = \sum_{k=1}^{k_{max}}\prod_{l=1}^{k-1}(1-\langle p_{j_l}\rangle)\,.
\end{equation*}
We want to boost the average success probability $1-\langle p_{fail}\rangle$ by repeating the protocol for applying $Q^j$ with a bounded while loop, as established above.
%
%Finally, we want to boost the average success probability $1-\langle p_{fail}\rangle$ by repeating the protocol for applying $Q^j$ with a bounded while loop, as established above. 
%The average failure probability after $s_{max}$ applications should be $\langle p_{fail, s_{max}}\rangle=\prod_{s=1}^{s_{max}}$ (we do not change any of the parameters $m, k, \lambda$ in the $l$ iteration). 
%Using $\langle p_{j_k}\rangle$ found in \cref{eq:avgSuccessProb}, and consequently $\langle p_{fail}\rangle$ as in \cref{avg_p_fail_innerloop}, we get
%$$\langle p_{fail, s_{max}}\rangle\leq\prod_{s=1}^{s_{max}}\left(\frac{3}{4}\right)^{k_{max}-k_*}\leq \varepsilon.$$
%Simple algebra leads to
%\begin{equation}
%s_{max}=\lceil\frac{1}{4}\frac{\log(1/\varepsilon)}{\log(4/3)}\rceil.
%\end{equation}
%\begin{equation}
%s_{max}=\lceil \frac{1}{4}\log_3(1/\varepsilon)\rceil
%\end{equation}
%Then the expected cost of Qsearch with the outer loop is simply $s_{max}\langle\cost[\text{QSearch}]\rangle$.
However, to obtain a simple bound, we count the cost of one iteration.
\end{proof}

\subsection{Quantum Minimum Finding}
The quantum minimum finding algorithm of Høyer and Dürr (see \cite{durr1999quantum}) takes as input 
an unsorted list $L$ of $N$ elements each holding a value from an ordered set which is assigned via a function 
$T: L \longrightarrow \mathbb{R}$ and outputs the element associated to the minimum of this set. 
The algorithm's main subroutine uses a generalization of Grover search, namely the \emph{quantum exponential searching algorithm}, 
which we replace by $\qsearch$, an algorithm that is discussed in \cref{app:QSearch}. 
We denote the running time by $t$ and assume that we have oracle access to the function $\chi: L \longrightarrow \lbrace 0, 1 \rbrace$ with:
\[ 
 \chi_i(j) = \left\{
\begin{array}{ll}
  1, & \text{if} \ \  T(j) < T(i)\\
   0,  & \text{else.}  
\end{array} 
\right. 
\]
\begin{algorithm}
\caption{Quantum Minimum Finding}\label{alg:qmf}
\begin{algorithmic}
\Function{$\qmin$}{$L$, $\chi: L \longrightarrow \lbrace 0, 1 \rbrace$}
\State  Choose threshold index $y \in [N]$ uniformly
at random.
\While{True}
\State Initialize the memory as: $\ket{\psi_0}=\frac{1}{\sqrt{N}}\sum_{x=0}^{N-1} \ket{x}\ket{y}$.
\State Mark every item $x$ for which $\chi_{y}(x) = 1$.
\State Apply $\qsearch (L, \chi)$ on the first register of $\ket{\psi_0}$.
\State Measure the first register: let $y^{\prime}$ be the outcome. If $\chi_{y} (y^{\prime})=1$, set $y = y^{\prime}$.
\EndWhile\\
\Return $y$.
\EndFunction
\end{algorithmic}
\end{algorithm}

The gate complexity of this algorithm is comprised of the following:
\begin{align*}
    \cost[\qmin] =  \langle n_{\qmin} \rangle ( \cost[Q]  +  \cost[\psi_0] + \cost[\text{meas}])
\end{align*}
Here $\langle n_{\qmin} \rangle$ is the expected number of calls to the oracle $O_{\chi_{i}}$ required for finding the minimum, 
%$\langle n_Q \rangle$ corresponds to the expected number of times we need to apply the Grover operator $Q$ in each iteration of $\qsearch$($L$, $\chi$),
$\cost[Q]$ denotes the cost of the Grover operator, $\cost[\psi_0]$ is the the cost for preparing the state $\ket{\psi_0}$ and $\mathcal{C}[\text{meas}]$ is the cost for performing a measurement. Typically, the measurement process requires one call, whereas the initialization of the memory state can be achieved using $n$ Hadamard gates. The cost of $Q$ and the expected number of applications of $Q$ per iteration can be retrieved from the discussion on $\qsearch$. We follow the work of \cite{cade2022grover} for the following cost analysis:
\begin{lemma}[Cost of $\qmin$ I]  \label{lem:nqmin without timeout}
Given a list $L$ and an oracle $\chi$, the expected number of queries to $\chi$ by $\qmin$ in order to find the minimum is given by:
\begin{align*}
    \langle n_{\qmin} \rangle =  \sum_{s=1}^{N-1} \frac{\langle n_{Q} \rangle(s)}{s+1},
\end{align*}
where $\langle n_{\qsearch} \rangle$ is the expected number of oracle calls made by $\qsearch$. 
\end{lemma}
The detailed proof can be found in Lemma 6 from the appendix of \cite{cade2022grover}.
Note that we consider calls to the quantum oracle rather than to the classical function $\chi$, which would include an extra factor. 
Introducing a timeout to make the algorithm a finite-time algorithm, the expected number of iterations needed to find the minimum changes.
%a little
\begin{lemma}[Cost of $\qmin$ II]  \label{lem:nqmin}
Given a list $L$ and an oracle $\chi$, the expected number of queries to $\chi$ by $\qmin$ in order to find the minimum with probability of success at least $1-\varepsilon$ is given by
\begin{align*}
    \langle n_{\qmin_{finite}}\rangle = \lceil \mathrm{log}_{3}(1/\varepsilon) \rceil 3 \langle n_{\qmin}\rangle,
\end{align*}
%where $\langle n_{\qsearch} \rangle$ is the expected number of iterations of $\qsearch$. 
with $\langle n_{\qmin}\rangle$ as in \cref{lem:nqmin without timeout}. 
\end{lemma}

\subsection{Quantum Amplitude Estimation}
\label{app:Quantum_ampl_estimation}
 \cref{alg:qae} returns an estimate for the probability of measuring a \emph{good} subset of solutions after preparing a given quantum routine, herein we follow an explanation from \cite{Brassard_2002}. 
Let \alg\ be a quantum algorithm that prepares a superposition of $\ket{x}$'s, $\alg \ket{0} = \sum_{x\in X}a_x\ket{x}$. 
Let $\chi:X\rightarrow \{0,1\}$ be a Boolean function, with $X=\{0,1\}^n$, which splits $X$ in a set of \emph{good} solutions, $G=\{x\in X\vert \chi(x)=1\}$, and \emph{bad} solutions, $B=\{x\in X\vert \chi(x)=0\}$. 
Let $p=\sum_{x\in G}\lvert a_x\rvert^2$ be the probability of finding this state in $G$. 
Quantum amplitude estimation returns an estimate $p'$ which, with probability at least $1-\delta$, is $\varepsilon$-close to $p$. 

\begin{algorithm}
\caption{Quantum Amplitude Estimation}\label{alg:qae}
\begin{algorithmic}
\Function{QAE}{unitary \alg, oracle $\chi:X\rightarrow\{0,1\}$, $\delta>0$, $\varepsilon >0$}
\State Prepare $\ket{\psi}=\alg \ket{0}$
\State Run $\qpe(Q, \varepsilon, \delta)$
\Comment QPE is quantum phase estimation,
Q is the Grover operator\\
\Return $\tilde{\theta}$, such that  $p'=\sin^2\tilde{\theta}$
\EndFunction
\end{algorithmic}
\end{algorithm}
\begin{lemma}[Cost of QAE]\label{lem:cost_qae}
The cost of quantum amplitude estimation is given by 
\begin{equation}
\label{cost_qae}
    \cost[\qae(\alg, \chi, \varepsilon, \delta)] = \cost[\alg]+\cost[\qpe(Q, \varepsilon, \delta)]\,.
\end{equation}
%The cost of quantum phase estimation is explained further below, see eq.\ \eqref{eq:qpeCost}.
 $\cost[\qpe(Q, \varepsilon, \delta)]$ is bounded in \cref{eq:qpeCost}, and the cost of an action of $Q$ is
\begin{equation}\label{eq:costQ}
    \cost[Q] = \cost[\chi]+2\cost[\alg]+2\cost_1+\cost_2+2(n-1)\cost_T\,.
\end{equation}
\end{lemma}
\begin{proof}
Let \alg\ be an algorithm that prepares a superposition of $x$'s, $\alg \ket{0} = \sum_{x\in X}a_x\ket{x}$. 
Define $\sin^2{\theta}=p$, where $p$ is the (unknown) probability with which \alg\ prepares a good solution, $p=\sum_{x\in G}\abs{a_x}^2$. 
We can rewrite $\ket{\psi}=\alg\ket{0}$ as 
\begin{equation*}
    \ket{\psi} = \sin{\theta}\ket{\psi_G}+\cos{\theta}\ket{\psi_B}\,,
\end{equation*}
where we have defined 
\begin{equation*}
    \ket{\psi_G}=\frac{1}{\sin{\theta}}\sum_{x\in G}a_x \ket{x}\,, \quad
    \ket{\psi_B}=\frac{1}{\cos{\theta}}\sum_{x\in B}a_x \ket{x}\,.
\end{equation*} 

Consider the operator $Q=R_\psi R_B$ from QSearch in \cref{alg:qsearch}.
$Q$ acts on the states $\ket{\psi}_{G,B}$ as 
\begin{equation*}
   \begin{split}    Q\ket{\psi_G}&=\cos{2\theta}\ket{\psi_G}-\sin{2\theta}\ket{\psi_B}\,,\\
    Q\ket{\psi_B}&=\sin{2\theta}\ket{\psi_G}+\cos{2\theta}\ket{\psi_B}\,.
\end{split} 
\end{equation*}
%from which we can show that $Q$ maps the subspace spanned by $\ket{\psi_G}$ and $\ket{\psi_B}$ onto itself. 
%So in the plane made out of real linear combination of $\ket{\psi_G}$ and $\ket{\psi_B}$, $Q$ simply generates rotations around the origin.
In the plane formed by the real linear combinations of $\ket{\psi_G}$ and $\ket{\psi_B}$, the oracle $Q$ generates rotations around the origin.
To see this, we diagonalize $Q$,
\begin{equation*}
\begin{split}
    Q\ket{\psi_{\pm}}=e^{\mp 2i\theta}\ket{\psi_{\pm}}\,, \quad 
    \ket{\psi_\pm}=\frac{1}{\sqrt{2}}\Bigl(\ket{\psi_B}\pm i \ket{\psi_G}\Bigr)\,.
\end{split}
\end{equation*}
%This establishes that $\theta$ is computable from the eigenvalue of $Q$, and hence quantum phase estimation on $\ket{\psi}, Q$ will produce an estimate of $p$. 
From this follows that we can estimate $\theta$ by applying quantum phase estimation (QPE) on the state $\ket{\psi}$ and the operator $Q$.

The cost of quantum amplitude estimation is then given by 
\begin{equation*}
    \cost[\qae(\alg, \chi, \varepsilon, \delta)] = \cost[\alg]+\cost[\qpe(Q, \varepsilon, \delta)]\,.
\end{equation*}
The first term accounts for the cost of preparing the state $\ket{\psi}$, parameters $\varepsilon$ and $\delta$ set the precision of the estimate of $\theta$ and the probability of success of the algorithm. 
Let $\tilde{\theta}$ be the estimate outputted by QAE, then with probability $p\ge 1-\delta$, we have $\abs*{\tilde{\theta}-\theta}\le \varepsilon$. 

The QPE routine and gate cost are given in \cref{app:Quantum_phase_estimation},  see \cref{lem:qpeCost}. 
It remains to compute the cost of a call to the Grover operator $Q$, where $R_B$ has cost given in \cref{eq:qsearch_Rbcost}, and now $\alg$ used in \cref{eq:qsearch_Rpsicost} to compute $R_{\psi}=\alg R_0\alg$ is given as an arguement. 
%For $R_B$, we assume that the function $\chi$ can be implemented as a quantum oracle (also denoted by $\chi$) with action $\chi\cdot\ket{x}\ket{a}=\ket{x}\ket{a\oplus \chi(x)}$. 
%$a$ is a qubit ancilla we prepare in $\ket{-}$, and $\oplus$ denotes addition mod 2. 
%Then we can implement $R_\chi$ with a phase query to $\chi$. 
%$$\chi\cdot\ket{x}\ket{-}=(-1)^{\chi(x)}\ket{x}\ket{-},$$ 
%so after tracing out the ancilla, we are left with the same action as $R_B$, with cost 
%\begin{equation}
%\cost[R_B]=\cost[\chi]+\cost_1+\cost_2.
%\end{equation}
Altogether, we arrive to \cref{eq:costQ}.
\end{proof}

\subsection{Quantum Phase Estimation}\label{app:Quantum_phase_estimation}
 \cref{alg:qpe} implements quantum phase estimation, a standard result we take from \cite{cleve1998quantum}, computing an $T$-bit estimate of a phase $\phi$.
Let $U$ be a unitary acting on $n$ qubits, and let $\ket{\psi}$ be one of its eigenstates: $U\ket{\psi}=e^{i2\pi\phi}\ket{\psi}$. 
Quantum phase estimation returns an estimate $\phi'$, which with probability at least $1-\delta$ is $\varepsilon$-close to $\phi$.
\begin{algorithm}
\caption{Quantum Phase Estimation}\label{alg:qpe}
\begin{algorithmic}
\Function{QPE}{unitary U, eigenstate $\ket{\psi}$, $\delta>0$, $\varepsilon >0$}
\State Prepare the clock register in state $\ket{\Omega}$
\Comment{$\ket{\Omega}\propto\sum_{\tau=0}^{2^{n_{c}}-1}\ket{\tau},\;n_{c}=\lceil \log_21/\varepsilon+\log_2(1+1/2\delta)\rceil$}{}
\State Apply controlled unitary $\mathcal{U}$
\Comment{$\mathcal{U} = \sum_{\tau=0}^{2^{n_c}-1}\ketbra{\tau}\otimes U^\tau$}{}
\State Apply QFT$^\dagger$
\State Measure the clock register\\
\Return $\tilde{\phi}$
\EndFunction
\end{algorithmic}
\end{algorithm}

\begin{lemma}[Cost of QPE]\label{lem:qpeCost}
The cost of QPE is bounded by
\begin{equation}\label{eq:qpeCost}
    \cost[\qpe(U,\varepsilon,\delta)]\geq n_c\cost_1 +\left(2^{n_c}-1\right)\cost[U]\,.
\end{equation}
%Here $\cost[\qft_{n_c}]$ is the cost of the quantum Fourier transform on $n_c$ qubits, see \cref{eq:cost_qft}, and
where
\begin{equation*}
n_c=\Bigl\lceil\log_2\frac1\varepsilon+\log_2\Bigl(1+\frac{1}{2\delta}\Bigr)\Bigr\rceil\,.
\end{equation*}
\end{lemma}
\begin{proof}
Let be $U$ a unitary acting on $n$ qubits and $\ket{\psi}$ one of its eigenstates, $U\ket{\psi}=e^{i2\pi\phi}\ket{\psi}$. 
%We will first consider the special case in which we know that $\phi=M/T$ holds for some unknown $M\in \mathbb{Z}$ and known $T\in\mathbb{N}$. In this case we can determine $\phi$ exactly with the algorithm we will explain in the following. 
Consider the unitary
\begin{equation*}
    \mathcal{U} = \sum_{\tau=0}^{T-1}\ketbra{\tau}\otimes U^\tau\,.
\end{equation*}
Each power of $U$ acts on an $n$ qubit register and  is controlled by a $T$-qubit register called the \emph{clock}-register. 
Preparing the clock-register in uniform superposition, $\ket{\Omega}=\frac{1}{\sqrt{T}}\sum_{\tau=0}^{T-1}\ket{\tau}$ and acting on $\ket{\Omega}\ket{\psi}$ with $\mathcal{U}$, we obtain
\begin{equation*}
    \mathcal{U}\ket{\Omega}\otimes\ket{\psi} = \Bigl(\frac{1}{\sqrt{T}}\sum_{\tau=0}^{T-1}e^{i2\pi \tau \phi}\ket{\tau}\Bigr)\otimes\ket{\psi}\,.
\end{equation*}
The term in parentheses is the quantum Fourier transform of $\ket{M}$, where $M\in\mathbb{Z}$.
Applying the inverse QFT to the clock register, we get
\begin{equation*}
    \qft^\dagger\cdot \frac{1}{\sqrt{T}}\sum_{\tau=0}^{T-1}e^{i2\pi \tau \phi}\ket{\tau}=\sum_{\tau=0}\Bigl(\frac{1}{T}\sum_{\tau'=0}e^{i2\pi \tau'(\phi-\tau/T)}\Bigr)\ket{\tau}\,.
\end{equation*}
Applying an inverse QFT to the clock register gives us $M$, and then our estimate for $\phi$ is $M/T$.
The maximal precision of our estimate is order $O(1/T)$. 
However there is a non-zero probability of finding an estimate far from $\phi$, which we need to take into account.
Consider the probability of measuring the clock register in state $\tau$
\begin{equation}
    p_\tau = \frac{1}{T^2}\frac{1-\cos[2\pi T(\phi-\tau/T)]}{1-\cos[2\pi(\phi-\tau/T)]}\,.
    \label{eq:probtau}
\end{equation}
The probability $p_\tau$, as a function of $\phi$, is periodic with period 1, $p_\tau(\phi)=p_\tau(\phi+1)$, and even around $\phi=\tau/T$. 
Moreover, we have $p_{\tau+1}(\phi)=p_\tau(\phi-1/T)$, such that the probabilities for different $\tau$'s are equivalent up to a shift in $\phi$.
%In \cref{fig:qpeProb} this probability is plotted as a function of $\phi$, for $\tau=1$, $T=10$.
%\begin{figure}[H]
%    \centering
%    \includegraphics[width=0.75\textwidth]{figures/QPEprob.pdf}
%    \caption{Probability of measuring $\tau=1$ for $T=10$, as a function of $\phi$.}
%    \label{fig:qpeProb}
%\end{figure}
%As expected, we find that the probability of measuring $\tau=1$ is 1, if $\phi=1/T$. Similarly, we find that if $\phi=M/T$ for $M=2,3,\dots$, the probability of measuring $\tau=1$ is 0. If $\phi=1/T+\Delta\phi$ for  a non-zero $\Delta\phi$ with $\abs{\Delta \phi}\ll1/T$, we find that the probability of measuring $\tau=1$, i.e.\ the best approximation given $T$, is strictly smaller than 1. Moreover, using the relation between $p_\tau$ for different $\tau$, we see that we have a small but nonzero probability of measuring any other $\tau\ne 1$.

Intuitively, we expect that selecting a sufficiently large $T$ will ensure a precise estimate with high probability. Let $\phi=M/T+\Delta \phi$, 
for $M\in \mathbb{Z}$, $N\in \mathbb{N}$ and $\Delta\phi\le 1/(2T)$, 
such that $M/N$ is the best approximation for $\phi$ given $T$. Given $\varepsilon$ and $\delta$, 
we want to find $T$ such that we find an estimate $\tilde{\phi}$ with $\abs*{\tilde{\phi}-\phi}<\varepsilon$ with probability $p\ge 1-\delta$. 

Following \cite{cleve1998quantum}, we bound the probability of measuring a $K$ with $\abs{K-M}>E$, for some given $E$. This is the probability of having an estimate differing more than $E/T$ from the optimal estimate, 
\begin{equation}\label{eq:perr}
    p_{err}=\sum^{K=-(M+E-N+1)}_{K=-(M-E)}p_K+\sum_{K=M+E-N+1}^{K=M-E}p_K\,.
\end{equation}
Each probability $p_K$ is given by \cref{eq:probtau}, which we upperbound using the expression $1-\cos{x}\le 2$ for the upper part, respectively $1-\cos{x}\ge 2x^2/\pi^2$ for the lower part of the ratio, obtaining
\begin{equation*}
    p_K \le \frac14 \frac{1}{(M-K+T\Delta\phi)^2}\,.
\end{equation*}
Plugging this in the equation \cref{eq:perr} and changing the summation index to $D=M-K$, we find
\begin{equation*}
\begin{split}
    p_{err}\le&\frac14\sum^{D=-E}_{D=-(T-1-E)}\frac{1}{(D-T\Delta\phi)^2}\\
&+\frac14\sum_{D=E}^{D=T-1-E}\frac{1}{(D-T\Delta\phi)^2}\,.
\end{split}
\end{equation*}
Finally, using $2\Delta\phi<1/T$ and upper bounding the sum with an integral we arrive at
%\AR{I get a factor 2 different from \cite{cleve1998quantum}. To recheck.}\SESS{Cleve et al consider E<D<T-1-E, and also -(T+1)<D<E. (They're also missing a factor of two in (C7)that you can argue cancels out if you do the $t\rightarrow t/2$ substitution in the integral.) But over the larger integration range, our answers match.}
\begin{equation*}
    p_{err}\le \frac{1}{2E-1}\,.
\end{equation*}
Let $\tilde{\phi}=K/T$ be our estimate, then 
\begin{equation*}
    \abs*{\tilde{\phi}-\phi}=\frac{1}{T}\abs{M-K+\Delta\phi}\le \frac{\abs{M-K}+1/2}{T}\,.
\end{equation*}
Thus we have showed that with probability $p=1-p_{err}$, $\abs{K-M}<E$. Setting $p_{err}=\delta$ and $(\abs{M-K}+1/2)/T=\varepsilon$, we find 
\begin{equation}
    T \leq \frac{1}{\varepsilon}\Bigl(1+\frac{1}{2\delta}\Bigr)\,.
    \label{eq:qpe_Tbound}
\end{equation}
Now that we have a bound on $T$, it remains to bound the cost of preparing the clock state and implementing $\mathcal{U}$. 

We specialize to the case where $T=2^{n_c}$ for some $n_c\in\mathbb{N}$, and then set
\begin{equation*}
n_c=\lceil\log_2\frac1\varepsilon+\log_2(1+\frac{1}{2\delta})\rceil\,.
\end{equation*}
The state $\ket{\Omega}$ can be prepared by acting with $H^{\otimes n_c}$ on $\ket{0}$. The unitary $\mathcal{U}$ can be implemented by acting with $U^{2^k}$ on the target qubit controlled on the $k$-th qubit of the clock register, $k=0,1,\dots, n_c-1$. In this case, we have
\begin{equation*}
    \cost[\mathcal{U}]=\sum_{k=0}^{n_c-1}\cost[\contr{1}U^{2^k}]\geq \left(2^{n_c}-1\right)\cost[\contr{1} U].
\end{equation*} 
Where to obtain a lowest cost bound, we neglect the overhead required by the control, i.e. use $\cost[\contr{1} U]\geq \cost[U]$. We also neglect the cost of $\qft$ - the inverse Fourier transform can be replaced with a series of Hadamard gates in the $\qae$ context \cite{Brassard_2002} and more generally can be cheaply approximated \cite{Nam_2020}.
%Taking all of the controlled unitaries into account, we arrive at
%\begin{equation}\label{eq:qpeLow}
%    \cost[\mathcal{U}]\ge \cost[\contr{1}-U]\,.
%\end{equation}
Putting everything together, 
\begin{equation*}%\tag{\cref{eq:qpeCost}}
    \cost[\qpe(U,\varepsilon,\delta)]\geq n_c\cost_1 +\left(2^{n_c}-1\right)\cost[U].
\end{equation*}
%\SESS{I took out all discussion of the upperbound..happy to discuss if needed}
%One strategy to implement $\contr{1}U^{2^k}$ is to act $2^k$ times with $\contr U$, which leads to the following upper bound

%where we have calculated the sum over $k$. 
%Notice that in certain cases, e.g.\ when $U$ is given by time evolution, there might be better way to implement the controlled unitary.
%Counting the sum of each $\cost[\contr U]$ would obtain an upperbound.

%Finally, the cost of QFT is given by \cref{eq:cost_qft}.
\end{proof}

\subsection{Linear Combination of Unitaries}\label{sec:lcu}
Linear combination of unitaries (LCU) is an algorithm originally from \cite{Berry_2015a, Berry_2015} which implements a linear combination of unitary operations on a quantum circuit. 
\cref{alg:lcu} is a version of LCU from \cite{childs2017quantum}, specialized to the quantum linear system problem. We state the result with a fixed success probability. 
\begin{algorithm}
\caption{LCU for QLSA}\label{alg:lcu}
\begin{algorithmic}
\Function{LCU}{state $\ket{b}$, set of coefficients $\{\alpha_i\}$,  corresponding set of unitaries $\{U_i\}$, $\varepsilon >0$, $\Delta$ }
\State $U\gets\sum_{i=0}^{\Delta}\ketbra{i}\otimes U_i$, 
\State Apply $\mathcal{P}_b$, preparing $\ket{b}$ on register $\ket{0^n}$ 
\Comment Neglect step when $\lcu$ is run within $\qaa$ or a similar routine
\State Apply $V$ to register $\ket{0^m}$
\Comment $V\ket{0}=\frac{1}{\sqrt{\alpha}}\sum_{i}\sqrt{\alpha_i}\ket{i}$, $\alpha=\sum_i\alpha_i$
\State Apply $U$
\State Apply $V^{\dag}$\\
\Return $\sum\alpha_iU_i\ket{b}$ prepared on the device
\EndFunction
\end{algorithmic}
\end{algorithm}
\begin{lemma}[LCU, \cite{childs2017quantum}]
\label{C10}
Let $A$ be a Hermitian operator with eigenvalues in a domain $\mathcal{D}\subseteq\mathbb{R}$. 
Suppose function $f: \mathcal{D}\rightarrow \mathbb{R}$ satisfies $\norm{f(x)}\geq1$ for all $x\in\mathcal{D}$  and is $\varepsilon$-close to $\sum_i\alpha_iT_i$ indexed $i=0, \dots, \Delta$ on $\mathcal{D}$ for some $\varepsilon\in(0, 1/2)$, 
$\alpha_i>0$, and $T_i:\mathcal{D}\rightarrow\mathbb{C}$. 
Let $\{U_i\}$ be a set of unitaries such that 
$$U_i\ket{0^t}\ket{\phi}=\ket{0^t}T_i(A)\ket{\phi}+\ket{\Psi_i^{\perp}}$$
for all states $\ket{\phi}$ where $t\in\mathbb{Z}_+$ and $\ket{0^t}\bra{0^t}\otimes \mathcal{I}\ket{\Psi_i^{\perp}}=0.$ 
Given an algorithm $\mathcal{P}_b$ for preparing state $\ket{b}$, 
there is a quantum algorithm that prepares a quantum state $4\varepsilon$-close to $f(A)\ket{b}/\left|\left|f(A)\ket{b}\right|\right|$ with success probability $\frac{1}{\alpha^2}$ and outputs a bit indicating whether it was successful or not. 
The cost is bounded by
\begin{equation}
    \cost[\lcu(\{U_i\}, \{\alpha_i\})]\geq \cost[U]=\sum_{i=0}^{\Delta}\ketbra{i}\otimes U_i.
    \label{eq:LCU_cost}
\end{equation}
%where
%\begin{equation}
%    U=\sum_{i=0}^{\Delta}\ketbra{i}\otimes U_i\,.
%\end{equation}
\end{lemma}

\begin{proof}
Given for coefficients $\alpha_i>0$, unitary operations $U_i$, and index $i=0,\dots, \Delta$, define operation
\begin{equation*}
    M=\sum_{i=0}^{\Delta}\alpha_i U_i\,,
\end{equation*}
Where here $M$ is $\varepsilon$-close to $f(A)$ by assumption.
In general $M$ is not a unitary, hence we can apply this transformation to state $\ket{0^m}\ket{b}$ probabilistically. 
To this end, we define a unitary $V$ acting on an ancillary register such that
\begin{equation*}
V\ket{0}=\frac{1}{\sqrt{\alpha}} \sum_{i=0}^{\Delta}\sqrt{\alpha_i}\ket{i}\,,
\end{equation*}
where $\alpha=\sum_i\alpha_i$, and the controlled unitary 
\begin{equation*}
    U=\sum_{i=0}^{\Delta}\ketbra{i}\otimes U_i.
\end{equation*}
We show that:
\begin{equation*}
    V^{\dagger}UV\cdot (\ket{0^t}\otimes \ket{0^m}\ket{b})=\frac{1}{\alpha}\ket{0}\otimes M\ket{0^m}\ket{b}+\ket{\Psi^{\perp}}\,,
\end{equation*}
where  $(\ketbra{0}\otimes\mathds{1})\ket{\Psi^\perp}=0$. 
The algorithm produces the wanted state $f(A)\ket{b}/\left|\left|f(A)\ket{b}\right|\right|$ with probability $1/\alpha^2$. 
The control register remains in state $\ket{0^t}$, and hence we can use oblivious amplitude amplification (OAA) to boost the success probability to $\mathcal{O}(1)$.
We defer the costs of raising the probability in our routine to algorithms that call LCU as a subroutine.
%This is because for specific problems, one might raise the probability with another technique, such as variable time amplitude amplification or classical postselection.
We assume that the costs of preparing $\ket{b}$ with $\mathcal{P_b}$ are known. 
Likewise, $\alpha$, and $\cost[U]$ should have bounds specific to the given $f$. 
$V$ is interpreted as a state preparation map \cite{childs2017quantum} on $m=\lceil\log (\Delta+1)\rceil$ qubits 
(note that the indexing starts at $0$) and \cite{Shende_2006} gives a procedure for computing this with $(\Delta+1)-2$ 2-qubit gates. 
\begin{equation*}
  \cost[V]\leq \left(\Delta-1\right)\cost_2.
\end{equation*}
Altogether, the cost of LCU is given by 
\begin{equation*}
    \cost[\lcu(\{U_i\},\{\alpha_i\})]=2\left(\Delta-1\right)\cost_2+\cost[U]+\cost{\mathcal{P}_b}\,.
\end{equation*}
We have not argued this is an optimal state preparation scheme for $V$, but instead given a reasonable heuristic allows us to include a cost estimate.
This is in any case not expected to be the dominant term, so it can be dropped to obtain a lowest gate cost bound.
\end{proof}

\subsection{Quantum Linear Solver}\label{sec:qlsa_fourier}
A quantum linear solver (QLS) is a quantum algorithm that can prepare the solution of a system of linear equations as the amplitude of a quantum state, herein an algorithm originally presented in \cite{childs2017quantum}.
QLS is a crucial subroutine within this work, and hence we give an overview of the general strategy for quantum linear system algorithms. Then, we will state the algorithm used along with its lowest bounded cost.
Consider a linear system of equations $Ax=b$, where $A \in \mathbb{C}^{N\times N}$ is a $N \times N$ invertible Hermitian matrix, with $\norm{A}=1$, and $b \in \mathbb{C}^N$ a vector. 
Up to normalization, we can encode the vector $b$ and the solution $x=A^{-1}b$ in quantum states,
\begin{equation*}
    \ket{b}:=\frac{\sum_i b_i|i\rangle}{\| \sum_i b_i|i\rangle \|}\,, \quad \quad|x\rangle:=\frac{\sum_i x_i|i\rangle}{\| \sum_i x_i|i\rangle \|} \,. 
\end{equation*}
The goal of a quantum linear solver is to output a state $\ket{\tilde{x}}$ such that $\norm{\ket{\tilde{x}}-\ket{x}} \leq \varepsilon$. 
We say that the matrix $A$ is $d$-sparse if it has at most $d$ nonzero entries in any row or column. We denote by $\kappa$ be the condition number of the matrix, and we say that the matrix is well-conditioned if $\kappa=\text{poly}(\log N)$. 
For hermitian matrices, the condition number is the ratio between the largest and smallest eigenvalues of the matrix. 
We assume that we have access to an oracle, $\bor$,  which prepares the quantum state $\ket{b}$,  
\begin{equation*}
\bor\ket{0}=\ket{b}.
\end{equation*}
%and two oracles, \sparsor\ and  \mator\ which give access respectively to the location and values of the entries of $A$, 
%\begin{equation}
%\sparsor \ket{i, l}=\ket{i, \nu(i,l)}\,,\quad \mator \ket{i, j}\ket{0}=\ket{i, j}\ket{A_{ij}}\,.
%\end{equation}
%Here $\nu(i,j)$ is the row of the $l$-th nonzero element of column $i$ of $A$. 
%See \cref{app:Grover_Rudolph} and \cref{sec:matrix_oracles} for more details on these oracles, and one possible implementation together with lower bounds on gate cost. 
%In this section, we provide lower bounds in terms of queries to this oracle. 
The authors of \cite{childs2017quantum} provide two approximations: one based on a Fourier representation, the other on Chebyshev polynomials. The basic idea of each is to decompose $A^{-1}$ as a sum of unitaries that can be efficiently implemented. 
We can then use LCU to apply this sum to the state $\ket{b}$, which we prepare invoking $\bor$. 
Herein, we consider only the Fourier version, where $A^{-1}$ is approximated with a sum of exponential terms $e^{iAt}$ which can be efficiently implemented using Hamiltonian simulation. 
\begin{algorithm}
\caption{QlsaFourier}\label{alg:qlsa_four}
\begin{algorithmic}
\Function{QlsaFourier}{matrix $A$, vector $b$, condition number $\kappa$, $\varepsilon >0$}
\State $\Delta_y\gets\frac{\varepsilon}{16}\Bigl[\log\Bigl(1+\frac{8\kappa}{\varepsilon}\Bigr)\Bigr]^{-1/2}$,
 $\Delta_z\gets\frac{2\pi}{\kappa+1}\Bigl[\log\Bigl(1+\frac{8\kappa}{\varepsilon}\Bigr)\Bigr]^{-1/2}$
\State $J\gets\frac{16\sqrt{2}\kappa}{\varepsilon}\log\Bigl(1+\frac{8\kappa}{\varepsilon}\Bigr)$, $
K\gets\frac{\kappa+1}{\pi}\log\Bigl(1+\frac{8\kappa}{\varepsilon}\Bigr)$\,,
\For{$i=0,\dots, J-1$ and $k=-K,\dots, K$}
\State $U_{kj}\gets e^{-iAt_{kj}}$ 
\State $\alpha_{jk}\gets\frac{1}{\sqrt{2\pi}}\Delta_y\Delta_z\abs{z_k}e^{-z_k^2/2}$
\Comment $t_{jk}=j\Delta_y\cdot k\Delta_z$
\EndFor
\State Prepare $\ket{b}$ with $\mathcal{P}_b$
\State  Apply LCU$(\{U_{kj}\}, \{\alpha_{kj}\}, J(2K+1))$
\State Apply OAA(LCU, $\alpha$, $\log_2(J(2K+1))$)\\
\Return $\ket{\psi}$ prepared on device, $\left|\left|\ket{\psi}-A^{-1}\ket{b}\right|\right|\leq \varepsilon$
\EndFunction
\end{algorithmic}
\end{algorithm}
\lemmaQLS*

\begin{proof}
We first find an approximate representation of $1/x$ as a linear combination of phases, $1/x\sim h(x)=\sum_n e^{i\omega_nx}h_n$. 
Then we replace the implementation of $h(A)$ using a LCU. 
The approximation of $1/x$ needs to be accurate only in the domain  $\domk\coloneqq[-1, -1/\kappa)\cup 1/\kappa, 1]$, from \cite{childs2017quantum} once such representation is
\begin{equation}\label{eq:qlsa_four_h}
h(x):=\frac{i}{\sqrt{2 \pi}} \sum_{j=0}^{J-1} \Delta_y \sum_{k=-K}^K \Delta_z z_k e^{-z_k^2 / 2} e^{-i x y_j z_k}\,,
\end{equation}
where $y_j:=j \Delta_y$, with $\Delta_y=y_J/J$, and $z_k:=k \Delta_z$, with $\Delta_z=z_K/K$. 
$h(x)$ is $\varepsilon$-close to $1/x$, for $x\in \domk$, provided we choose appropriate $\Delta_y, \Delta_z, K, J$ values.
%\begin{equation}\label{eq:qlsa_four_par}
%\begin{split}
%    \Delta_y=\frac{\varepsilon}{16}\Bigl[\log\Bigl(1+\frac{8\kappa}{\varepsilon}\Bigr)\Bigr]^{-1/2}\,,\quad J=\frac{16\sqrt{2}\kappa}{\varepsilon}\log\Bigl(1+\frac{8\kappa}{\varepsilon}\Bigr)\,,\\
%    \Delta_z=\frac{2\pi}{\kappa+1}\Bigl[\log\Bigl(1+\frac{8\kappa}{\varepsilon}\Bigr)\Bigr]^{-1/2}\,,\quad K=\frac{\kappa+1}{\pi}\log\Bigl(1+\frac{8\kappa}{\varepsilon}\Bigr)\,,
%\end{split}
%\end{equation}
We briefly summarize the main steps that lead to such choices; more details can be found in the original paper \cite{childs2017quantum}. The goal is to approximate $1/x$ by a Fourier series. Consider the identity
\begin{equation*}
    \frac{1}{x}=\frac{1}{\sqrt{2\pi}}\int_0^\infty \mathrm{~d} y \int_{-\infty}^{\infty} \mathrm{~d} z z e^{-z^2 / 2} e^{-i x y z}\,.
\end{equation*}
Truncating and discretizing the integrals, one arrives at \cref{eq:qlsa_four_h}.
The authors of \cite{childs2017quantum} prove that $h(x)$ is $\varepsilon$-close to $1/x$ in steps. We interpolate between between $1/x$ and $h(x)$ using 3 functions defined as 
\begin{equation*}
\begin{split}
  &h_3(x) \coloneqq \frac{1}{\sqrt{2\pi}}\frac{1}{x}\sum_{k=-\infty}^{+\infty}\Delta_z e^{-z_k^2/2}\,,\\
  &h_2(x) \coloneqq \frac{1}{\sqrt{2\pi}}\frac{1}{x}\sum_{k=-\infty}^{+\infty}\Delta_z e^{-z_k^2/2}\bigl(1-e^{-ixy_Jz_k}\bigr)\,,\\
  &h_1(x) \coloneqq \frac{1}{\sqrt{2\pi}}\frac{1}{x}\sum_{k=-K}^{K}\Delta_z e^{-z_k^2/2}\bigl(1-e^{-ixy_Jz_k}\bigr)\,,
\end{split}
\end{equation*}
such that 
\begin{align*}
    \Bigl\lvert \frac{1}{x}- h(x) \Bigr\rvert \le &\Bigl\lvert \frac{1}{x}-h_3(x)\Bigr\rvert + \lvert h_3(x)-h_2(x)\rvert \nonumber \\
 &+\lvert h_2(x)-h_1(x)\rvert+\lvert h_1(x)-h(x)\rvert\,.
\end{align*}
To have the $\abs{1/x-h(x)}\le \varepsilon$, it is sufficient that each term on the r.h.s.\ is smaller than $\varepsilon/4$. The following bounds for the terms on r.h.s.\ are proven in \cite{childs2017quantum}
\begin{equation*}
\begin{split}
    \Bigl\lvert \frac{1}{x}-h_3(x)\Bigr\rvert&\le 2\kappa\Bigl(\frac{1}{1-\exp\bigl(-2\pi^2/\Delta_z^2\bigr)}-1\Bigr)\,,\\
    \bigl\lvert h_3(x)-h_2(x)\bigr\rvert&\le \kappa\exp\Bigl(-\frac{(\kappa y_J)^2}{2}\Bigr) \\&+2\kappa\Bigl(\frac{1}{1-\exp\bigl[-\frac12\bigl(2\pi/\Delta_z-y_J\bigr)^2\bigr]}-1\Bigr)\,,\\
    \bigl\lvert h_2(x)-h_1(x)\bigr\rvert&\le \frac{4\kappa}{\sqrt{2\pi}}\int_{z_K}^\infty dze^{-z^2/2}\,,\\
    \bigl\lvert h(x)-h_1(x)\bigr\rvert&\le2\sqrt{2}\Delta_y\,.
\end{split}
\end{equation*}
These bounds are valid provided $\abs{\Delta_y z_K /k}=O(\varepsilon)$. Then to ensure that $\abs{1/x-h(x)}\le \varepsilon$, it is sufficient to take 
\begin{equation*}
\begin{split}
    \Delta_y =\frac{\varepsilon}{16}\Bigl[\log\Bigl(1+\frac{8\kappa}{\varepsilon}\Bigr)\Bigr]^{-1/2}\,,\\ J=\lfloor\frac{16\sqrt{2}\kappa}{\varepsilon}\log\Bigl(1+\frac{8\kappa}{\varepsilon}\Bigr)\rfloor\,,\\
    \Delta_z =\frac{2\pi}{\kappa+1}\Bigl[\log\Bigl(1+\frac{8\kappa}{\varepsilon}\Bigr)\Bigr]^{-1/2}\,,\\ K=\lfloor\frac{\kappa+1}{\pi}\log\Bigl(1+\frac{8\kappa}{\varepsilon}\Bigr)\rfloor\,.
\end{split}
\end{equation*}
We can implement $h(A)$ as an LCU,
\begin{equation*}
    h(A) = \sum_{j=0}^{J-1}\sum_{k=-K}^{K}\alpha_{jk}\cdot \sign(k)e^{-iAt_{kj}}\,,
\end{equation*} 
where we have defined
\begin{equation*}
    \alpha_{jk}=\frac{1}{\sqrt{2\pi}}\Delta_y\Delta_z\abs{z_k}e^{-z_k^2/2}\,,\quad t_{jk}=y_jz_k\,.
\end{equation*}
A short calculation leads to 
\begin{equation*}
    \alpha = 2\sqrt{\pi}\frac{\kappa}{\kappa+1}\sum_{k=-K}^{K}\abs{k}\Delta_z e^{-(k\Delta_z)^2/2}\,.
\end{equation*}
The last sum can be calculated numerically, provided we know $\kappa$; it scales as $O(\Delta_z^{-1})$. Using this, we see that $\alpha=O(\kappa\sqrt{\log(\kappa/\varepsilon)})$.

To bring the probability of success from $\frac{1}{\alpha^2}$ to $\mathcal{O}(1)$, we use 
quantum amplitude amplification (QAA), with costs given in \cref{lem:QAA_cost}. 
Note that because the solution we care about is coupled to $\ket{00...0}$ on the ancillary register $V$ is applied on, one can simplify the reflection operator by rotating only around this register. It contains $\lceil\log_2(\Delta+1)\rceil$ qubits (since $i$ index runs $t=0, \dots, \Delta$), where $\Delta=\abs{\{\alpha_i\}}=J(2K+1)$ is the number of terms used in the sum over $i$.
Altogether,
\begin{align*}
    \cost[\qlsa]=&4l\cost_1+2l\cost_2+(2\log_2(J(2K+1))-1)\cost_T \nonumber \\
&+\left(2l+1\right)\cost[\text{LCU}(\{U_i\}, \{\alpha_i\}, J(2K+1))],
\end{align*} 
To lower bound the cost of QLSA, we consider only the last term. The cost of using the LCU is given in \cref{eq:LCU_cost}, specifically 
\begin{align*}
\cost[\text{LCU}(\{U_i\}, \{\alpha_i\})]&=2\left(J(2K+1)\right)\cost_2\nonumber\\
&+\cost[\{U_i\}]+\cost{\mathcal{P}_b},
\end{align*}
%\cost[\mathbf{U}]&\ge\cost[e^{iAy_Jz_K}].
The cost of LCU is dominated by the cost of implementing the controlled unitary $\mathbf{U}$ which comprises all $\{U_i\}$ steps,
\begin{equation*}
    \mathbf{U} = \sum_{kj}\ketbra{kj}\otimes e^{-iAt_{kj}}\,,
\end{equation*}
We lower bound the cost of implementing $\mathbf{U}$ with
\begin{equation*}
    \cost[\mathbf{U}] \geq \cost{(e^{-iAy_Jz_K})}\,,
\end{equation*}
where 
\begin{align*}
    y_J\equiv J\Delta_y=\sqrt{2}\kappa \sqrt{\log(1+\frac{8\kappa}{\varepsilon})}\,,\nonumber \\ 
    z_K\equiv K\Delta_z=2 \sqrt{\log(1+\frac{8\kappa}{\varepsilon})}\,.
\end{align*}
The cost of QLSA can be lower bounded by the cost of Hamiltonian simulation for a time given by $t = y_Jz_K$, and finally
\begin{align}
\cost[\mathbf{U}]&\ge\cost[e^{iAy_Jz_K}]
\end{align}
For Hamiltonian simulation we use the bound given in \cref{lemma:ham_sim_2}.
 With $\gamma  = \varepsilon/ \sqrt{2}d^3t$ and $w$ computed from \cref{eq:k_implicit_def},
\begin{equation*}
\begin{split}
\cost[e^{iAy_Jz_K}]\geq& 10 t (\norm{A}_1-d^2\gamma)w \\
&\cdot \Bigl(\bigl\lceil\log(\norm{A}_{1}/\gamma-d^2)\bigr\rceil -1\Bigr)\cost{_T}.
\end{split}
\end{equation*}
Finally a Toffoli gate can be decomposed in $C_1, C_2$ gates at the expense of small linear factors, hence we set $\cost_T=1$ as a coarse bound.
\end{proof}

\subsection{Quantum Amplitude Amplification}
\label{sec:qaa}
Given a quantum algorithm $\mathcal{A}$ preparing a desired state with bound probability, \cref{alg:oaa} is used to increase the probability of preparing a desired state $\ket{\psi_G}$ to $\mathcal{O}(1)$ in a procedure using the Grover oracle. We mainly follow \cite{Brassard_2002} hierein.

\begin{algorithm}
\caption{Quantum Amplitude Amplification}\label{alg:qaa}
\begin{algorithmic}
\Function{QAA}{unitary \alg, probability $p$, $\mu\in\mathbb{Z_+}$}
\State Prepare $\ket{\psi}$ 
\Comment Ignore step if $\qaa$ is a subroutine 
\State Apply $S^m(\alg, \mu)$
\Comment $m=\lfloor\frac{\pi}{4\arcsin\sqrt{p}}\rfloor$
%\State Measure the state\\
\Return $\ket{\psi_m}=\ket{\psi_G}$ prepared on device with probability $\mathcal{O}(1)$
\EndFunction
\end{algorithmic}
\end{algorithm}

\begin{lemma}[Cost of QAA]\label{lem:QAA_cost}
Assume an initial $n$-qubit state $\ket{\psi}$ that can be prepared from some oracle $\mathcal{P}_b$, and consider unitary operation $\alg$ such that
\begin{equation}
    \alg\ket{\psi}=\sin\theta\ket{\psi_G}+\cos\theta\ket{\Psi_{\perp}}.
    \label{QAA_U_op}
\end{equation}
where $\theta\in(0, \pi/2)$ is determined by the known (or bounded) probability of finding this state in $\ket{0^{\mu}}V\ket{\psi}$, $p=\sin^2\theta$. 
The cost of quantum amplitude amplifiationtion is given by 
\begin{equation}
    \cost[QAA] = m\cost[\chi]+(2m+1)\cost[\alg]+2m\cost_1+m\cost_2+2m(n-1)\cost_T
    \label{eq:QAA_cost}
\end{equation}
where 
\begin{equation*}
    m=\Bigl\lfloor \frac{\pi}{4\theta}\Bigr\rfloor\,,
\end{equation*}
\end{lemma}

\begin{proof}
Let $\chi:X\rightarrow \{0,1\}$, \alg\, $\ket{\psi_G}$, and $\ket{\psi_B}$ be defined as in \cref{app:QSearch}, consider state 
\begin{equation*}
    \ket{\psi} = \alg\ket{0}= \sin{\theta}\ket{\psi_G}+\cos{\theta}\ket{\psi_B}.
\end{equation*}

Using the operators $R_\psi, R_{B}, Q$ defined in \cref{app:QSearch}, one can show $Q$ generates translations in $\theta$. Let $\ket{\psi_j}=Q^j\ket{\psi}$, and then
\begin{equation*}
    \ket{\psi_j}=\sin[(2j+1)\theta]\ket{\psi_G}+\cos[(2j+1)\theta]\ket{\psi_B}\,.
\end{equation*}
So we can repeatedly apply $Q$ to boost the probability with which we measure a good $x$. 

Let $m$ be the number of times we apply $Q$. We want to find the value such that $\ket{\psi_m}$ is close as possible to $\ket{\psi_G}$. If there exists $n\in \mathrm{N}$ such that $(2n+1)\theta=\pi/2$, then we can boost the success probability to 1 by choosing $m=n=\pi/4\theta-1/2$. More generally, if we pick 
\begin{equation*}
    m=\Bigl\lfloor \frac{\pi}{4\theta}\Bigr\rfloor\,,
\end{equation*}
we have that the probability of success is lower bounded by $1-p$. 
To see this consider the probability of failure $p_{fail}=\cos^2[(2m+1)\theta]$, and let $\mu\in\mathbb{R}$ such that $(2\mu+1)\theta=\pi/2$, then
\begin{equation}\label{eq:qaa_pfail}
\begin{split}
    p_{fail}&=\cos^2[2(m-\mu)\theta+(2\mu+1)\theta]\\
    &= \sin^2(2\abs{m-\mu}\theta)\\
    &\le \sin^2\theta =p\,,
\end{split}
\end{equation}
where, in going to the last line, we have used $\abs{m-\mu}\le 1/2$. Notice that if $p\ge 1/2$, $m=0$ and we don't apply $Q$ at all. So the algorithm works with probability of success lower bounded by $\max(1-p, p)$. 

The cost of the algorithm is given by
\begin{equation}
    \cost[QAA] = \cost[\alg]+m\cost[Q]\,,
    \label{QAA_cost}
\end{equation}
where the first terms accounts for the preparation of the initial state, and the second term for the repeated action of $Q$.
%
%We begin by the reflection around the bad solutions, $R_B$. We assume that the function $\chi$ can be implemented as a quantum oracle, which we will also denote by $\chi$, with action $\chi\cdot\ket{x}\ket{a}=\ket{x}\ket{a\oplus \chi(x)}$, where $a$ is a qubit ancilla, and $\oplus$ denotes addition mod 2. Then we can implement $R_\chi$ with a phase query to $\chi$, namely we prepare the ancilla in the state $\ket{-}$. A simple calculation shows that $\chi\cdot\ket{x}\ket{-}=(-1)^{\chi(x)}\ket{x}\ket{-}$. After tracing out the ancilla, we are left with the same action as $R_B$.
%
%The reflection around $\ket{\psi}$ can be implemented as $R_{\psi}=\alg R_0\alg$, where $R_0=2\ketbra{0}-\mathds{1}$. We can implement $R_0$ the same way we implemented $R_B$, provided we have an oracle $\chi_0$ that maps $x=0$ to 0 and all the other $x$'s to 1. This is simply a multi-controlled Toffoli gate followed by an X gate acting on the target qubit. \SESS{Recall \ref{{CU_toff_etc}} for our choice of Toffoli gate decomposition.}

Putting everything together from \cref{eq:qsearch_Rbcost} and \cref{eq:qsearch_Rpsicost}, we find that 
\begin{equation}\label{QAA_oracle_cost}
    \cost[Q] = \cost[\chi]+2\cost[\alg]+2\cost_1+\cost_2+2(n-1)\cost_T\,.
\end{equation}
\end{proof}

\subsection{Oblivious Amplitude Amplification}
\label{sec:oaa}
There are many variants of quantum amplitude amplification, here we consider oblivious quantum amplitude amplification (OAA) from \cite{Berry_2014} along with bound cost. 
To utilize OAA, we assume initial state $\ket{0^{\mu}}\ket{\psi}$, and consider a quantum routine $\mathcal{A}$ that prepares 
$$\alg\ \ket{0^{\mu}}\ket{\psi}=\sin\theta\ket{0^{\mu}}V\ket{\psi}+\cos\theta\ket{\Psi_{\perp}}$$
where  $p=\sin^2\theta$ is the probability of finding this state in $\ket{0^{\mu}}V\ket{\psi}$. 
Oblivious amplitude amplification gives a good solution with probability greater than $\max(1-p,p)$.

\begin{algorithm}
\caption{Oblivious Amplitude Amplification}\label{alg:oaa}
\begin{algorithmic}
\Function{OAA}{unitary \alg, probability $p$, $\mu\in\mathbb{Z_+}$}
\State Prepare $\ket{\psi}$
\Comment Ignore step if $\oaa$ is a subroutine 
\State Apply $S^m(\alg, \mu)$
\Comment{$m=\lfloor\frac{\pi}{4\arcsin\sqrt{p}}\rfloor$}{}
%\State Measure the state\\
\Return $\ket{\psi_m}=\ket{0^{\mu}}V\ket{\psi}$ prepared on device with probability $\mathcal{O}(1)$
\EndFunction
\end{algorithmic}
\end{algorithm}

\begin{lemma}[Cost of OAA]\label{lem:OAA_cost}
Assume an initial $n$-qubit state $\ket{\psi}$ that can be prepared from some oracle $\mathcal{P}_b$, and consider unitary operations $\alg, V$ such that
\begin{equation}
    \alg\ \ket{0^{\mu}}\ket{\psi}=\sin\theta\ket{0^{\mu}}V\ket{\psi}+\cos\theta\ket{\Psi_{\perp}}.
    \label{OAA_U_op}
\end{equation}
where $\ket{\Psi_{\perp}}$ satisfies $\left(\ket{0^{\mu}}\bra{0^{\mu}}\otimes \mathds{I}_n\right)\ket{\Psi_{\perp}}=0$, $V$ is a unitary operator on $n$ qubit space of $\ket{\psi}$, and $\theta\in(0, \pi/2)$ is determined by the known (or bounded) probability of finding this state in $\ket{0^{\mu}}V\ket{\psi}$, $p=\sin^2\theta$. 
The cost of oblivious amplitude estimation is given by 
\begin{equation}
    \cost[\oaa] = 4m\cost_1+2m\cost_2+(2\mu-1)\cost_T+\left(2m+1\right)\cost[\alg]\,,
    \label{eq:OAA_cost}
\end{equation}
where 
\begin{equation*}
    m=\Bigl\lfloor \frac{\pi}{4\theta}\Bigr\rfloor\,,
\end{equation*}
\end{lemma}

\begin{proof}
As in Lemma 3.6 of \cite{Berry_2014}, define two operators 
\begin{equation*}
R=2\left(\ket{0^{\mu}}\bra{0^{\mu}}\otimes \mathbb{I}_n\right)-\mathbb{I}_{\mu+n}
\end{equation*}
and 
\begin{equation*}
S=-\mathcal{A}R\mathcal{A}^{\dag}R.
\end{equation*}
Then for any $l\in\mathbb{Z}_+$, 
\begin{equation*}
    S^l\alg\ \ket{0^{\mu}}\ket{\psi}=\sin\left[(2l+1)\theta\right]\ket{0^{\mu}}V\ket{\psi}+\cos\left[(2l+1)\theta\right]\ket{\Psi_{\perp}}.
\end{equation*}
This result follows by demonstrating that $S$ is a rotation in the subspace spanned by $\{\Psi, \Psi_{\perp}\}$ when $\Psi$ has the form $\ket{0^{\mu}}V\ket{\psi}$.
The success probabiliy is boosed by choosing $m=\Bigl\lfloor \frac{\pi}{4\theta}\Bigr\rfloor$.
%Then, as with other forms of quantum amplitude amplification, we can boost the success probability to 1 by choosing $m=n=\pi/4\theta-1/2$. More generally, set 
%\begin{equation}
  %  m=\Bigl\lfloor \frac{\pi}{4\theta}\Bigr\rfloor\,,
%\end{equation}
We introduce $\nu\in\mathbb{R}$ such that $(2\nu+1)\theta=\pi/2$.
The probability of failure is, following the same reasoning as in \cref{eq:qaa_pfail}, $ p_{fail}\le \sin^2\theta =p$
%\begin{equation}
%\begin{split}
%   p_{fail}&=\cos^2[2(m-\nu)\theta+(2\nu+1)\theta]\\
 %   &= \sin^2(2\abs{m-\nu}\theta)\\
 %   &\le \sin^2\theta =p,
 %  \label{QAA_prob}
%\end{split}
%\end{equation}
%where in the last line we have used $\abs{m-\nu}\le 1/2$. 

Therefore, the algorithm works with probability of success lower bounded by $\max(1-p, p)$. 
The costs are adapted from \cref{sec:qaa}. We get
\begin{equation*}
    \cost[R] = 2\cost_1+\cost_2+(2\mu-1)\cost_T
\end{equation*}
leading to
\begin{equation*}
    \cost[\oaa] = 4m\cost_1+2m\cost_2+(2\mu-1)\cost_T+\left(2m+1\right)\cost[\alg].
\end{equation*}
To put the result in a more standard convention where the algorithm begins in initial state $\ket{0^{\mu+n}}$, we need only to redefine $\alg'=\alg\mathcal{P}_{\psi}$ and $\ket{\psi'}=\ket{0^n}$. 
\end{proof}

\subsection{Hamiltonian Simulation}\label{sec:hamiltonian-simulation}
The objective of Hamiltonian simulation is to approximate the unitary evolution of a $d$-sparse matrix $H$, denoted by $e^{-iHt}$. 
Berry at al. \cite{Berry_2014} provide an algorithm to simulate the Hamiltonian efficiently by reducing it to the fractional-query model. 
We refer to reader to Lemma 3.5 in \cite{Berry_2014} for further details.

\begin{lemma}
\label{lemma:ham_sim_2}
Given a $d$-sparse hermitian matrix $H$ and $\varepsilon \ge 0$, the algorithm from \cite{Berry_2014} simulates $e^{-iHt}$ up to error $\varepsilon$, with cost lower bounded by
\begin{equation}
\label{eq:low_bound_ham_sim}
\cost{(e^{iHt})} \geq 5 t (\norm{H}_1-d^2\gamma)  w \cost{[Q]} ,
\end{equation}
where $t$ is the simulation time and $\gamma  = \varepsilon/ \sqrt{2}d^3t$, and $w$ is the smallest integer such that the following bound holds
\begin{equation}
    \frac{e^w}{w^w} \leq \frac{\varepsilon_{seg}^2}{2}\,,\quad \varepsilon_{seg}=\dfrac{1}{3}\frac{1}{6d^2\lceil \norm{H}_{max}/\gamma\rceil}\dfrac{\varepsilon}{5\gamma t}\,,
    \label{eq:k_implicit_def}
\end{equation} 
with norm $\norm{H}_{max}=\max_{ij}\abs{H_{ij}}$. Finally, $Q$ is an oracle whose
% is given by
%\begin{equation}
%\label{equation:Q_G_jk}
%    Q = \sum_{j,k} \ketbra{j, k} \otimes e^{iG_{jk}}
%\end{equation}
%with $\varepsilon_{seg} = \dfrac{\varepsilon}{5\tau} \cdot \dfrac{1}{3} $. 
cost is lower bounded by
%We lower bound to cost of implementing $Q$ as
\begin{equation}\label{eq:low_bound_Q_ham_sim}
\cost{[Q]} \geq 2 \Bigl(\bigl\lceil\log(\norm{H}_{1}/\gamma-d^2)\bigr\rceil -1\Bigr)\cost{_T}\,,
\end{equation}
where $\cost{_T}$ is the cost of implementing one Toffoli gate. 
\end{lemma}

\begin{proof}
The algorithm relies on reducing Hamiltonian evolution to the fractional-query model. 
We start by explaining how this linkage is established.

To begin with, it is necessary to decompose the Hamiltonian into a sum of $1$-sparse Hamiltonians, $H_j$, 
\begin{equation*}
    H = \sum_{j=1}^{n_c} H_j\,.
\end{equation*}
To accomplish this, one solves an auxiliary edge-coloring problem, as explained in Lemma 4.4 from \cite{Berry_2014}. 
The number of $1$-sparse Hamiltonians required, which corresponds to the number of necessary colors, is bounded as: $d \le n_c \le d^2$. 
%dependent on the classical pre-processing method used to color the edges of the graph, and it is 
%We fix that $n_c = d$. 
%Additional details on this method are available in \cite{Berry_2014}. 
The next step involves decomposing each $H_j$ into 1-sparse Hamiltonians that have eigenvalues of $0$ and $\pi$.
We refer to these Hamiltonians as $G_{jk}$
\begin{equation*}
    \sum_{j=1}^{n_c} H_j = \gamma\sum_{j=1}^{n_c} \sum_{k=1}^{\eta_j}G_{jk} .
\end{equation*}
The exact number $\eta_j$ of matrices $G_{jk}$ necessary to decompose matrix $H_j$ is dependent on the details of the construction, which are outlined in Lemma 4.3 of \cite{Berry_2014}.

In the following analysis, we provide a lower bound for the total number of matrices $G_{jk}$, denoted as $\Omega\equiv\sum_j \eta_j$. 
However, for the time being, we keep this expression in a general form.

The basic Lie product formula can then be applied to approximate the Hamiltonian evolution
\begin{equation*}
    e^{-iHt} \approx (\prod_{j,k} e^{-iG_{jk} \gamma t/r})^r .
\end{equation*}
As shown in Theorem 4.1 from \cite{Berry_2014} this is equivalent to a fractional-query algorithm utilizing the oracle 
\begin{equation}\label{equation:Q_G_jk}
   Q = \sum_{j,k} \ket{j,k} \bra{j, k} \otimes e^{iG_{jk}} .
\end{equation}
The fractional-query cost of this algorithm is $\gamma t\Omega$.
For a definition of fraction-query algorithm and cost, we refer the reader to Definition 1 in \cite{Berry_2014}. 

%multiple oracles $Q_j = e^{-\sum_j \sum_{k=1}^{\eta_j}G_{j,k} }$ for time $t$.
%More generally, this is equivalent to the fractional-query algorithm 

Lemma 3.5 from \cite{Berry_2014} proves that a fractional-query algorithm can be simulated up to an error $\varepsilon_{seg}$ with fractional-query complexity less than $1/5$ using $w$ queries to the oracle $Q$.

In this context, $w$ represents the smallest integer that satisfies the following inequality
\begin{equation*}
\dfrac{e^w}{w^w} \leq \frac{\varepsilon_{seg}^2}{2}.
\end{equation*}
One can show that $w$ is of order $\mathcal{O}(\log 1/\varepsilon_{seg})$.

In order to apply Lemma 3.5 to Hamiltonian simulation, it is necessary to divide the time evolution into smaller segments, 
each with a fractional-query complexity of no more than $1/5$. 
This implies that $5\gamma t\Omega$ segments must be simulated.

To implement the Hamiltonian simulation to a precision of $\varepsilon$, each segment must be approximated to a precision of 
\begin{equation*}
\varepsilon_{seg}=\dfrac{1}{3}\frac{1}{6d^2\lceil \norm{H}_{max}/\gamma\rceil}\dfrac{\varepsilon}{5\gamma t}\,.
\end{equation*}
Here, the factor of $1/3$ is necessary due to the oblivious amplitude amplification step (see \cref{alg:oaa}); and we have used $\Omega\le 6d^2\lceil \norm{H}_{max}/\gamma\rceil$. 
See the proof of Lemma 4.3 in \cite{Berry_2014} for more details. 

Hence, we deduce that the cost of implementing the unitary $(e^{iHt})$ is lower bounded by
\begin{equation}\label{eq:low_bound_ham_sim_1}
\cost{(e^{iHt})} \geq 5 \gamma t \Omega w \cost{[Q]}.
\end{equation}
The next step is to determine a lower bound for both the cost of one query to $Q$ and the value of $\Omega$.

Following the reasoning of the proof of Lemma 4.3 in \cite{Berry_2014}, each $H_j$ is decomposed in at least
\begin{equation*}
\eta_j\ge\frac{\norm{G_X}_{max}+\norm{G_Y}_{max}+\norm{G_Z}_{max}}{\gamma}-1\,.
\end{equation*}
Here $G_X$, $G_Y$, $G_Z$ are matrices obtained selecting from $H_j$ respectively the real off-diagonal entries, the imaginary off-diagonal entries, and the diagonal entries; 
$\norm{M}_{max}$ denotes the largest entry of $M$ in absolute value; and $\gamma$ is a precision parameter which needs to be set to $\gamma  = \varepsilon/ \sqrt{2}d^3t$. 
Clearly we have $\norm{G_X}_{max} + \norm{G_Y}_{max} + \norm{G_Z}_{max} \geq \norm{H_j}_{max}$.
%, because $\norm{G_X}_{max}$ is the largest integer off-diagonal value, $\norm{G_Y}_{max}$ is the largest imaginary value, and $\norm{G_Z}_{max}$ is the largest diagonal value in the matrix.
So we have the following bound 
\begin{equation*}
    \Omega \geq \sum_j\Bigl( \frac{\norm{H_j}_{max}}{\gamma}\ \Bigr)-n_c\,.
\end{equation*}
We can lower bound the term in the sum as follows.
Every element of $H$ that occupies the same column will be in distinct $H_j$ matrices, hence we have 
%Consequently, we can lower bound the number of $H_j$ matrices as
\begin{equation*}
    \sum_{j=1}^{n_c} \norm{H_j}_{max} \ge \norm{H}_1\,,
\end{equation*}
where the 1-norm is defined as the largest column-sum in the matrix, $\norm{H}_1= \max_j \sum_{i} \norm{H_{ij}}$. 
The number of colors $n_c$
is upper bounded by $d^2$. Finally, we obtain
\begin{equation*}
    \Omega \geq  \frac{\norm{H}_{1}}{\gamma}-d^2\,.
\end{equation*}
Plugging this in \cref{eq:low_bound_ham_sim_1}, we get \cref{eq:low_bound_ham_sim}.

The matrices $G_{jk}$ are, up to signs, given by permutation matrices. 
Without carrying out the decomposition explicitly, the best we can do is lower bound their cost by one gate. 
To implement $Q$, we need to take into account the cost of implementing controlled versions of $G_{jk}$. 
The number of controlling qubits is given by $\ceil{\log \Omega}$. 
Following the discussion in \cref{sec:controlled_unitaries}, we arrive to \cref{eq:low_bound_Q_ham_sim}.
\end{proof}

\end{document}